\documentclass{jfm}
\usepackage{lineno,hyperref}
\usepackage{amsmath}
\usepackage{subfigure}
\usepackage{graphicx}
\usepackage[export]{adjustbox}
\usepackage[table,usenames,dvipsnames]{xcolor}
\usepackage[normalem]{ulem}
\usepackage{graphicx}
\usepackage{epstopdf, epsfig}
\usepackage[ruled,vlined]{algorithm2e}

\def\lline{\vrule width14pt height2.5pt depth -2pt}

\def\m1line{\vrule width3pt height2.5pt depth -2pt}

\def\thickline{\vrule width14pt height3.5pt depth -2pt}
\def\bdot{\raise.2em\hbox to .15em{.}}
\def\grayline{\textcolor{gray}{\lline}}

\def\dashed{\m1line\hskip3.5pt\m1line\hskip3.5pt\m1line\thinspace}

\def\dashedgray{\textcolor{gray}{\dashed}}

\def\dotted{\bdot\ \bdot\ \bdot\ \bdot\thinspace}

\usepackage{todonotes}
\def\outline#1{\textcolor{Gray}{}}
\long\def\comment#1{}

\shorttitle{State estimation in turbulent channel flow}
\shortauthor{M. Wang \& T. A. Zaki} 

\title{State estimation in turbulent channel flow from limited observations}
\author{Mengze Wang \and Tamer A. Zaki\corresp{\email{t.zaki@jhu.edu}}}
\affiliation{Department of Mechanical Engineering, Johns Hopkins University, Baltimore, MD 21218, USA}

\begin{document}
	
	\maketitle
	
	\begin{abstract}
		Estimation of the initial state of turbulent channel flow from limited data is investigated using an adjoint-variational approach.
		The data are generated from a reference direct numerical simulation (DNS) which is sub-sampled at different spatiotemporal resolutions.
		When the velocity data are at 1/4096 the spatiotemporal resolution of DNS, the correlation coefficient between the true and adjoint-variational estimated state exceeds 99 percent. 
		The robustness of the algorithm to observation noise is demonstrated.
		In addition, the impact of the spatiotemporal density of the data on estimation quality is evaluated, and a resolution threshold is established for a successful reconstruction.  
		The critical spanwise data resolution is proportional to the Taylor microscale, which characterizes the domain of dependence of an observation location.
		Due to mean advection, either the streamwise or temporal data resolution must satisfy a criterion based on the streamwise Taylor microscale.
		A second configuration is considered where the sub-sampled data are comprised of velocities in the outer layer and wall shear stresses only.  
		The near-wall flow statistics and coherent structures, although not sampled, are accurately reconstructed, which is possible because of the coupling between the outer flow and near-wall motions.
		Finally, the most challenging configuration is addressed where only the spatiotemporally resolved wall stresses are observed.
		The estimation remains accurate within the viscous sublayer and deteriorates significantly with distance from the wall.
		In wall units, this trend is nearly independent of the Reynolds number considered, and is indicative of the fundamental difficulty of reconstructing wall-detached motions from wall data.
	\end{abstract}

	
	\section{Introduction}
	\label{sec:intro}

	Estimation of instantaneous turbulent flows from the assimilation of limited observations is a challenging problem due to the chaotic nature of turbulence \citep{Bewley_review}.
	Given flow-field information with limited resolution, such as PIV data or pressure measurements, there are potentially multiple solutions that satisfy the Navier-Stokes solutions and match the observations.
	In addition, a small error in the initial state or boundary conditions will amplify exponentially in time, and thus the estimated state will diverge from the true one \citep{Deissler1986chaotic}. 
	Adjoint-variational methods address the state estimation problem by constructing an optimal initial condition that generates a trajectory in state space as close to the observations as possible. 
	In this work, we evaluate the accuracy of the adjoint-variational approach for estimating turbulent channel flow and the dependence of estimation quality on the locations and resolution of the observations. 
	
	Three classes of state estimation techniques have been applied to flow problems: linear stochastic estimation (LSE) \citep[e.g.][]{Adrian1988LSE,Naguib_Morrison_Zaki_2010,Jimenez2019LSE}, filtering and smoothing \citep{Stuart2015}.
	LSE utilizes prior knowledge of two-point correlation to estimate the flow state from observations.
    The correlation, which is not always available in practice, sets an upper bound on the estimation accuracy.
    In addition, LSE does not satisfy the Navier-Stokes equations and is not suitable for accurate forecasting beyond the observation time.
	
	Filtering, or sequential, techniques consist of a prediction step which involves marching the governing equations until the observation time, and an update step where the prediction is augmented with observations \citep{Evensen1994EnKF}.
	When the governing equations are linear, the optimal weight for the observations can be analytically derived, and the corresponding method is the so-called Kalman filter which has been adopted for estimating the disturbance of laminar flows \citep{Bewley_part1}.
	For non-linear problems, the weight can be calculated by either linearizing the governing equations (extended Kalman filter) or marching an ensemble of different states in time (ensemble Kalman filter). 
	Both of these methods have been evaluated for estimating turbulent channel flow from wall observations \citep{Bewley_part2,Bewley_part3,Suzuki2017}. 
	Ultimately the accuracy of the filtering techniques is limited because they only focus on fitting the data at one moment rather than a time interval \citep{Bewley_review}.
	Also, the filtered state may not satisfy the Navier-Stokes equations due to observation noise and the difference between estimation and observations.
	
	Smoothing techniques utilize a time series of data to search for the optimal initial condition, boundary conditions and model parameters which ensure that the evolution of the predicted state reproduces available data.
	Therefore, an accurate forecast of the flow evolution beyond observation window is possible.
	This class of techniques is also capable of optimizing sensor placement and weighting in order to achieve the best prediction accuracy \citep[see e.g.][]{Mons2017sensor}.
	\citet{Mons2016} compared three of the most popular smoothing techniques: the adjoint-variational method (refereed to as 4DVar in numerical weather prediction \citep{Dimet1986_4dvar}), the ensemble Kalman smoother, and the ensemble variational method. 
	The objective was to estimate the unsteady free-stream condition for laminar flow around a cylinder, and 4DVar achieved the lowest estimation error, for a specified computational cost. 
	Adjoint techniques were also demonstrated to be viable in transitional \citep{Mao2013,Mao2017} and turbulent flows \citep{Bewley2004,Vishnampet2015}, including for example for estimating scalar sources from remote observations  \citep{Wang_hasegawa_zaki_2019,Cerizza_Zaki_2016}.
	\citet{Wang2019} derived the discrete adjoint of the incompressible Navier-Stokes equations in general curvilinear coordinates and applied it to estimating the turbulent state of circular Couette flow; they demonstrated accuracy of the forward-adjoint relation to within eight significant figures.
	We herein adopt the adjoint-variational approach to examine the influence of available observations on the accuracy of the estimated turbulent fields in channel flow.
	
	Previous efforts in the context of channel flow have all attempted to estimate the entire state from wall observations only, namely the wall stresses and pressure \citep{Bewley2004,Bewley_part1,Bewley_part2,Bewley_part3,Suzuki2017,Hasegawa2020}.
	No matter which method was adopted, the estimated state was only correlated with the true state up to the buffer layer. 
	The literature on wall-bounded turbulence has not, however, examined how the accuracy of turbulence reconstruction changes with spatiotemporal resolution and placement of the observations, e.g.\, if more information about the flow state is available from PIV data.
	Recent state estimation tests in homogeneous isotropic turbulence \citep{Yoshida2005,Eyink2013,PCDL2019,Li2020} demonstrated that the reconstruction of turbulence is successful only when the highest wavenumber $k_m$ of velocity data satisfies $k_m \eta > 0.2$, where $\eta$ is the Kolmogorov scale of the flow.
	In wall-turbulence, however, flow inhomogeneity in the wall-normal direction, the wall-normal dependence of mean advection and the turbulence production all preclude adopting the same criterion from the homogeneous case.  For the same reasons, it is also anticipated that the critical data resolution for reconstructing the turbulence is anisotropic\textemdash a matter that we will explore herein.  
	Our focus is on reconstruction of turbulence at all scales using the nonlinear Navier-Stokes equations, and thus the critical data resolution is more restrictive than the one for designing reduced-order model for flow control \citep{Jones2011,Jones2015}.

	In \S \ref{sec:4dvar}, we introduce the adjoint-variational state estimation algorithm, and provide the details of the flow configuration and problem setup.
	The state estimation results are presented in \S \ref{sec:results}. 
	A benchmark case with sub-sampled volume-data of velocity is analyzed, followed by the effect of observation noise. 
	Then a range of streamwise, spanwise and temporal data resolutions are explored.
	We propose criteria for minimal data required to successfully reconstruct the turbulent state. 
	The possibility of estimating near-wall structures, which are difficult to measure experimentally, from data in the outer region and at the wall is subsequently investigated. 
	At the end of \S \ref{sec:results}, the Reynolds number effect on state estimation is discussed in the context of wall observations.
	The main conclusions that are drawn from these tests are summarized in \S \ref{sec:conclusion}.
	
	\begin{figure}
		\centering
		\includegraphics[width=0.4\textwidth]{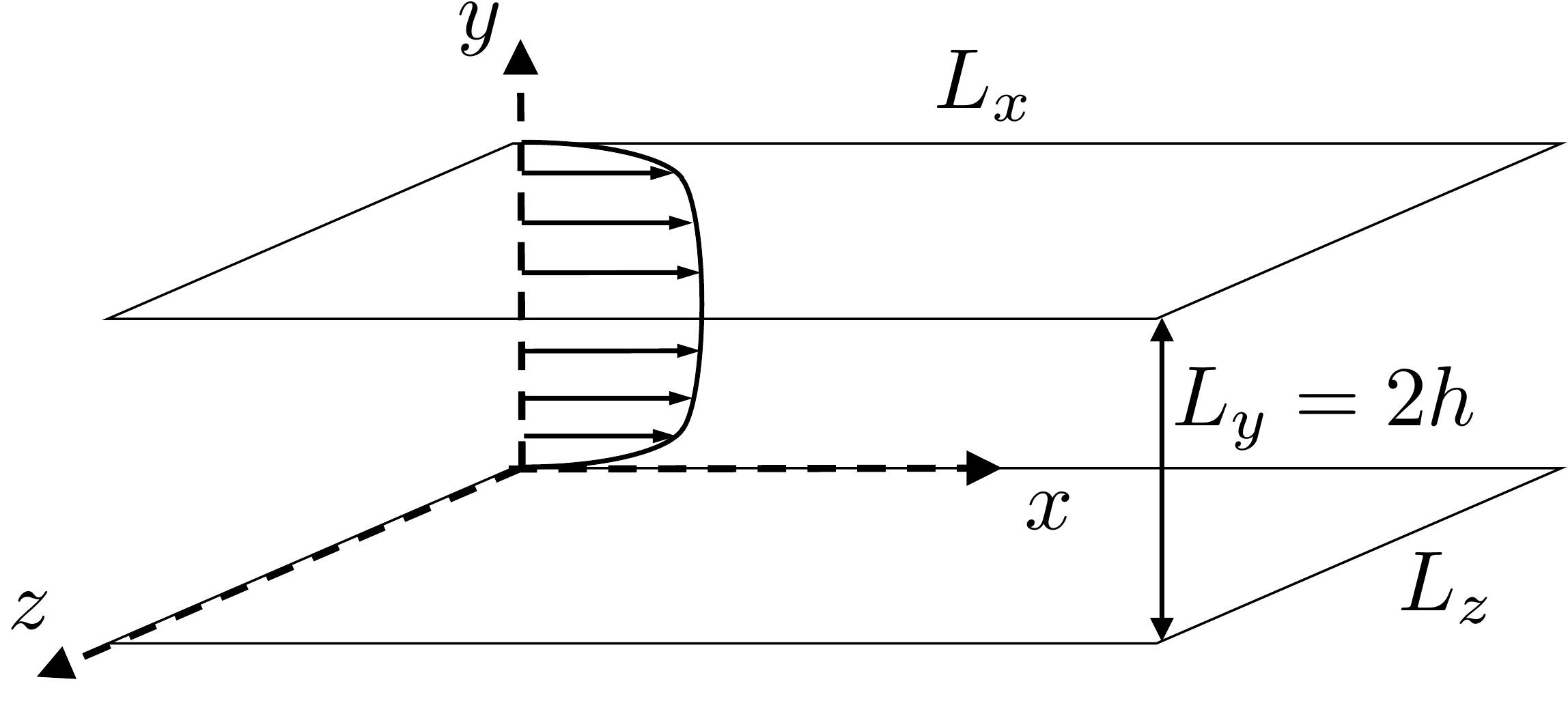}
		\caption{Schematic of channel flow and the coordinate system.}
		\label{fig:setup}
	\end{figure}

	\section{Adjoint-variational state estimation}
	\label{sec:4dvar}
	A schematic of the channel-flow configuration is shown in figure \ref{fig:setup}. 
	The domain is periodic in the streamwise and spanwise directions, and bounded by two no-slip surfaces in the vertical direction.  
	The relevant Reynolds numbers are $Re = U_b h / \nu$ and $Re_{\tau} = u_{\tau} h / \nu$, where $U_b$ is the bulk velocity, $u_{\tau}$ is the friction velocity, $h$ is the half channel height, and $\nu$ is the kinematic viscosity.
	
	The adjoint-variational state estimation is formulated as a constrained optimization problem. 
	The constraint is the numerical model $\boldsymbol u^{n+1} = \mathcal N (\boldsymbol u^n)$, which governs the evolution of the velocity field $\boldsymbol u$ from one time instant $n$ to the next $n+1$.
	The control vector, or the subject of the optimization, is the initial condition $\boldsymbol u^0$. 
	Given the observation data $\{\boldsymbol m \}_{n=0}^N$, we define a cost function,
		\begin{equation}
		\label{eq:cost}
		J(\boldsymbol u^0) = \sum_{n=0}^N \frac 12 \|\boldsymbol m^n - \mathcal M (\boldsymbol u^n) \|_O^2
		\end{equation}
	which is the L-2 norm of the difference between the observation data and their estimation from an initial condition $\boldsymbol u^0$.
	The subscript $O$ represents the observation space, and $\mathcal M$ is an observation operator, which generates the measured quantity from any velocity field. 
	The adjoint model is invoked to calculate the gradient of the cost function, which is necessary for its minimization procedure.   
	The minimizer is the estimated initial condition, and the velocity field marched from this initial condition is the estimated flow. 
	A detailed derivation and validation of the adjoint variational method is provided by \citet{Wang2019}.
	In the following, we briefly summarize the forward model, adjoint equations, and the optimization procedures.

	\subsection{Forward equations and data acquisition}
	\label{sec:forward}
	The flow evolution is governed by the incompressible Navier-Stokes equations, 
	\begin{eqnarray}
	\label{eq:cont_div}
	\nabla \cdot \boldsymbol{u} &=& 0  \\
	\label{eq:cont_mom}
	\frac{\partial \boldsymbol{u} }{\partial t} +  \nabla \cdot \boldsymbol{(uu)} &=& - \nabla p + \frac{1}{Re} \nabla^2 \boldsymbol{u} 
	\end{eqnarray}
	where $t$ is time and $p$ is pressure. These equations are also referred to as the forward model because they are adopted to advance the flow state in time. 
	
	The Navier-Stokes equations are solved using a fractional-step method with a local volume flux formulation on a staggered grid \citep{Rosenfeld1991}.
	The advection terms are discretized by the Adams-Bashforth scheme, and the Crank-Nicolson scheme is adopted for the diffusion terms. 
	The pressure Poisson equation is solved using Fourier transform in the periodic directions and tri-diagonal inversion in wall-normal direction. 
	The algorithm has been applied in a number of direct numerical simulations of transitional and turbulent flows \citep{Zaki2010,Zaki2013,zaki_durbin_2005,Sangjin}.
	For simplicity, the discretized Navier-Stokes equations will be denoted as,
	\begin{equation}
	\label{eq:disc_for}
	\mathbf q^{n+1} = \mathbf G^n \mathbf q^n
	\end{equation}
	where $\mathbf q^n$ is the state vector, including the velocity and pressure at every grid points, and $\mathbf G^n$ is a matrix that represents the discretized Navier-Stokes operator. 
	Note that $\mathbf G^n$ is also a function of $\mathbf q^n$ because the equations are nonlinear. 
	
	\begin{table}
		\centering
		\begin{tabular}{c c c}
			Domain Size & \ Grid points \ & Grid resolution \\
			\begin{tabular}{c c c}
				\hline 
				$L_x/h$ & $L_y/h$ & $L_z/h$ \\
				4$\pi$   &   2         &  2$\pi$
			\end{tabular} & 
			\begin{tabular}{c c c}
				\hline
				$N_x$ & $N_y$ & $N_z$ \\
				384     & 256    & 320
			\end{tabular} & 
			\begin{tabular}{c c c c}
				\hline
				$\Delta x^+$ & $\Delta y^+_{min}$ & $\Delta y^+_{max}$ & $\Delta z^+$ \\
				5.89             &   0.20           &   2.95                     &  3.53
			\end{tabular} \\
		\end{tabular}
		\caption{Domain size and grid resolution.}
		\label{table:setup}
	\end{table}

	The true state is statistically stationary turbulence, and is sustained by a known constant pressure gradient in the streamwise direction.
	Except in \S \ref{sec:Re} where we explore the effect of Reynolds numbers, we set $Re_{\tau} = 180$. 
	While the forward model at these conditions has been extensively studied \citep{Moin_1987,Jelly2014}, this Reynolds number is higher than previously attempted in the context of adjoint-variational state estimation in channel flow.  
	The domain size and the grid resolution are summarized in table \ref{table:setup}. 
	The computational domain is the same as the one adopted by \citet{Moin_1987} who used a pseudo-spectral algorithm. 
	For our finite-volume scheme, we have doubled the resolution in each direction and performed extensive validation \citep[see e.g.][]{Jelly2014}.  
	The grid resolution is also reported in viscous units, denoted by superscript $(\cdot)^+$, $\Delta x^+ \equiv (\Delta x/h)Re_{\tau}$.  The time step size is $\Delta t^+ \equiv (\Delta t U_b/h) (Re_{\tau}^2/Re) = 0.058$ such that the Courant--Friedrichs--Lewy (CFL) number is lower than half. 
	
	\begin{figure}
		\centering
		\includegraphics[width=0.9\textwidth]{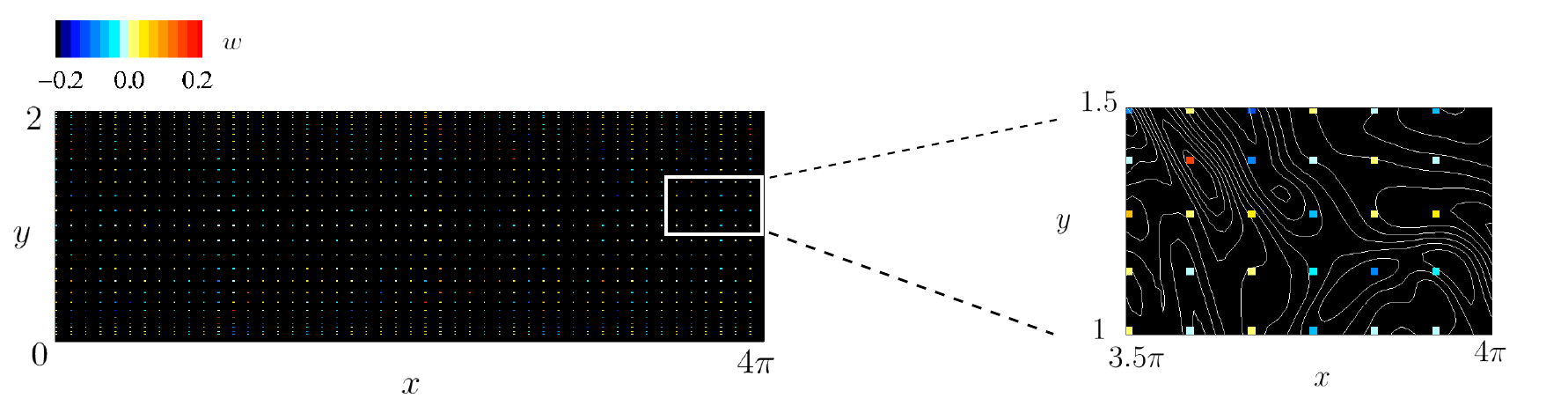}
		\caption{Visualization of observation data of spanwise velocity in $x$-$y$ plane, sub-sampled at one-eighth the DNS resolution.
		Colored regions represent the observation locations, and line contours are the full DNS field.}
		\label{fig:data}
	\end{figure}

	We consider two types of observations: sub-sampled velocity data and stresses on both channel walls.
	The observations setup is summarized in table \ref{table:obs}.
	In all cases, the estimation window is $T = 4.5$ ($T^+ = 50$).
    This choice is motivated by the following considerations: The duration $T$ should be sufficiently long that each point in the fluid is within the domain of dependence of observations.  It should also be longer than the time to ``fill'' the turbulence energy spectra starting solely form observations. Finally, $T$ should not appreciably exceed the Lyapunov timescale \citep[$\tau^+_{\sigma} = 48$ at $Re_{\tau} = 180$ according to][]{Nikitin2018}.
    If $T \gg \tau_{\sigma}$, any infinitesimal perturbation in the initial condition will exponentially amplify and the accuracy of the state estimation deteriorates \citep{Li2020,Chandramouli2020}.
	We start with a benchmark case (\S \ref{sec:benchmark}, case B1), where the velocity field is observed every eighth point in each dimension, including space and time.   
	The velocity at the observation locations is assumed to be known precisely, without measurement noise.
	Subsequently (\S\ref{sec:noise}, cases N1-N2), the data will be contaminated with Gaussian random noise with standard deviation proportional to the local velocity, for example for the streamwise velocity,
	\begin{equation}
	\label{eq:gaussian}
	u_m = u_{true} + \eta, \quad \eta \sim N(0, \sigma |u_{true}|).
	\end{equation}
	The effect of spatial and temporal data resolution is explored in \S\ref{sec:resolution} (cases RZ1$-$RZ4, RX1$-$RX4, RT1$-$RT4), and in \S\ref{sec:wall_layer} the velocity observations in the viscous sublayer and buffer layer are removed.
	Finally, in \S\ref{sec:Re}, we evaluate the influence of Reynolds number when the only observations are the wall stresses.

	\begin{table}
		\centering
			\begin{tabular}{c c c c c c c c c c c c}
				\S & Case & $\Delta M_x $ & $\Delta M_y$ & $\Delta M_z$ & $\Delta M_t$ & $\Delta x_{m}^+$   & $\Delta y^+_{m}$  & $\Delta z_{m}^+$ & $\Delta t^+_{m}$  & $T^+$ & $\sigma(\%)$  \\
				\hline 
				\rowcolor{blue!10}	3.1 & B1 & 8 & 8 & 8 & 8  & 47 & [1.8, 24] & 28 & 0.46 &  50 & 0 \\
				3.2 & N1, N2 & 8 & 8 & 8 & 8  & 47 & [1.8, 24] & 28 & 0.46 & 50 & 5, 10 \\
				3.3 & RZ1$-$4 & 1 & 1 & 4$-$32 & 8  & 5.9 & [0.2, 2.95] & 14$-$112& 0.46 & 50 & 0 \\
				3.3 & RX1$-$4 & 4$-$32 & 1 & 1 & 8  & 24$-$192 & [0.2, 2.95] & 3.53 & 0.46 & 50 & 0 \\
				3.3 & RT1$-$4 & 32 & 1 & 1 & 16$-$128  & 192 & [0.2, 2.95] & 3.53 & 0.92$-$7.4 & 50 &0 \\
		\end{tabular}
		\caption{Parameters of observations relative to the reference DNS. 
		    The sampling rate is $\Delta M$, with a subscript that denotes the spatial or temporal dimension; subscript $m$ denotes the resolution of the observation grid, superscript `+' denotes viscous scaling, $T$ is the observation time horizon, and $\sigma$ is the standard deviation of the Gaussian noise. }
		\label{table:obs}
	\end{table}

	\subsection{Adjoint equations and the state-estimation algorithm}
	\label{sec:adjoint}
	In order to minimize the cost function (\ref{eq:cost}) while satisfying the Navier-Stokes constraint, we introduce the Lagrangian,
	\begin{equation}
	\label{eq:lagrangian}
	L = J - \sum_{n=0}^{N-1} \left(\mathbf q^{\dag (n+1)} \right)^T \left(\mathbf q^{n+1} - \mathbf G^n \mathbf q^n \right).
	\end{equation}
	Note that the Lagrangian is a function of $\left\{ \mathbf q\right\}_{n=0}^N$ and $\left\{ \mathbf q^{\dag}\right\}_{n=1}^N$.
	Taking the derivative of the Lagrangian with respect to $\mathbf q^{\dagger(n)}$ and setting it to zero, we obtain the forward Navier-Stokes equations (\ref{eq:disc_for}).
	By setting the derivatives of the Lagrangian with respect to $\mathbf q^n$ to zero ($1 \leq n \leq N-1$), we obtain the discrete adjoint equations,
	\begin{equation}
	\label{eq:disc_adj}
	\mathbf q^{\dag (n)} = \left(\mathbf G^n\right)^T \mathbf q^{\dag (n+1)} + \frac{\partial J}{\partial \mathbf q^n}, \quad 1 \leq n \leq N-1,
	\end{equation}
	which are marched backward in time, and are forced by the gradient of the cost function with respect to the state. 
	The forward operator $\mathbf G^n$ contains the forward variables $\mathbf q^n$, which means that the full spatiotemporal evolution of the forward fields are required and must be stored for the solution of the adjoint equations;
	The second term on the right-hand side of (\ref{eq:disc_adj}) can be analytically derived from the expression of the cost function.
	When the adjoint equations are marched back to $n=0$, the following relation is obtained, 
	\begin{equation}
	\label{eq:grad}
	\frac{\partial L}{\partial \mathbf q^0} = \left( \mathbf G^0 \right)^T \mathbf q^{\dag 1} + \frac{\partial J}{\partial \mathbf q^0} \equiv \mathbf q^{\dag 0}.
	\end{equation}
	The initial adjoint field is therefore the gradient of the Lagrangian, and also the gradient of the cost function when both the forward and adjoint equations are satisfied,
	\begin{equation}
	    \label{eq:gradJ}
	    \nabla_{\mathbf q^0} J = \frac{\partial L}{\partial \mathbf q^0} = \mathbf q^{\dag 0}.
	\end{equation}
	Note that $\mathbf q^0$ contains the velocity field $\mathbf u^0$ only, because the initial pressure field is not required to solve the incompressible Navier-Stokes equations.
	Similarly, the initial adjoint field $\mathbf q^{\dag 0}$ is comprised of $\mathbf u^{\dag 0} $only.
	Since the above derivation starts from the discrete Navier-Stokes equations, the gradient obtained using the discrete adjoint in equation (\ref{eq:grad}) is accurate to machine precision.
	Detailed expressions of the discrete adjoint and verification the forward-adjoint relation are provided in \citep{Wang2019}.

	With the gradient of the cost function, we adopt the Limited-memory Broyden--Fletche--Goldfar--Shanno (L-BFGS) optimization algorithm to minimize the cost function \citep{LBFGS}. 
	In order to guarantee that the estimated initial condition is divergence free, we slightly modify the L-BFGS algorithm by introducing a symmetric projector.
	The basic idea is to update the new estimate of the initial condition using, 
	\begin{equation}
	\label{eq:update}
	\mathbf u^0_{k+1} = \mathbf P\left(\mathbf u^0_k + \alpha_k \mathbf d_k \right),
	\end{equation}
	where subscript $k$ denotes the $k$-th iteration of the optimization procedure, and $\mathbf d_k = - \mathbf B_k \mathbf P \mathbf u^{\dag 0}$ is the search direction;
	the matrix $\mathbf B_k$ is a rank-2 approximation of the inversion of the Hessian matrix of the cost function;
	and the matrix $\mathbf P$ is a symmetric projector, which projects any velocity field $\mathbf u^0$ or gradient $\mathbf u^{\dag 0}$ onto the divergence-free space.
	The symmetry of $\mathbf P$ ensures that $\mathbf P \mathbf u^{\dag 0}$ is the gradient of the cost function with respect to $\mathbf u^0$ when $\mathbf u^0$ is projected and before being advanced by the forward equations.
	The step size $\alpha_k$ is computed by the line search routine CVRSCH \citep{linesearch}, which enforces the strong Wolfe condition and adopts cubic interpolation to update $\alpha_k$.

	Combining the forward solver (\S\ref{sec:forward}) with the adjoint solver and optimization algorithm (\S\ref{sec:adjoint}), we obtain the adjoint-variational state estimation algorithm.  A summary is provided in Algorithm \ref{alg:adjoint}.  	In all the examined configurations, the algorithm is always performed for one hundred L-BFGS iterations, and as such comparisons are made using the same computational cost.  
	It is important to emphasize that, in addition to the storage requirements associated with saving the forward fields at full spatiotemporal resolution, the computational cost is also substantial because each of the one hundred L-BFGS iterations comprises at least one forward and one adjoint computations.
	
	\begin{algorithm}[h]
		\SetAlgoLined
		\textbf{Step 1}: Forward model\;
		\Indp
		\textbullet~Start with an estimate of the initial condition $\mathbf u^{0^-}$ and project it onto a divergence-free space, $\mathbf u^0 = \mathbf P \mathbf u^{0^-}$ \;
		\textbullet~March $\mathbf u^0$ using the forward equations (\ref{eq:disc_for}) from $n=0$ to $n=N-1$ and store the forward velocity fields at every time step\;
		\textbullet~Evaluate the cost function (\ref{eq:cost})\;
		\Indm
		\textbf{Step 2}: Adjoint model\;
		\Indp
		\textbullet~Solve the discrete adjoint equations (\ref{eq:disc_adj}) from $n=N-1$ to $n=1$\;
		\textbullet~Obtain the gradient of the cost function, $\mathbf u^{\dag 0}$, defined in (\ref{eq:grad})\;
		\textbullet~Project the gradient onto solenoidal space $\mathbf P \mathbf u^{\dag 0}$\;
		\Indm
		\textbf{Step 3}: Update the estimated initial state\;
		\Indp
		\textbullet~Compute the search direction using L-BFGS algorithm\;
		\textbullet~Find an appropriate step size along the search direction\;
		\textbullet~Update the estimate of the initial state and repeat Steps 1-3 until a prescribed maximum number of iterations is reached.\ 
		\caption{Adjoint-variational state estimation.}
		\label{alg:adjoint}
	\end{algorithm}

	\section{State estimation results}
	\label{sec:results}

	\subsection{Performance of the algorithm: the benchmark case}
	\label{sec:benchmark}

\begin{figure}
		\centering
		\includegraphics[width=1.0\textwidth]{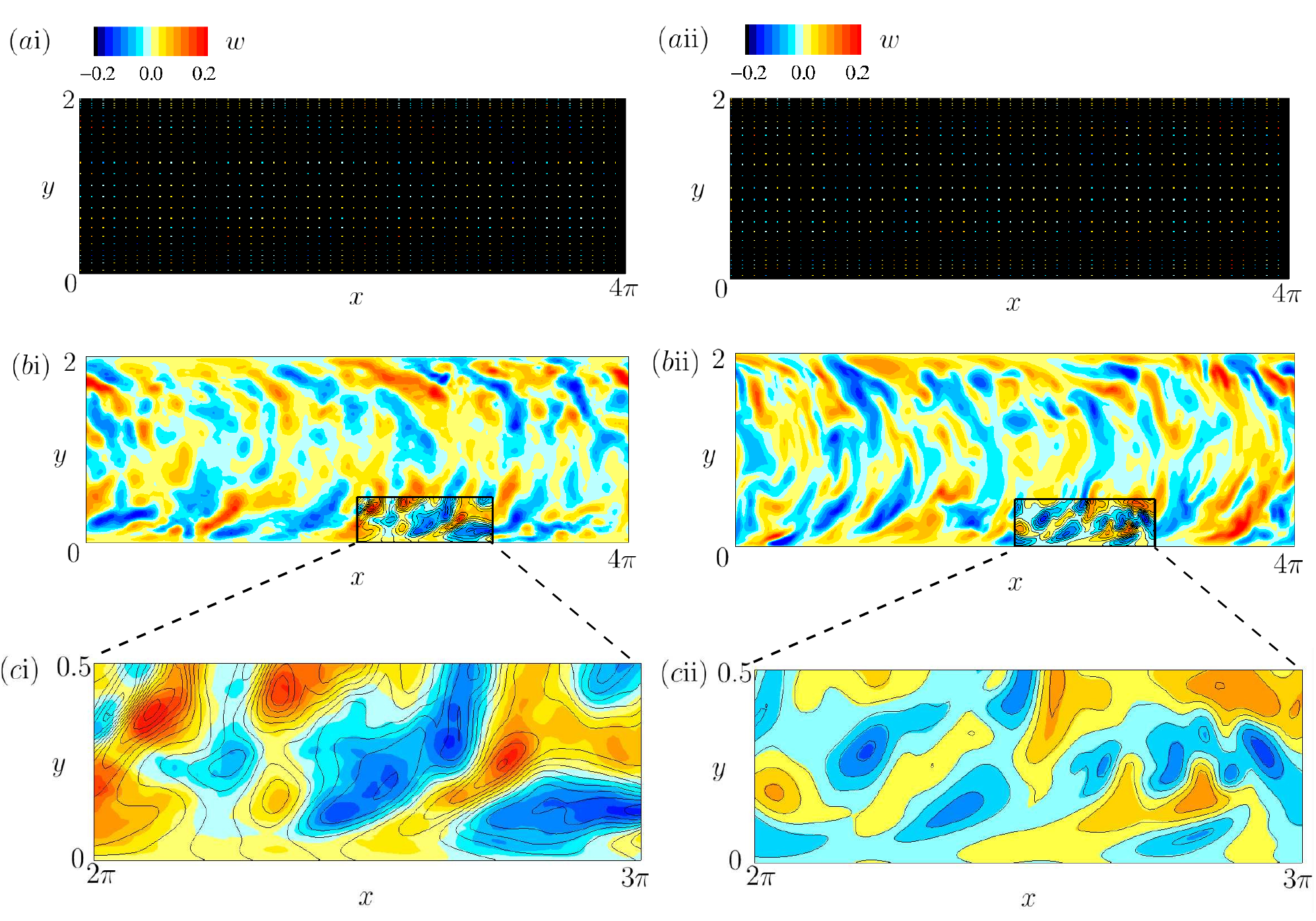}
		\caption{
			Instantaneous spanwise velocity $w$ at $z=L_z/2$ visualized using ($a$) observation data; ($b$) (lines) true fields and (colours) adjoint-variational prediction at (i) $t=0$; (ii) $t=T$ ($T^+ = 50$); ($c$) zoom-in views of ($b$).
		}
		\label{fig:dx8_instant}
	\end{figure}

	In order to provide a qualitative perspective on the performance of algorithm, figure \ref{fig:dx8_instant} shows realizations of the flow from case B1 (see table \ref{table:obs}), at both the (i) initial and (ii) final ($t^+ = T^+ = 50$) times  within the assimilation window.  
	At each instant, the field is visualized using ($a$) the observations data only, ($b$) the adjoint-variational prediction, and ($c$) a detailed comparison of the predicted (colour) and true (lines) states.  
	Recall that the observations from benchmark case B1 are at 1/4096 the resolution of the the simulation, since the velocity is observed at every eighth point in each spatial dimension and in time.
	The quality of the reconstruction is evident in the figure, with the predictions at the initial time capturing the large scales of the flow but appear to contain some small-scale, or high-wavenumber, deviations (panel $c$i).  
	At the final time, in contrast, these small-scale deviations are mostly vanished and the reconstruction quality is, qualitatively, improved (panel $c$ii).
	
	\begin{figure}
		\centering
		\subfigure{
			\includegraphics[width=0.35\textwidth]{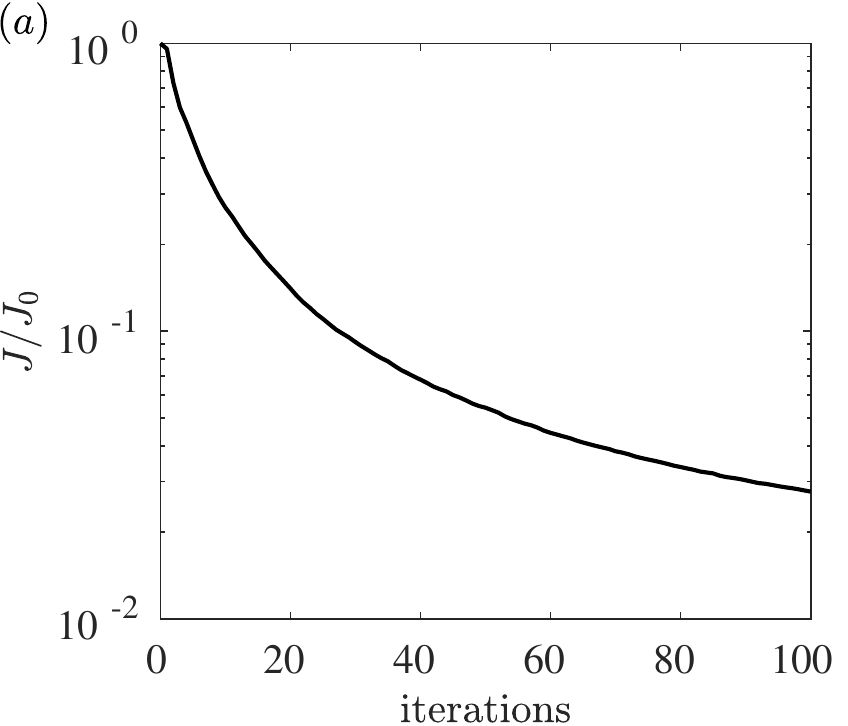}
		}
		\subfigure{
			\hbox{\hspace{0.05\textwidth}} 
			\includegraphics[width=0.35\textwidth]{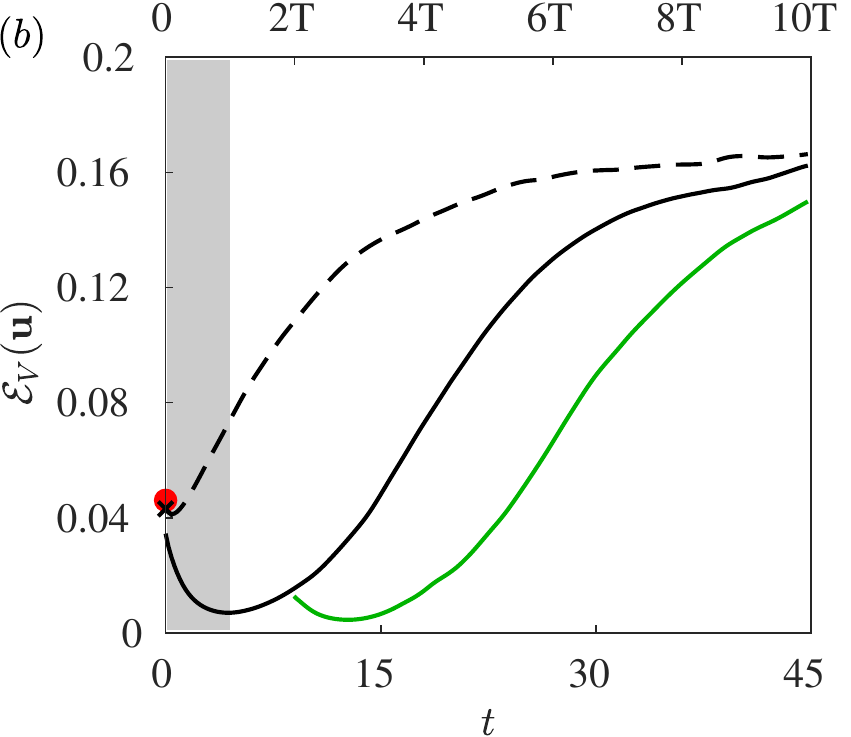}
		}
		\caption{
			($a$) Convergence history of the cost function, normalized by its initial value;
			($b$) Volume-averaged error (\ref{eq:err_t}) of instantaneous velocity field estimated using different strategies:
			(\dashed) Interpolated initial condition (red circle) is projected onto divergence-free space (black cross) and advanced using the Navier-Stokes equations;
			(\lline) Estimated initial condition using data within $t \in [0,T]$ (gray region) is evolved in time;
		    (\textcolor{Green}{\lline}) Estimated state at $t=2T$ using data within $t \in [2T,3T]$ (green region) is evolved in time.
		}
		\label{fig:dx8_error_t}
	\end{figure}
	
	The convergence history of the cost function that generated the assimilated initial condition is shown in figure \ref{fig:dx8_error_t}$a$.  
	The normalization is performed using the cost function associated with advancing the initial guess which was obtained by spline-interpolation of the observations and their projection onto the divergence-free space. 
	The monotonic decrease of the cost function is ensured by the accurate evaluation of its gradient using the discrete adjoint.
	After 100 L-BFGS iterations, the cost function is reduced to $2.8\%$ of its initial value.

	A quantitative assessment of prediction accuracy starts with an evaluation of the root-mean-square errors between the predicted and true states,
		\begin{equation}
		\label{eq:err_t}
		{\mathcal E}_V({\mathbf u})  = \langle \| \mathbf u  -  \mathbf u_{\mathrm {true}} \|^2 \rangle_{V}^{1/2}, 
		\end{equation}
where $\langle \bullet \rangle$ denotes averaging, and the subscript indicates the averaging domain, here over the volume $V$; the same convention is adopted throughout the work for errors $\mathcal E$ and correlation coefficients $\mathcal C$.
The error is plotted in figure \ref{fig:dx8_error_t}$b$ versus time.  
	When errors and correlations coefficients between the predicted and true fields are reported, they are evaluated throughout the domain, and not only at observation locations.  
	At the initial time, the spline-interpolation of the observations (red circle), projection onto the solenoidal space (cross) and the adjoint-variational state all have seemingly similar errors\textemdash the last being obtained after 100 L-BFGS iterations.  
	A mild reduction in the initial errors is achieved by the projection, and a further modest improvement is achieved by the variational approach, but the initial errors remain approximately $4\%$ of the bulk velocity, which is of the same order of magnitude as the root-mean-square fluctuations.  
	The important difference arises, however, when the initial conditions are advanced in time. 
	When the interpolated observations are projected onto the divergence-free field and marched using the Navier-Stokes equations (dashed line), without any data assimilation, the errors amplify as expected due to the chaotic nature of the flow. 
    At long times, after a transient divergence, non-linear effects become dominant and thus the estimation error saturates.
	Simply performing spatio-temporal interpolation of the observations would maintain lower errors, similar to the red circle, although would not be a solution to the governing equations.
	Now consider the errors when the initial condition is obtained from the adjoint-variational state estimation (solid line).
	The initial errors in the assimilated field decay with time, and the flow more closely tracks the trajectory of the true field in state space during the observation window (shaded region, $0 \le t \le 4.5$).
	At $t=T$, the errors are an order of magnitude smaller than those from interpolating and advancing the initial condition or performing spatio-temporal interpolation of observations. 
	
	Beyond the observation window, the estimated state again diverges from the truth but remains more accurate than evolving the interpolated initial field during the interval $[T, 3T]$.  
	These results demonstrate the potential of the algorithm to provide a better prediction of the future state $t>T$, when observations are not available. 
	If new data do become available at later times, the estimated state can be adopted as the initial guess, and the same variational procedure can be applied to drive the estimated state towards the true state, again.  
	This idea is demonstrated in the figure:  new observation data were provided in the interval $t \in [2T, 3T]$, which is marked by the light green region.  The adjoint variational algorithm was applied in that new interval, and the predicted state at $t=2T$ now yields a new trajectory that more closely follows the true flow (green line).
	While this point was noteworthy, the focus of the present effort is on characteristics of the state estimation in the first window $t \in [0, T]$, which are generally applicable to any subsequent interval of new observations.   
	
	\begin{figure}
		\centering
		\subfigure{
			\includegraphics[width=0.32\textwidth]{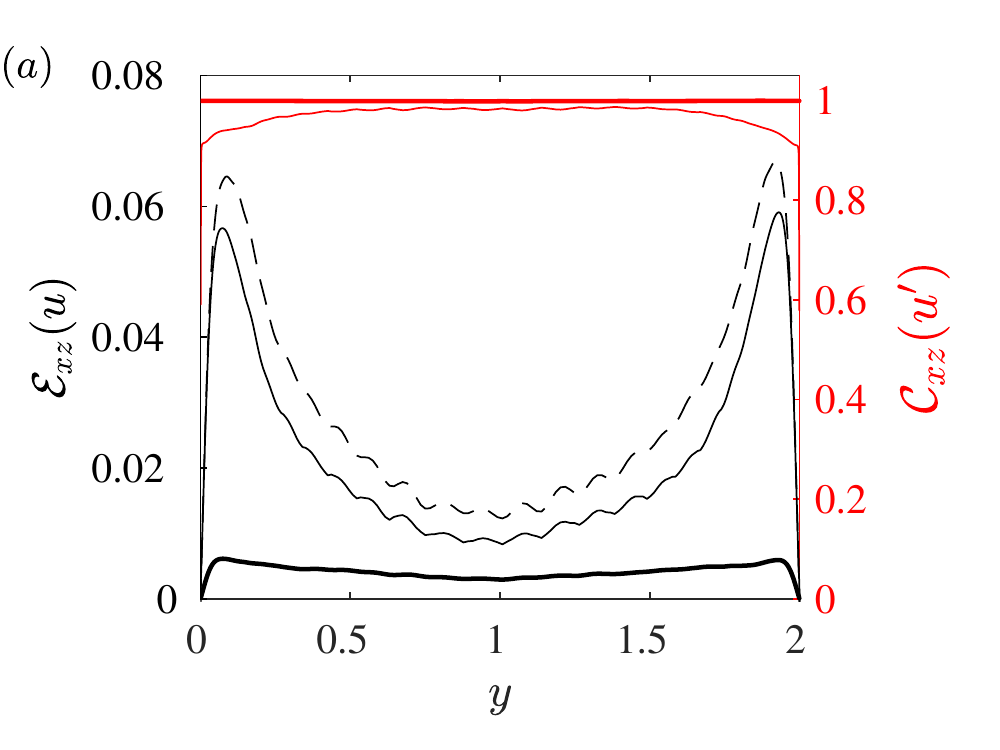}
		}
		\subfigure{
			\hbox{\hspace{-0.25in}} 
			\includegraphics[width=0.32\textwidth]{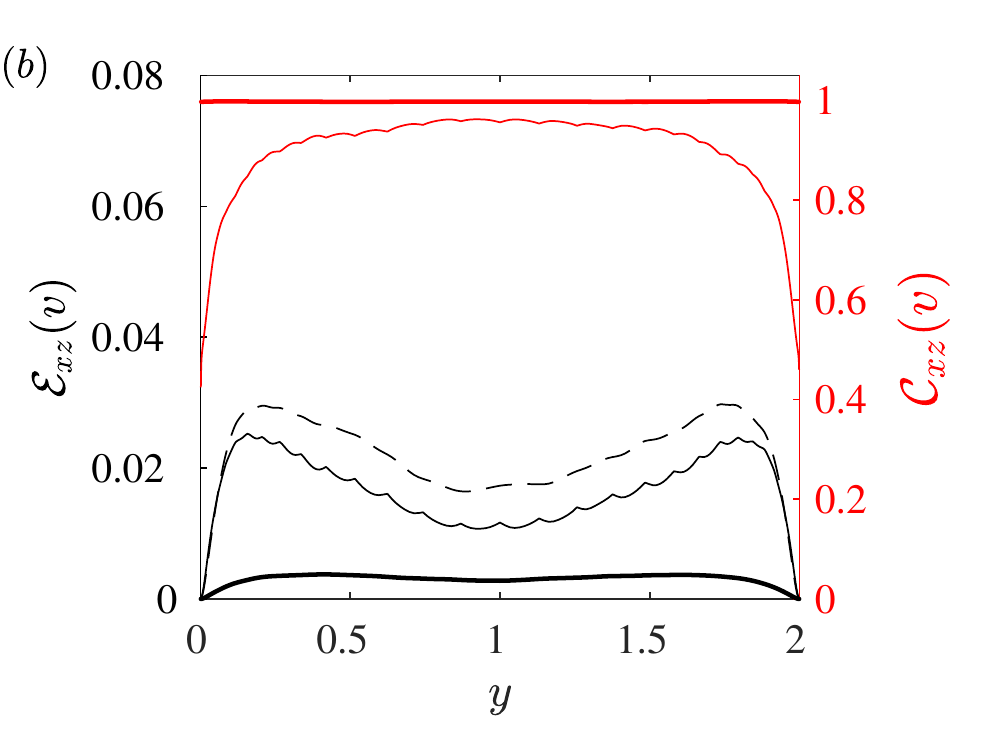}
		}
		\subfigure{
			\hbox{\hspace{-0.25in}} 
			\includegraphics[width=0.32\textwidth]{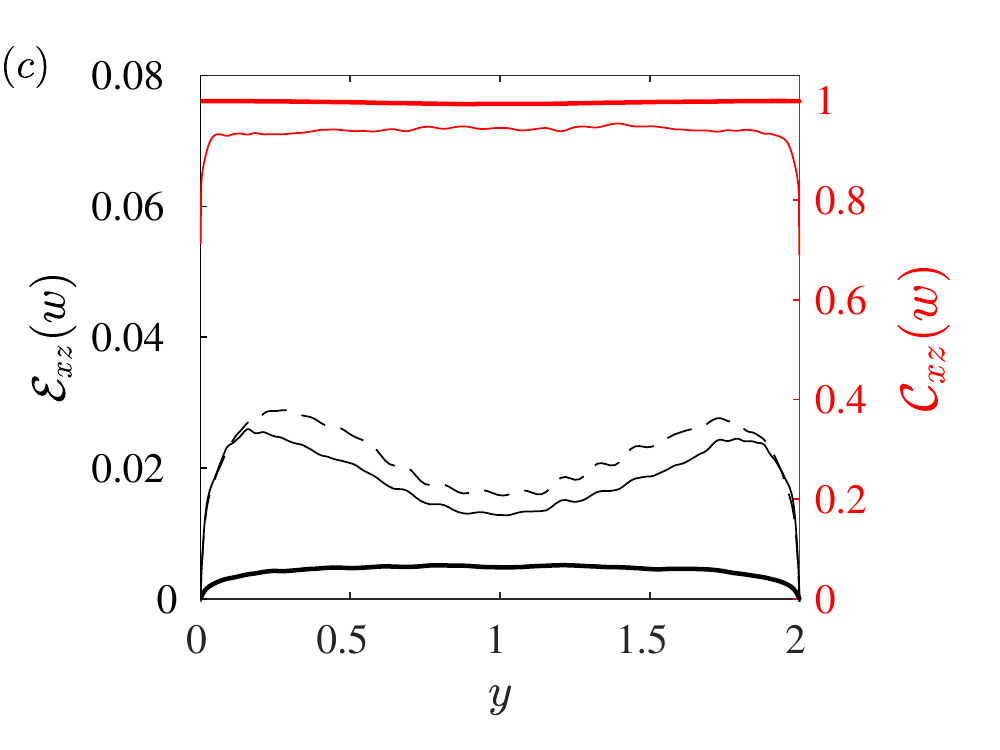}
		}
		\caption{
			Wall-normal profiles of (black curves) horizontally averaged errors $ \mathcal E_{xz} (q) $ and (red curves) correlation coefficient $\mathcal C_{xz}(q)$ (\ref{eq:cc_y}) between the estimated and true flow states: 
			($a$) streamwise, ($b$) wall-normal and ($c$) spanwise components. 
			(\dashed) Spline interpolation of observed velocity field at $t=0$;
			(\lline, \textcolor{Red}{\lline}) adjoint-variational estimation at $t=0$; 
			(\thickline, \textcolor{Red}{\thickline}) estimation at $t=T$ ($T^+ = 50$).}
		\label{fig:dx8_error_y}
	\end{figure}

	The root-mean-square estimation error is evaluated in the horizontal plane, $ \mathcal E_{xz} (q)$, are plotted as a function of wall-normal direction  in figure \ref{fig:dx8_error_y}.
	The errors in the interpolated initial guess (dashed lines) are proportional to the level of physical fluctuations in the velocity field and, as a result, the errors in the streamwise component are most dominant especially in the near-wall region where $u$-perturbations are most energetic.  
	The estimated initial condition (thin black line) is slightly improved relative to the interpolated state.
	The key observation is, however, at $t=T$ where all three components of errors in the estimated state are an order of magnitude more accurate than advancing the interpolated state using the forward model.
	For the streamwise component, the error is lower than $0.5 \%$ of the bulk velocity, or equivalently, $8\%$ of the peak value of root-mean-square streamwise fluctuation.
	Figure \ref{fig:dx8_error_y}$b$ also shows the correlation coefficient, 
		\begin{equation}
		\label{eq:cc_y}
		\mathcal C_{xz}(q)  = \frac{ \langle q \ q_{\mathrm{true}}\rangle_{xz} } 
		{ {\langle q^2 \rangle_{xz}^{1/2}} { \langle q_{\mathrm{true}}^2 \rangle_{xz}^{1/2}} };
		\end{equation}
	At $t=T$, the estimated and the true fields are nearly perfectly correlated, which highlights the capacity of the assimilated field to reproduce the true trajectory of the flow in state space.  
	
	\begin{figure}
		\centering
		\includegraphics[width=\textwidth]{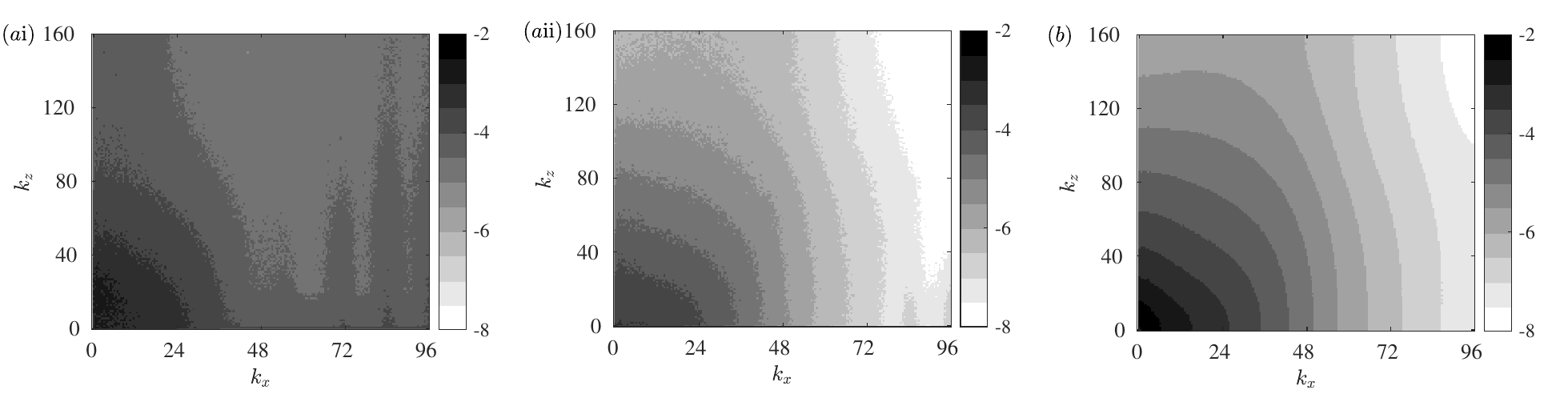}
    	\caption{
    		($a$) Fourier spectrum of errors from adjoint-variational estimation, averaged in $y$ and normalized by the bulk velocity, $\log_{10}\mathcal E_y (\hat{\mathbf u})$. 
    		($a$i) $t=0$ and ($a$ii) $t=T$ ($T^+ = 50$);
    		($b$) Spectrum of the true velocity field averaged in time and wall-normal direction, $\log_{10} \langle \| \hat {\mathbf u}_{\mathrm {true}} \|^2 \rangle^{1/2}_{yt}$. 
    		Wavenumbers $(k_x,k_z)$ are normalized by half-channel height.
    	}
		\label{fig:dx8_error_spectra}
	\end{figure}

	In order to explain the improvement in accuracy during the observation time-horizon, we evaluate the spectra of the estimation error, 
		\begin{equation}
		\label{eq:err_k}
		\mathcal E_y (\hat{\mathbf u}) = \langle \| \hat {\mathbf  u}  -  \hat {\mathbf u}_{\mathrm {true}} \|^2 \rangle_y^{1/2},
		\end{equation}
	where $\hat {\mathbf u}$ is the Fourier transform of $\mathbf u$ in the streamwise and spanwise directions. 
	The spectra of the errors are reported in figure \ref{fig:dx8_error_spectra}$a$ as a function of the horizontal $(k_x, k_z)$ wavenumbers; also shown in panel $b$ is the spectrum of the true velocity field.
	%
	Errors appear largest in the low-wavenumber modes, but they should be viewed relative to the high energy content in these modes in figure \ref{fig:dx8_error_spectra}$b$.
	Normalized by the spectral density in the true field, low wavenumbers are better reconstructed, since they are encoded in the sparse observations.
	A more important observation is the behaviour of the high-wavenumber components of the errors.
	The estimated initial condition (panel $a$i) has appreciable errors in those wavenumbers (higher than the interpolated state\textemdash not shown). 
	However, since most of these modes decay with time (panel $a$ii), they have little impact on the estimation quality later within the assimilation time horizon. 
	In addition, due to their rapid decay, these high-wavenumber initial errors do not appreciably affect the value of the cost function which is integrated over the entire observation window. 
	As a result, they persist in the initial condition during the optimization procedure. 
	An effective strategy to reduce these initial high-wavenumber errors is to incorporate time-dependent weights in the cost function, that amplify the contribution of early observations near $t=0$ \citep{Wang2019}.
	One should caution, however, that not all the initial high-wavenumber errors are stable and decaying;
	Small components of that error are unstable and amplify at the Lyapunov rate, and although they are not perceptible within the present time horizon, they will dominate at longer times.

	\begin{figure}
		\centering
		\includegraphics[width=\textwidth]{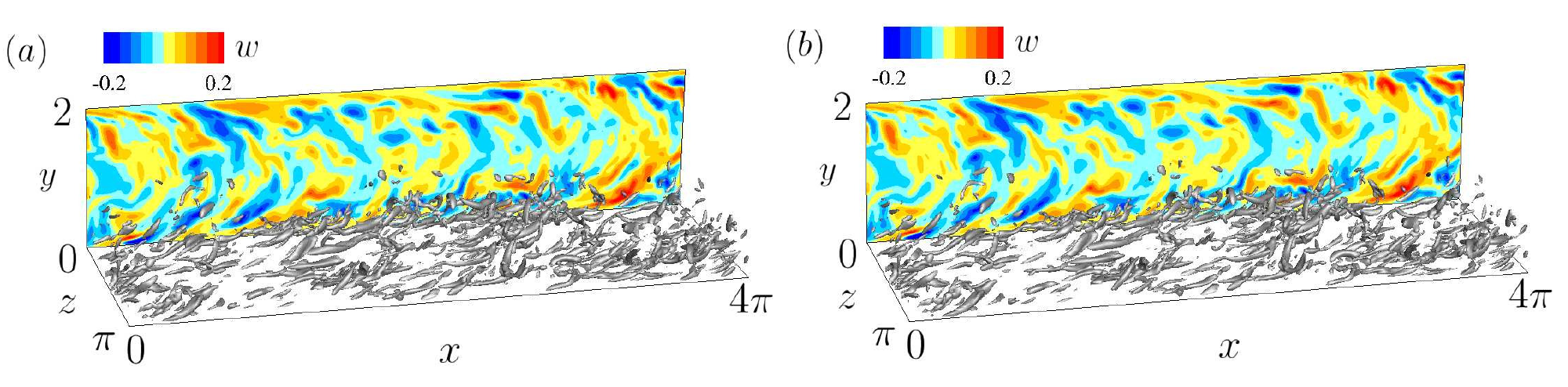}
		\caption{
				Comparison of ($a$) true and ($b$) adjoint-variational estimated state at $t=T$ ($T^+ = 50$).
				Gray isosurfaces: vortical structures visualized using the $\lambda_2$ vortex-identification criterion with threshold $\lambda_2 = -2$;
				Side-view: contours of spanwise velocity.
		}
		\label{fig:dx8_lambda}
	\end{figure}

	Consistent with the above integral and spectral measures of the estimation quality, the reconstructed velocity field at $t = T$ is almost identical to the true one (side views in figure \ref{fig:dx8_lambda}).
	An accurate estimation of vortical structures is, however, more challenging since the computation involves velocity gradients.
	These structures are visualized using the $\lambda_2$ vortex identification criterion \citep{lambda2} and compared in figure \ref{fig:dx8_lambda} (gray isosurfaces).
	%
	%
	The prediction of vortical structures is very compelling, both in the near-wall and outer regions. 
	In the former, the vortical tubes are attached to the wall and elongated in the streamwise direction; 
	Farther from the wall, the lifted vortical tubes break down and generate small-scale structures that are mostly captured by the estimated state.
	In practice, reconstructing the vorticity field from coarse-resolution experimental data is a challenge \citep{Suzuki2012}.
	The satisfactory estimation quality in panel ($b$) demonstrates the potential of our algorithm to augment under-resolved turbulent data from experiments.

	\subsection{The influence of noise in the observation data}
	\label{sec:noise}
	In the previous section, the observation data were free of any noise.
	In practice, however, experimental measurements invariably contain errors and, as such, may violate the governing equations, lead to statistical errors, and severely preclude accurate evaluation of derivatives especially in turbulent flows where strong velocity gradients are expected \citep[see e.g.][]{Bardet2010PIV}.
	For example, \citet{Abrahamson1995} assessed the ability of conventional circulation and least-square methods to reproduce the vorticity field from randomly perturbed DNS velocity field.
	When $5\%$ noise was superposed onto the fully-resolved velocity data, the uncertainty in the computed vorticity field reached $40\%$, and most of the small-scale structures were absent in the reconstruction.
	
	In order to evaluate the performance of adjoint-variational algorithm with noisy data, we contaminate the observed velocities by Gaussian random noise with standard deviation that is proportional to the local velocity component (\ref{eq:gaussian}).
	The spatiotemporal resolution of the data and the estimation window $T^+ = 50$ remain the same as in the benchmark case.
	We consider three levels of noise: $\sigma = \{0, 5, 10\}\%$, which correspond to cases B1, N1 and N2 in table \ref{table:obs}.
	The first guess of the initial condition is interpolation of of the noisy data at $t=0$.
	
	\begin{figure}
		\centering
		\subfigure{
			\includegraphics[width=0.32\textwidth]{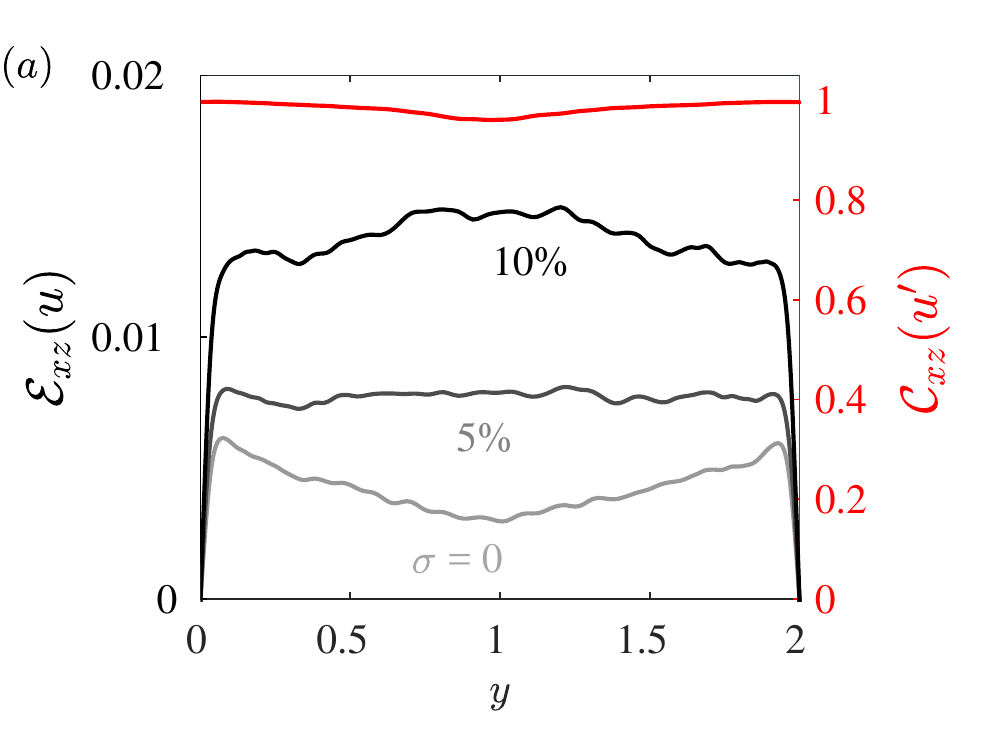}
		}
		\subfigure{
			\hbox{\hspace{-0.25in}} 
			\includegraphics[width=0.32\textwidth]{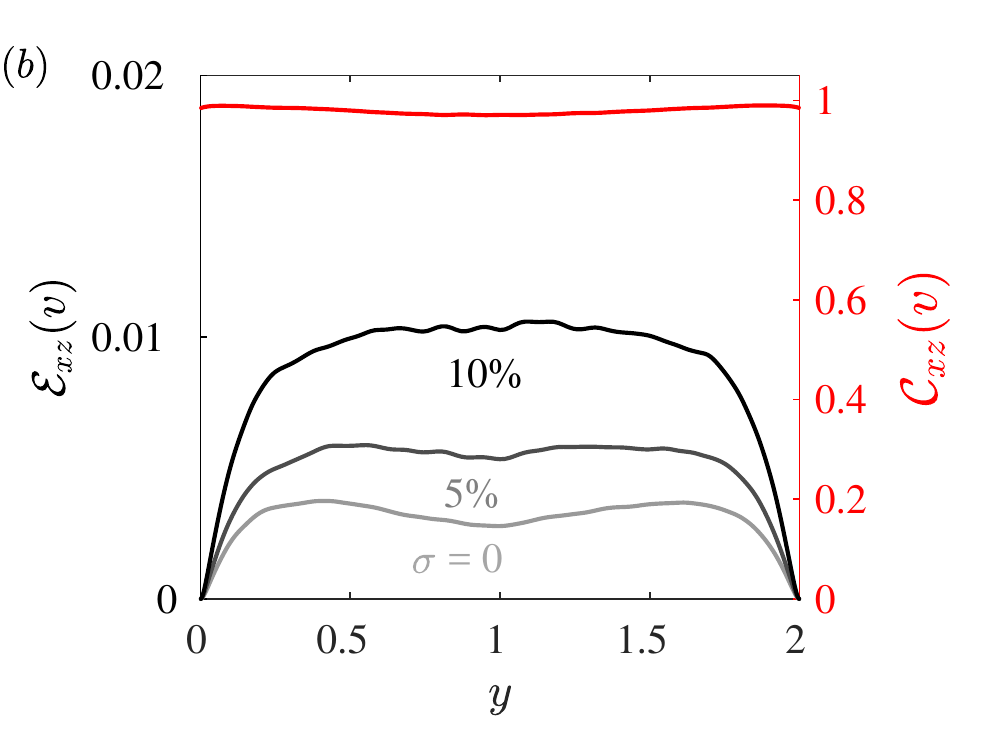}
		}
		\subfigure{
			\hbox{\hspace{-0.25in}} 
			\includegraphics[width=0.32\textwidth]{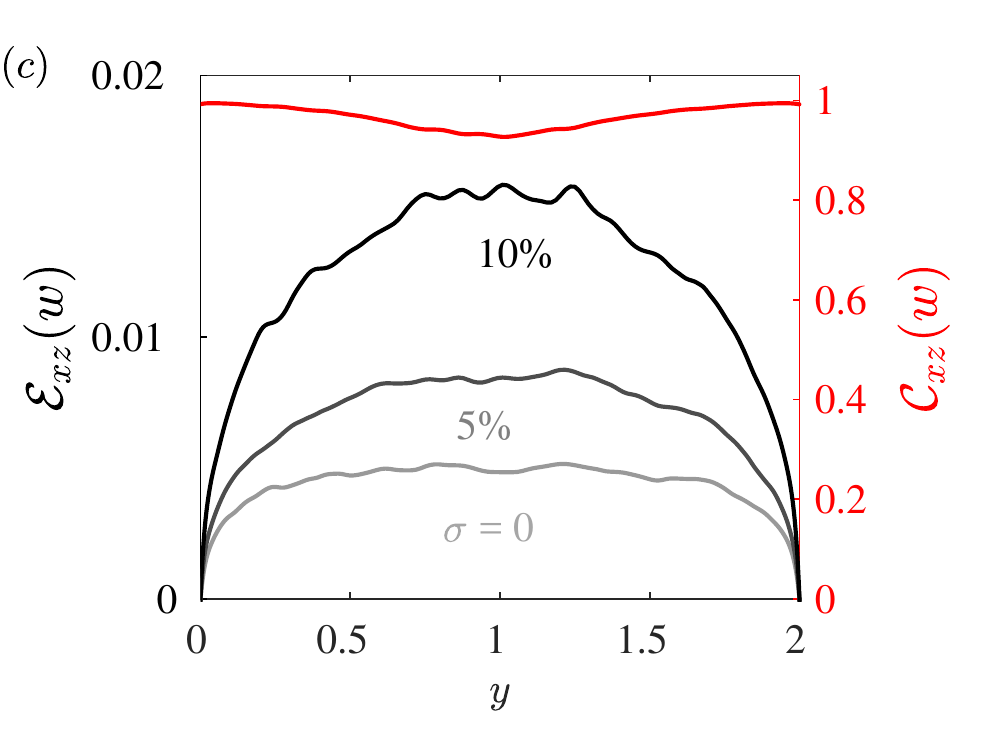}
		}
		\caption{
			Effect of observation noise level on (gray to black: $\sigma = \{0, 5, 10\}\%$) estimation error $\mathcal E_{xz}(q)$ and (red, $\sigma = 10\%$) correlation coefficient $\mathcal C_{xz}(q)$ at $t=T$ ($T^+ = 50$):
			($a$) streamwise, ($b$) wall-normal and ($c$) spanwise components. 
		}
		\label{fig:noise_error}
	\end{figure}
	
	\begin{figure}
		\centering
		\includegraphics[width=\textwidth,trim=0.1in 0.2in 0.1in 0.2in,clip]{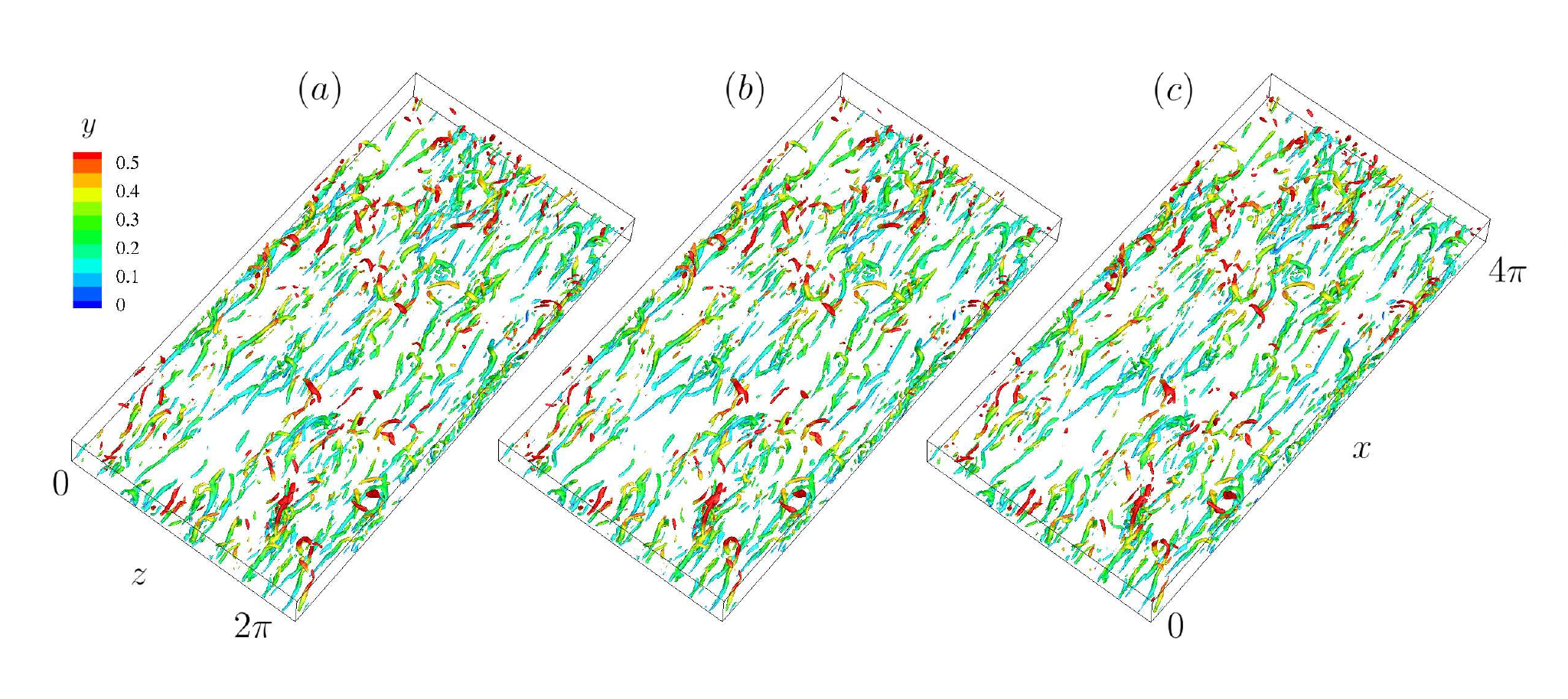}
		\caption{
			Comparison of ($a$) true vortical structures and ($b$,$c$) the adjoint-variational reconstruction when $\sigma = \{5, 10\}\%$ at $t=T$ ($T^+ = 50$) within the bottom half channel, visualized using the $\lambda_2$ vortex-identification criterion with threshold $\lambda_2 = -2$. 
		}
		\label{fig:noise_lambda}
	\end{figure}
	
	\begin{figure}
    	\centering
    	\subfigure{
    		\includegraphics[width=0.33\textwidth]{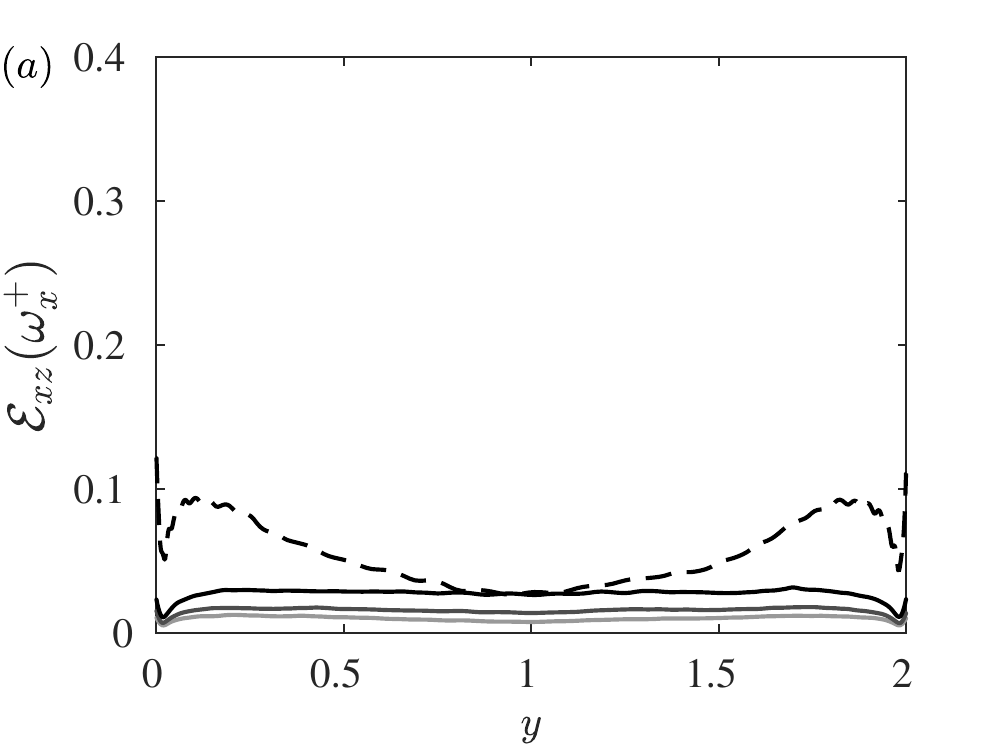}
    	}
    	\subfigure{
    		\hbox{\hspace{-0.25in}} 
    		\includegraphics[width=0.33\textwidth]{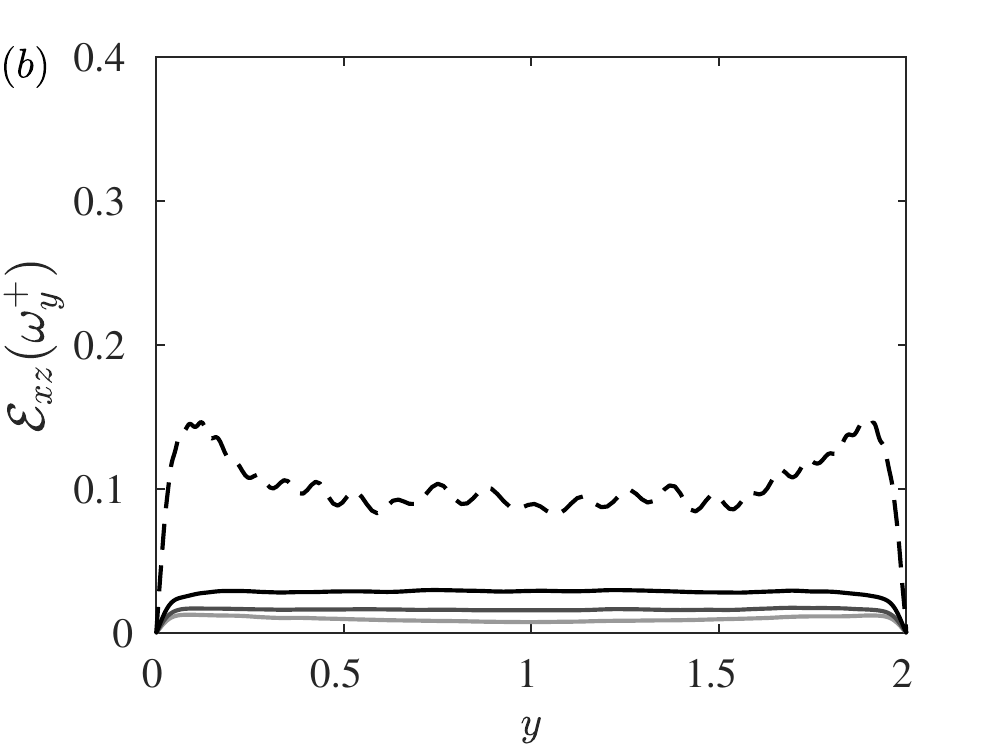}
    	}
    	\subfigure{
    		\hbox{\hspace{-0.25in}} 
    		\includegraphics[width=0.33\textwidth]{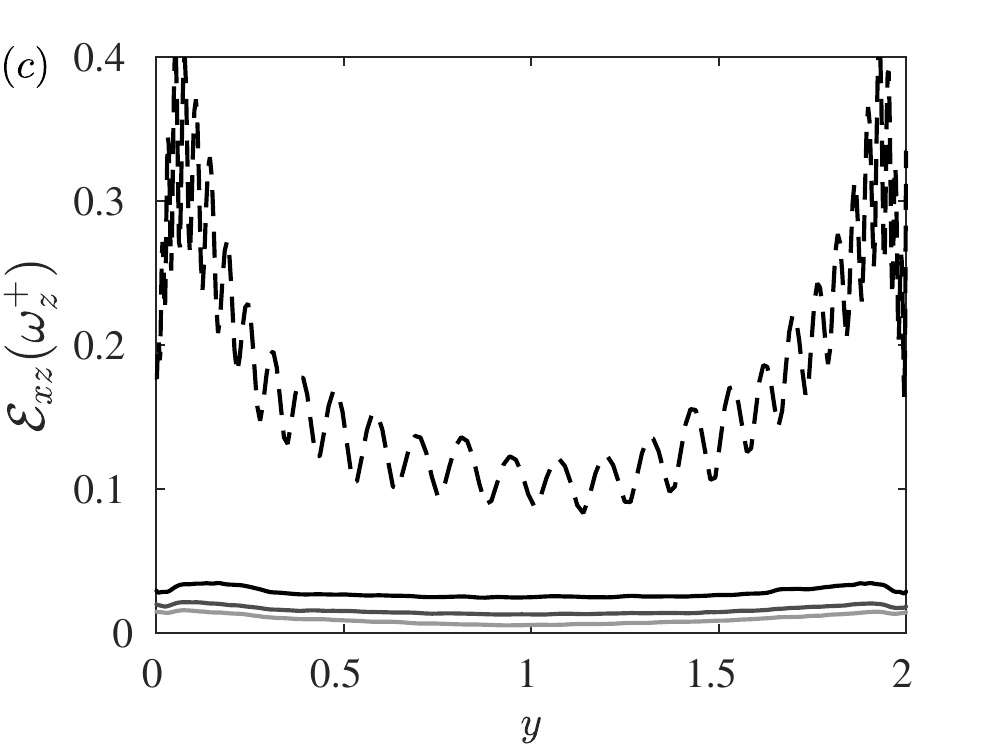}
    	}
    	\caption{
    		Horizontally averaged error $\mathcal E_{xz}(q)$ of vorticity field at $t=T$ ($T^+ = 50$), estimated by (\dashed) interpolating noisy velocity data and (\lline) adjoint-variational approach (gray to black: noise level $\sigma = \{0, 5, 10\}\%$):
    		($a$) streamwise, ($b$) wall-normal and ($c$) spanwise components. 
    		The error is normalized by mean vorticity $\langle du / dy\rangle $ at the wall.
    	}
    	\label{fig:noise_error_omega}
    \end{figure}

	After 100 L-BFGS iterations, the cost function is decreased to $\{2.8 , 27 , 45\}\%$ for the three levels of contamination.  
	Since the cost function is defined as the difference between the estimated and contaminated data, and the latter do not satisfy the Navier-Stokes equations, the cost function cannot decrease to zero.
	The root-mean-square error of the estimated state relative to the true, uncontaminated flow field was evaluated and shows a similar decay from the initial to the final time as in the benchmark case without noise.  
	Therefore, only the results at $t=T$ are examined here (figure \ref{fig:noise_error}).
	The estimation error increases with the noise level (from light gray to black lines), but remains within $2\%$ of the bulk velocity.
	Note that due to the mean flow, the observation noise in the streamwise direction can exceed $\sigma$ times the bulk velocity, which means that the estimation accuracy of the streamwise velocity actually exceeds the precision of observation data.
	Comparatively, the estimation errors of the wall-normal and spanwise velocity components are bounded by the observation noise, which is approximately $\sigma$ times the root-mean-square turbulence fluctuations.
	Overall, even with the highest noise level ($\sigma = 10 \%$), the correlation coefficient (\ref{eq:cc_y}) between the estimated and true state is close to unity at all the $y$ locations, as shown by the red lines in figure \ref{fig:noise_error}$b$.

	The reconstructed vortical structures are visualized in figure \ref{fig:noise_lambda}.
	Although some of the small-scale structures are not captured, most of the reconstructed wall-attached and detached vortical structures remain nearly identical to the true flow, and the estimation quality is almost independent of the noise level.
	A quantitative assessment of the quality of the vorticity field is provided in figure \ref{fig:noise_error_omega}.
	When noisy observations are interpolated and vorticity is evaluated, the error (dashed lines) in the near-wall region reaches $10-40\%$ of the mean vorticity at the wall. 
	The error of the vorticity field from adjoint-variational estimation (black solid lines) is within $4\%$ of the wall vorticity, and the estimation accuracy is robust against observation noise.
	These results demonstrate the superiority of adjoint-variational approach for evaluating velocity gradients and its robustness to observation noise.

	\subsection{The effect of spatial and temporal data resolution}
	\label{sec:resolution}
	
	\begin{figure}
		\centering
		\subfigure{
			\includegraphics[width=0.45\textwidth,trim=0.1in 0.55in 0.15in 2.5in,clip]{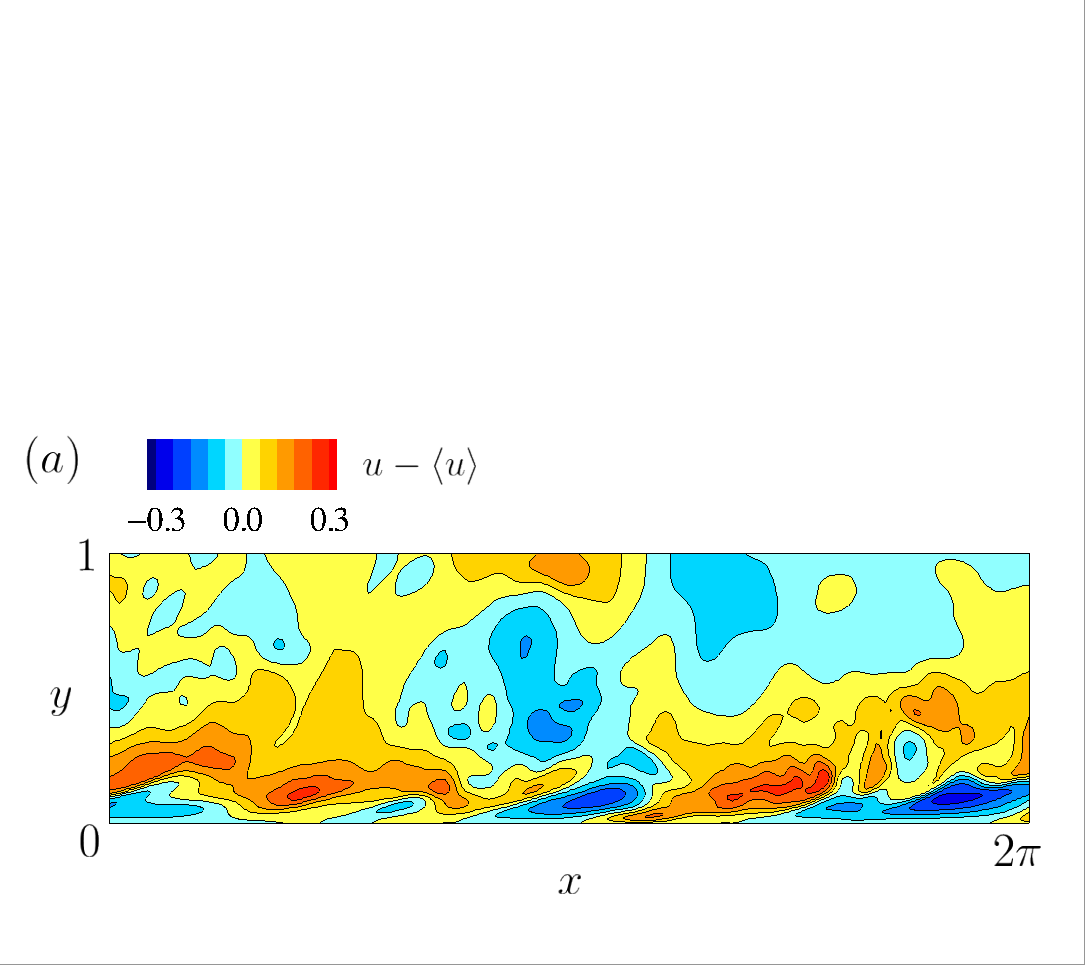}
		}
		\subfigure{
			\includegraphics[width=0.45\textwidth,trim=0.1in 0.55in 0.15in 2.5in,clip]{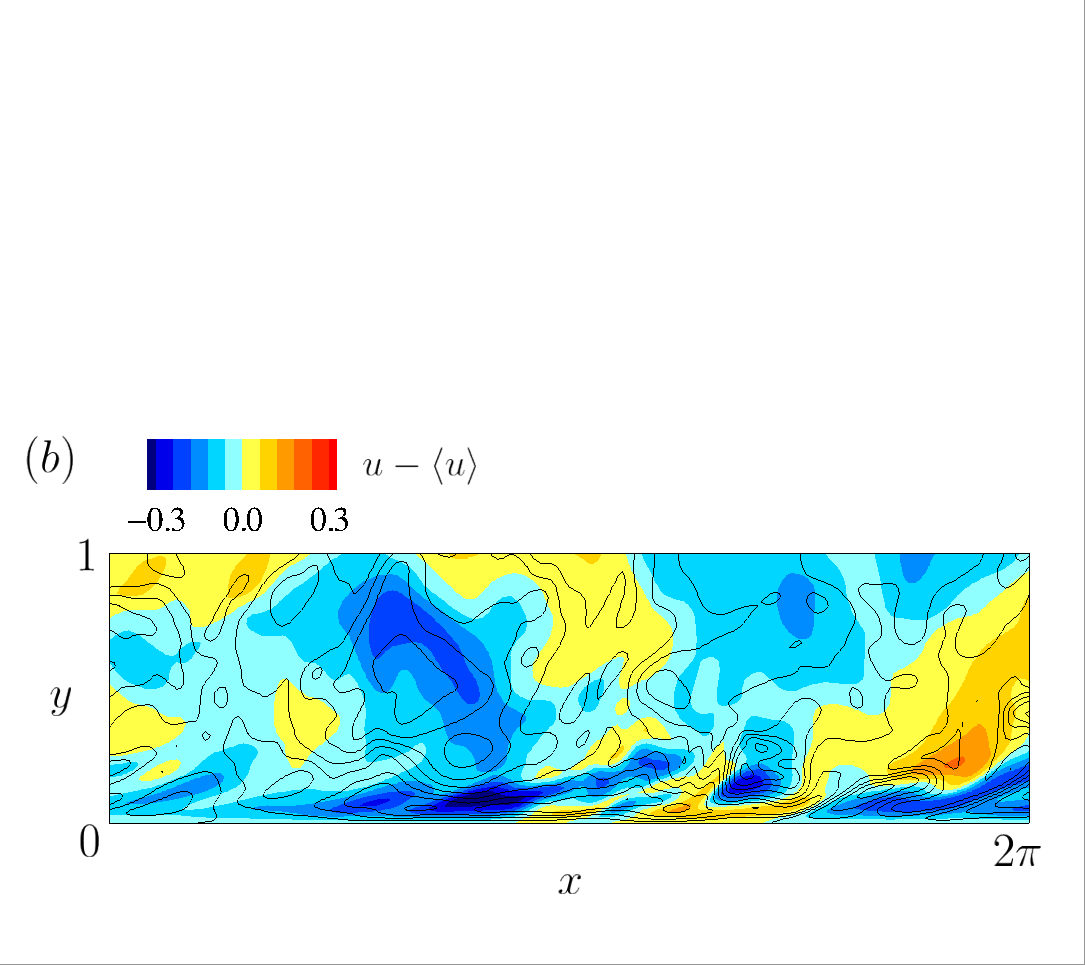}
		}
		\caption{
			(Lines) True streamwise fluctuation $u - \langle u \rangle $ at $t=T$ ($T^+ = 50$) and (colours) estimation with data resolution $\Delta z_m^+ = 112$ (case RZ4 in table \ref{table:obs}). The fields are visualized at ($a$) an observation plane, $z = z_m$ and ($b$) the midpoint between two observation planes, $z = z_m + 0.5 \Delta z_m$.
		}
		\label{fig:kskip32_u}
	\end{figure}

	Although the results thus far demonstrated the accuracy of the flow reconstructions, it is expected that the estimation quality depends on the spatiotemporal resolution of observations.
	And it is also of interest to query the lowest resolution requirement for accurate estimation.  
	In homogeneous isotropic turbulence, it has been reported that reconstruction of the full field is successful only if the resolution of spatial observations satisfies $\Delta _m < 15 \eta$ \citep{Li2020}, where $\eta$ is the Kolmogorov lengthscale. 
	An equivalent criterion has not, however, been proposed for anisotropic, wall-bounded turbulence where the effects of mean shear, advection and the no-slip boundary may alter the resolution requirements of observations.
	Hereafter, we revert to adopting noise-free data and focus on the influence of spatio-temporal resolution of observations on the accuracy of state estimation within the time horizon $T^+ = 50$.

	We first consider the impact of spanwise spacing of observations that are fully resolved in the $x$-$y$ plane (cases RZ1$-$RZ4 in table \ref{table:obs}). 
	With the most coarse observations ($\Delta z_m^+ = 112$, case RZ4), the estimated state is visualized in figure \ref{fig:kskip32_u} and compared to the true one.
	At the observation locations ($z = z_m$), velocity data are reproduced by the algorithm.
	At the midpoint between observation planes ($z = z_m + 0.5 \Delta z_m$), however, the estimation accuracy is notably compromised.

\begin{figure}
		\centering
		\subfigure{
			\includegraphics[width=0.22\textwidth,trim=0.2in 0.3in 5.15in 3in,clip]{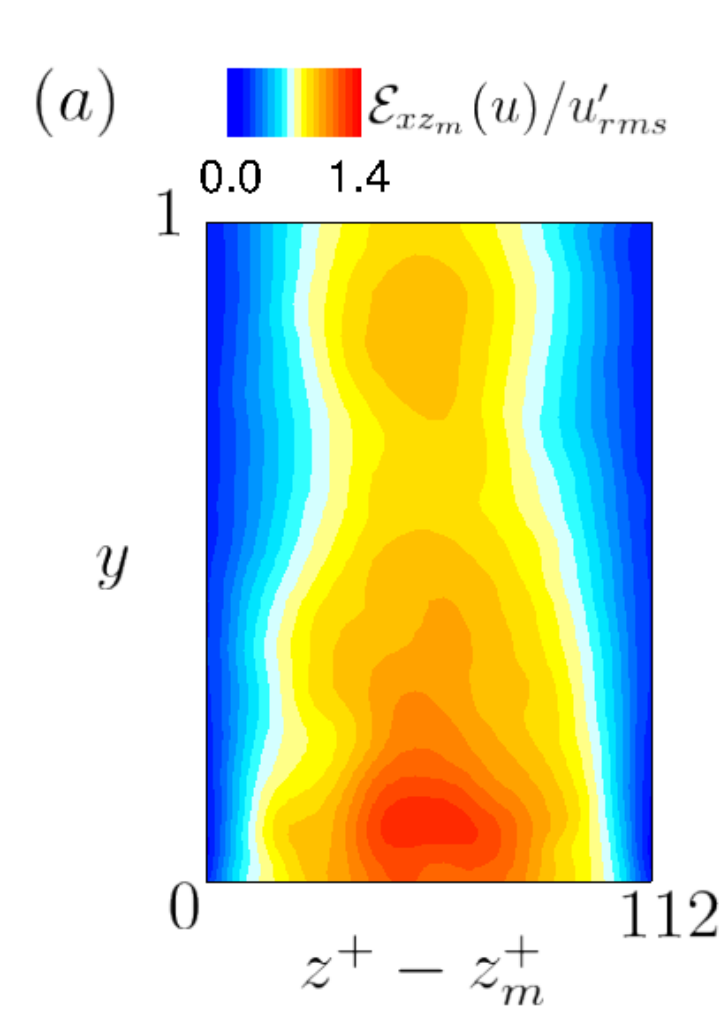}
		}
		\subfigure{
			\includegraphics[width=0.22\textwidth,trim=0.2in 0.3in 5.15in 3in,clip]{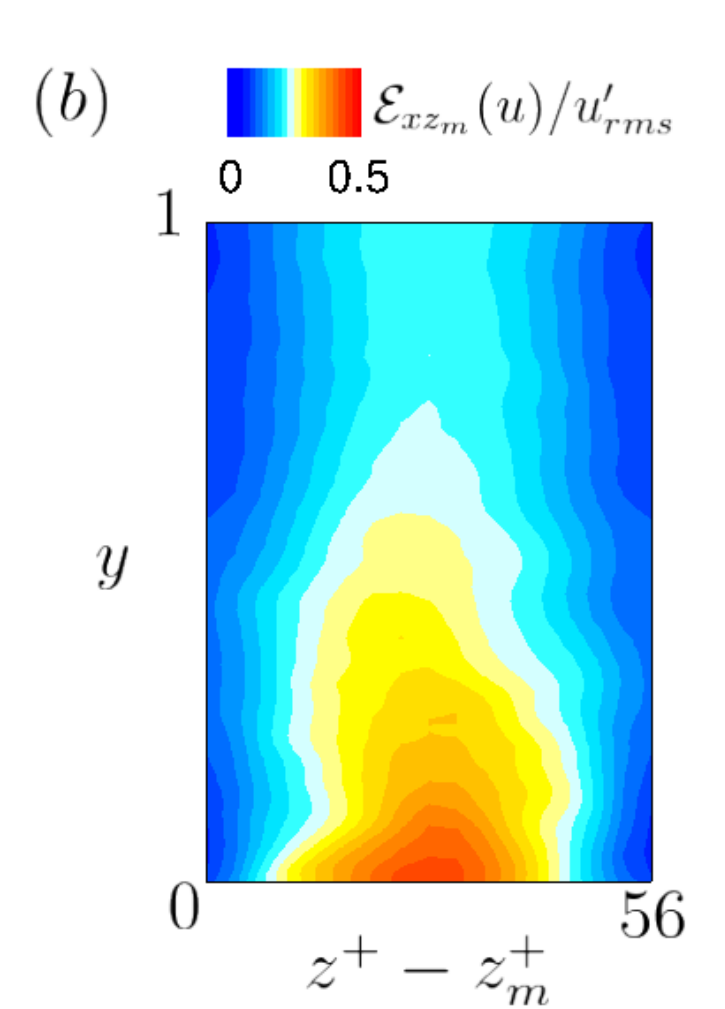}
		}
		\subfigure{
			\includegraphics[width=0.22\textwidth,trim=0.2in 0.3in 5.15in 3in,clip]{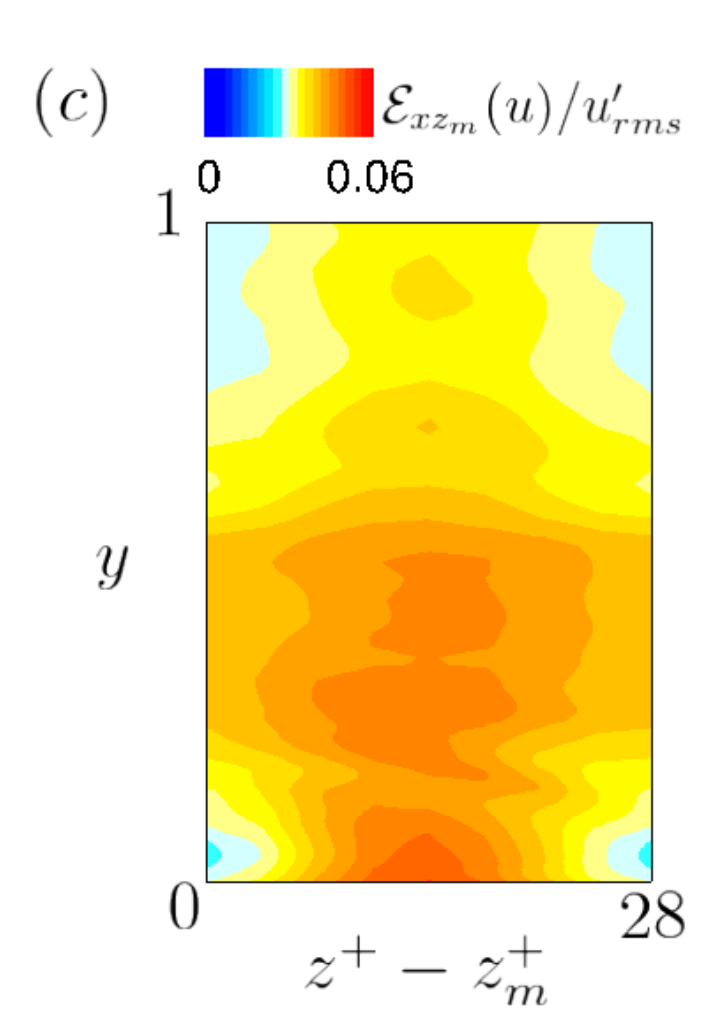}
		}
		\subfigure{
			\includegraphics[width=0.22\textwidth,trim=0.2in 0.3in 5.15in 3in,clip]{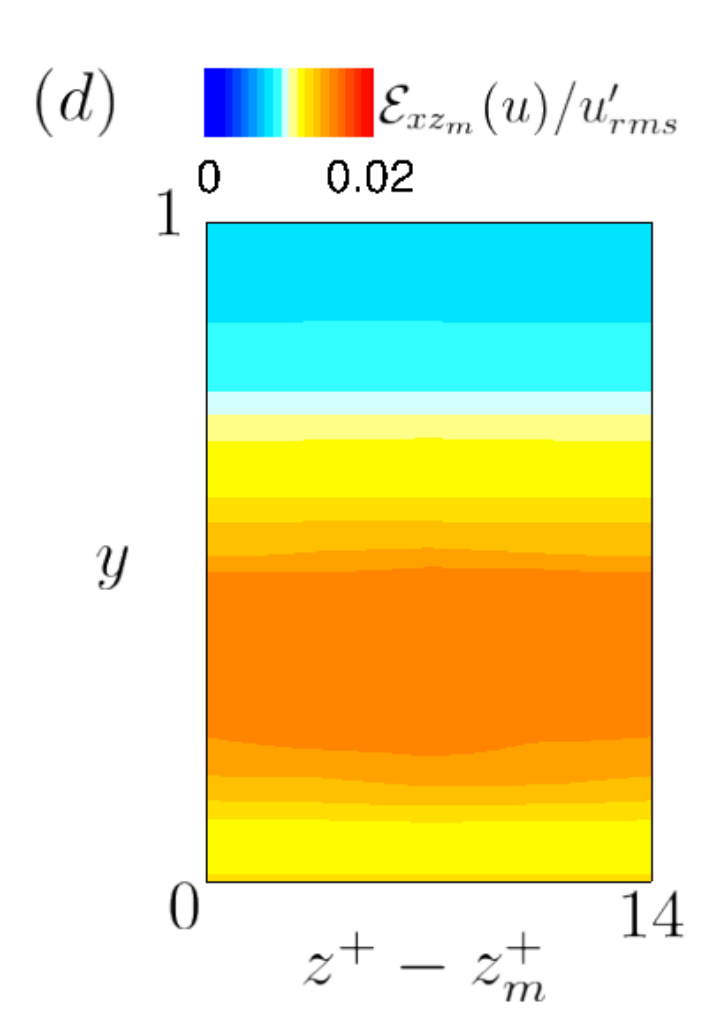}
		}
		\caption{
			Streamwise- and phase-averaged estimation error $\mathcal E_{xz_m}(u)$ between two $x$-$y$ observation planes at $t=T$ ($T^+ = 50$), normalized by the true local root-mean-square fluctuations $u^{\prime}_{rms}$. 
			($a$-$d$): $\Delta z^+_m = \{112,56,28,14\}$.
		}
		\label{fig:error_dzm}
	\end{figure}

	Since our interest is in the distribution of errors between observation planes, the estimation error is phase averaged in the span in addition to averaging in the streamwise direction, and is denoted $\mathcal E_{xz_m}(q)$.
	The results for cases RZ1$-$RZ4 are shown in figure \ref{fig:error_dzm}.  
	Only the error for $u$ component is plotted, and the results for $v, w$ components are similar.
	With the most poorly resolved data (panel $a$), the estimation error increases by an order of magnitude from the observation planes to the midpoint between them. 
	The error in the near-wall region is of the same order of magnitude as the local turbulence fluctuations, which is significantly higher than the error at the channel center.
	As better-resolved observations are adopted (panels $b$-$d$), the inhomogeneity of the error distribution in the spanwise direction becomes weaker. 
	
	\begin{figure}
		\centering
		\subfigure{
			\includegraphics[width=0.475\textwidth]{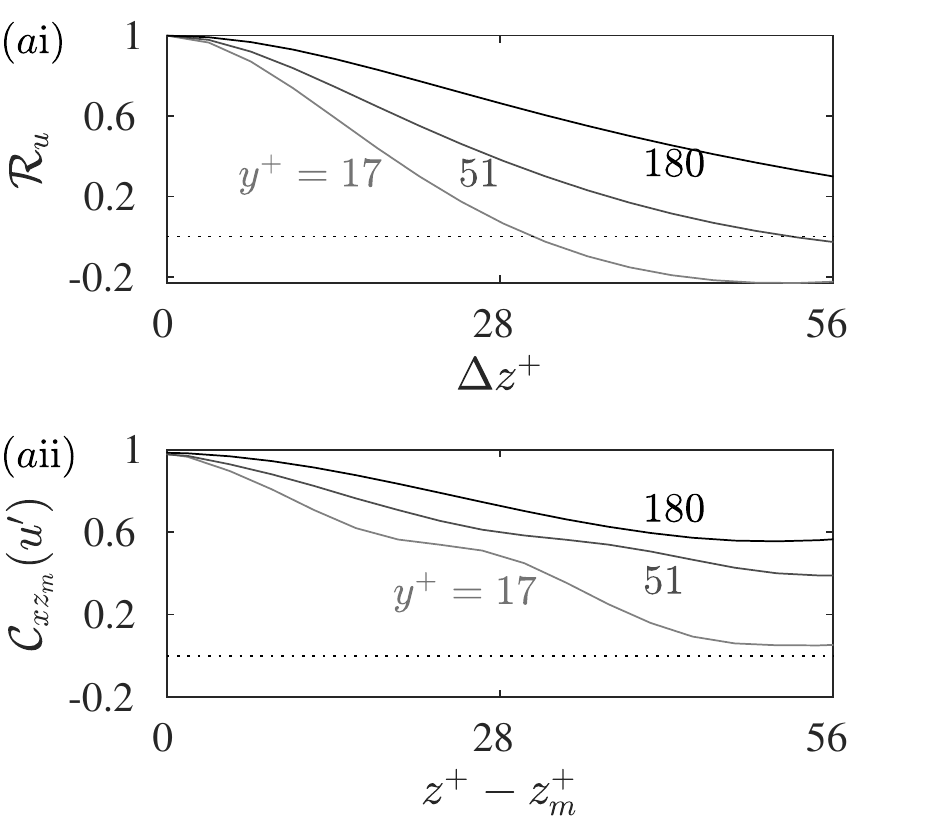}
		}
		\subfigure{
			\hbox{\hspace{0pt}} 
			\includegraphics[width=0.475\textwidth]{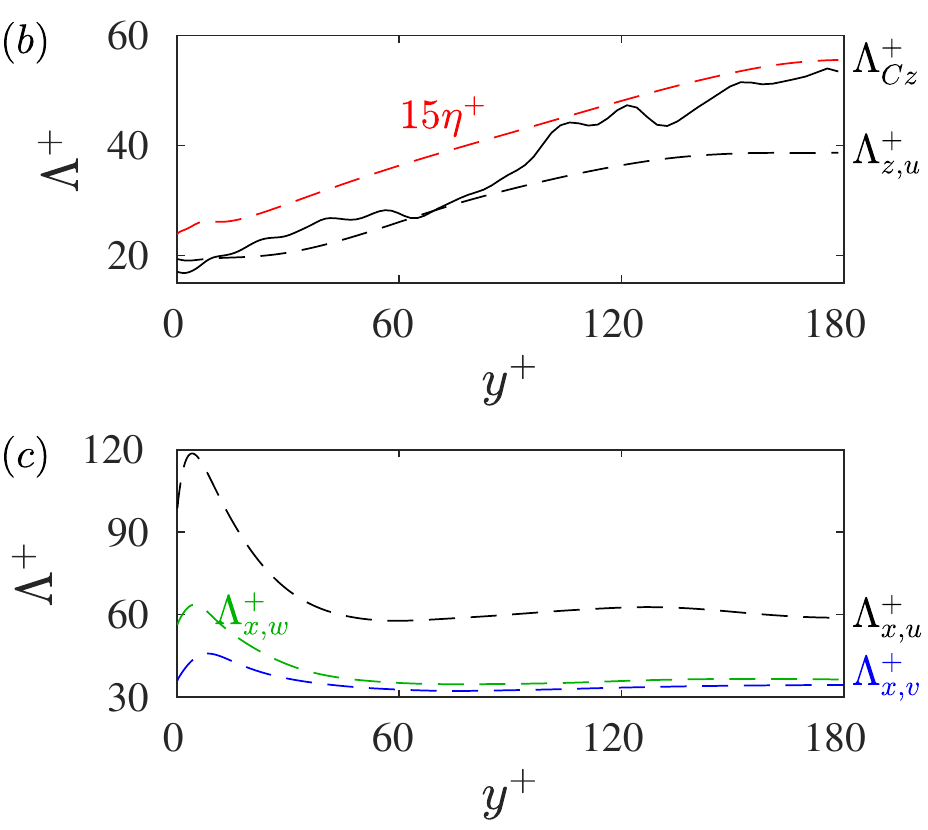}
		}
		\caption{
			($a$i) Two-point correlations $\mathcal R_u$ of the true streamwise fluctuation as a function of spanwise separation;
			($a$ii) correlation coefficient $\mathcal C_{xz_m}(u^{\prime})$ between estimated (case RZ4, $\Delta z_m^+ = 112$) and true streamwise fluctuations at $t = T$ ($T^+ = 50$) as a function of distance from observation planes.
			($a$i-$a$ii) Gray to black correspond to $y^+ = \{17, 51, 180\}$.  
			The horizontal line (\dotted) marks zero correlation.
			($b$) Wall-normal profiles of ({\textcolor{Red}\dashed}) Kolmogorov lengthscale and Taylor microscales based on
			(\lline) correlation coefficient $\mathcal C_{xz_m}(u^{\prime})$, and 
			(\dashed) spanwise two-point correlations.
			($c$) Streamwise Taylor microscales based on two-point correlations of (\dashed) $u$, (\textcolor{Blue}{\dashed}) $v$ and (\textcolor{Green}{\dashed}) $w$ components.
		}
		\label{fig:correlation_dz}
	\end{figure}

	The effect of spanwise data resolution can be explained by examining the spanwise two-point correlations $\mathcal R_u(\Delta z)$ in figure \ref{fig:correlation_dz}($a$i).
	Since the dominant structures in the wall layer are streaks and streamwise vorticies that are narrow in the span, the two-point correlation decays faster at locations closer to the wall.
	As a result, the domain of dependence of the observation planes shrinks from channel center to the wall, which explains the high estimation error in the wall layer.
	Figure \ref{fig:correlation_dz}($a$ii) shows the reconstruction quality at $t=T$ in terms of the correlation coefficient between the true and estimated state from the least-resolved data ($\Delta z_m^+ = 112$).
	The profiles are qualitatively similar to the two-point correlations within $\Delta z^+ < 20$. 
	However, while the turbulent structures decorrelate at larger distances,
	the accuracy of the reconstruction remains relatively higher, and $\mathcal C_{xz_m}(u^{\prime})$ returns to approximately unity as we approach the next observation plane at $z^+ - z^+_m = 112$.

	Similar to the notion of the Taylor microscale,
	\begin{equation}
	\label{eq:taylor}
	\Lambda_{z,u} = \left(-\frac 12 \frac{d^2 \mathcal R_u}{d (\Delta z)^2} \Big|_{\Delta z = 0} \right)^{-1/2},
	\end{equation}
	we introduce a lengthscale for the domain of dependence of one observation plane by replacing $\mathcal R_u$ in (\ref{eq:taylor}) by the correlation coefficient $\mathcal C_{xz_m}(u^{\prime})$.  
	The resulting lengthscale, which we denote $\Lambda_{Cz}$, is representative of the domain of accurate estimation.
	Both $\Lambda_{z,u}$ and $\Lambda_{Cz}$ are plotted in figure \ref{fig:correlation_dz}$b$, and have similar values across the height of the channel:  The spanwise size of the domain of dependence of observations is similar in the Taylor microscale.  
	Therefore, the criterion
	\begin{equation}
	\label{eq:crit_z}
	\Delta z_m \lesssim 2\Lambda_{z,u},
	\end{equation}
	must be satisfied 
	to guarantee an accurate estimation of the local $u$ field.
	Similar criteria can be adopted for accurate estimation of $v$ and $w$ components using the Taylor microscales $\Lambda_{z,v}$ and $\Lambda_{z,w}$, respectively. Since those lengthscales are commensurate with $\Lambda_{z,u}$, the condition (\ref{eq:crit_z}) suffices.  
	Using (\ref{eq:crit_z}), we can also interpret the estimation results with different spanwise data resolutions (figure \ref{fig:error_dzm}).
	When $\Delta z^+_m = 56$ (figure \ref{fig:error_dzm}$b$), the criterion (\ref{eq:crit_z}) is satisfied for the bulk of the channel ($2\Lambda^+_{z,u} \approx 80$, c.f.\,figure \ref{fig:correlation_dz}$b$) and starts to be violated for $y^+ < 70$.
	As such, while figure \ref{fig:error_dzm}$b$ reports high prediction accuracy in the bulk ($\mathcal C_{xz_m}(u') = 0.97$), errors increase in that near-wall region ($\mathcal C_{xz_m}(u') = 0.83$) and become inhomogeneous in the span.
	When $\Delta z^+_m = 28 < \min_y 2\Lambda_{z,u}^+$ (figure \ref{fig:error_dzm}$c$), every point in the flow is covered by the domain of dependence of observations, so the estimation quality becomes more accurate and uniform at all the $y$ locations ($\mathcal C_{xz_m}(u') = 0.99$). 
	
    We compare the criterion (\ref{eq:crit_z}) to the one for homogeneous isotropic turbulence \citep{Yoshida2005,Eyink2013,Li2020} that $\Delta_m \lesssim 15 \eta$.  
	The Kolmogorov scale in our case is $\eta \equiv (Re^3 \mathcal D)^{-1/4}$, where $\mathcal D = (2/Re) \langle s'_{ij} s'_{ij}\rangle$ is the viscous rate of dissipation and $s'_{ij} = (\partial_i u_j^{\prime} + \partial_j u_i^{\prime})/2$ is the fluctuating strain rate tensor. 
	The criterion $15 \eta$ is plotted as red dashed line in figure \ref{fig:correlation_dz}$b$.
	Since the average dissipation is affected by the streamwise elongated structures in the channel, $\eta$ is larger than the spanwise size of the smallest eddies.  
	Nonetheless, the trend is similar to the Taylor microscale condition provided above.  
	Physically, both criteria demonstrate that the critical data resolution for accurate estimation of entire flow is within the transition zone between inertial and viscous dissipation ranges. 
	
	\begin{figure}
		\centering
		\subfigure{
			\includegraphics[width=0.22\textwidth,trim=0.2in 0.3in 5.2in 3in,clip]{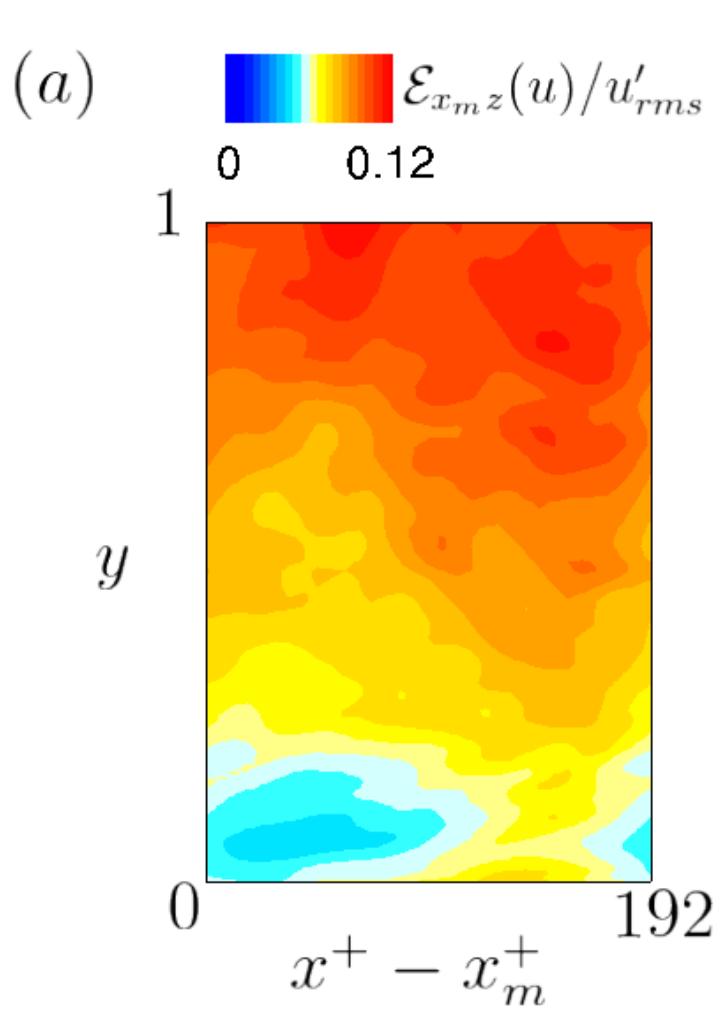}
		}
		\subfigure{
			\includegraphics[width=0.22\textwidth,trim=0.2in 0.3in 5.2in 3in,clip]{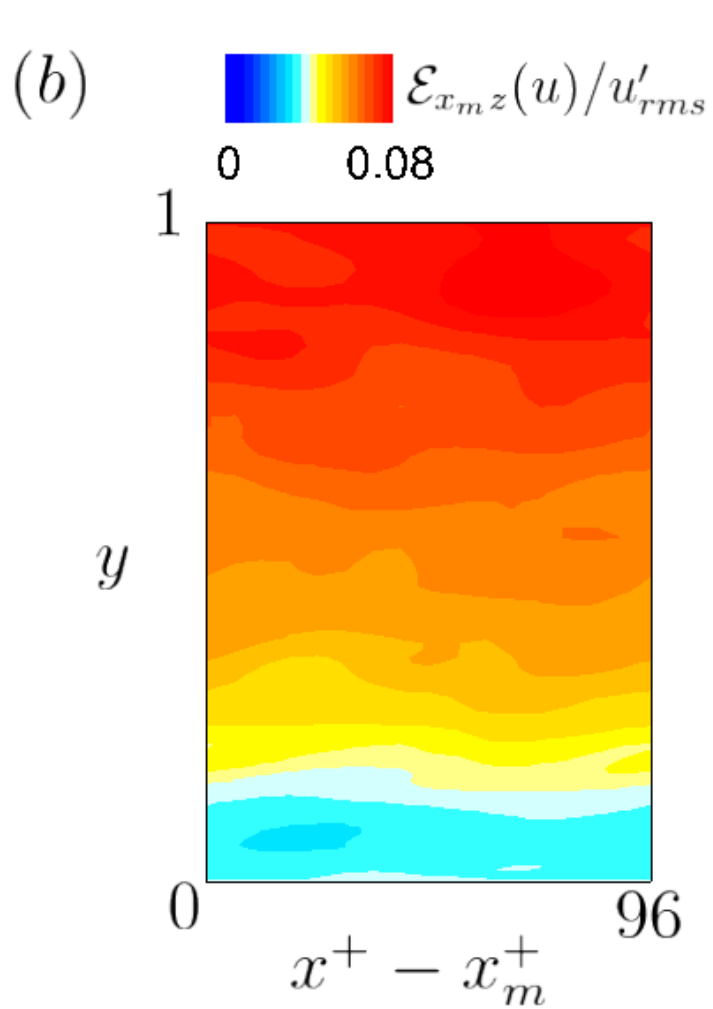}
		}
		\subfigure{
			\includegraphics[width=0.22\textwidth,trim=0.2in 0.3in 5.2in 3in,clip]{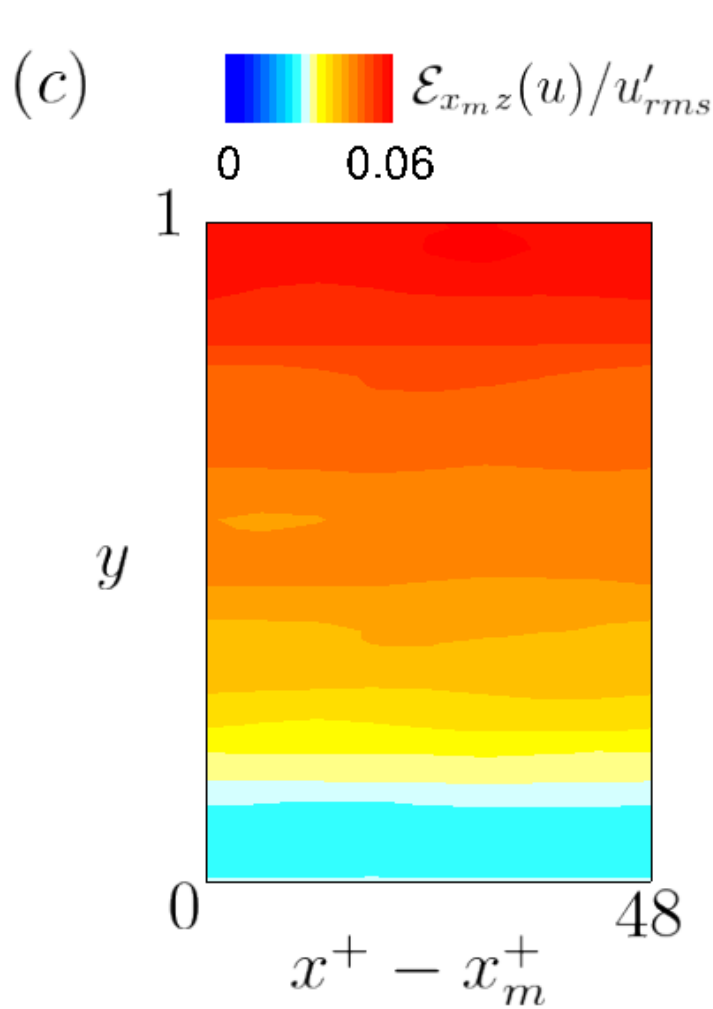}
		}
		\subfigure{
			\includegraphics[width=0.22\textwidth,trim=0.2in 0.3in 5.2in 3in,clip]{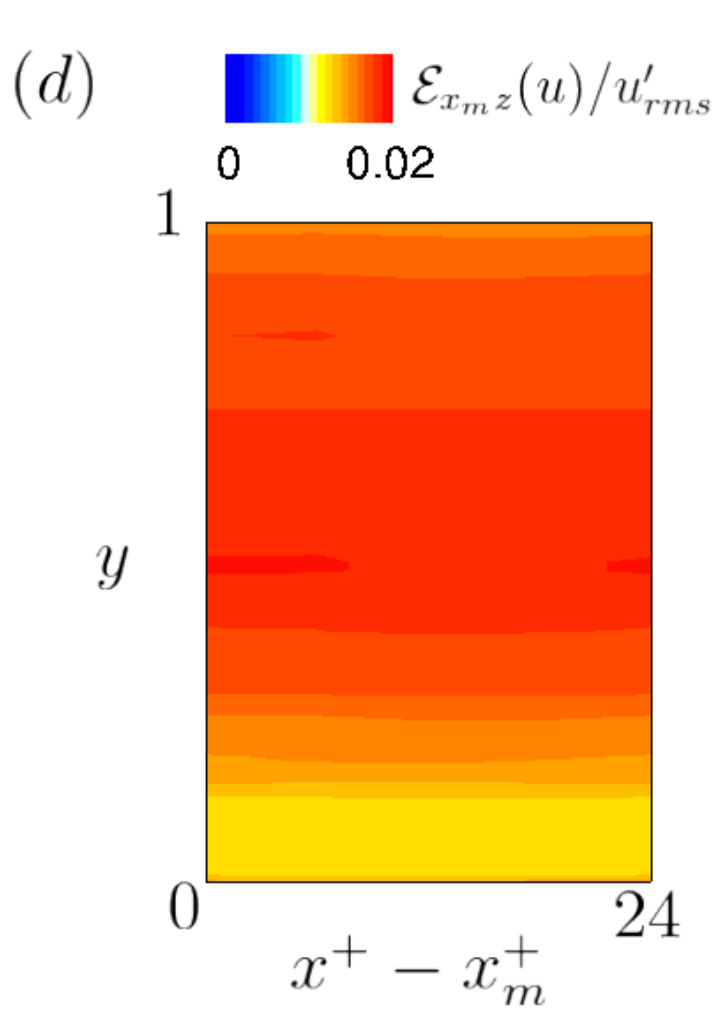}
		}
		\caption{
			Spanwise- and phase-averaged estimation error $\mathcal E_{x_m z}(u)$ between two $y$-$z$ observation planes at $t=T$ ($T^+ = 50$), normalized by the true local root-mean-square fluctuations $u^{\prime}_{rms}$. 
			($a$-$d$):  $\Delta x^+_m = \{192,96,48,24\}$.
		}
		\label{fig:error_dxm}
	\end{figure}
	
	Next, we consider fully resolved observations in the cross-flow $y$-$z$ planes, and under-resolved in the streamwise direction (cases RX1$-$RX4 in table \ref{table:obs}).
	Before examining the state estimation results, we report the Taylor microscales $\Lambda_{x,i}$ ($i=u,v,w$) in figure \ref{fig:correlation_dz}$c$, which is computed from the streamwise two-point correlations.
	Due to the near-wall elongated streaky structures, $\Lambda_{x,u}$ is largest among all three components, peaks near the wall and decays towards the channel center. We therefore expect that the estimation, particularly of the $u$ component, should be most accurate in the inner layer relative to the accuracy in the outer flow.
	
	The root-mean squared estimation error is plotted in figure \ref{fig:error_dxm}, where $\mathcal E_{x_mz}(u)$ is averaged in the span and phase-averaged in the streamwise direction, and also normalized by the r.m.s. fluctuations. 
	Overall, the estimation error decreases as better-resolved data are included (panels $a$-$d$).
	For each observation resolution, the estimation quality deteriorates from the wall to channel center, as expected.
	Two notable differences from the effect of spanwise resolution are observed:
	(i) the estimation error is not symmetric with respect to the midpoint between observation planes, especially in panels ($a$-$b$);
	(ii) although the separation of observation planes in figure \ref{fig:error_dxm}$a$ is larger than
	$2\Lambda^+_{x,u} (\approx 118)$ in the bulk (c.f.\,figure \ref{fig:correlation_dz}$c$), the estimation error does not increase appreciably between streamwise observation locations and remains within approximately $10\%$ of the local root-mean-squared fluctuations ($\mathcal{C}_{x_mz}(u^{\prime}) \ge 0.95$ while $\mathcal{R}_{u}(\Delta x^+=96) \approx 0.6$).
	Both points are caused by the mean advection in the streamwise direction.
	Conceptually, every instant when data are recorded in the cross-flow $y$-$z$ plane corresponds to an accurately estimated ``layer" in the spatiotemporal evolution of the flow, propagating by the advection velocity $U_a$. 
	The thickness of such layers is approximately the Taylor microscale $\Lambda_{x,u}$ and the distance between two adjacent ones is approximately $U_a \Delta t_m$.
	Since the observation data are temporally well resolved, $U_a \Delta t_m \ll 2\Lambda_{x,u}$, the accurately estimated layers overlap with one another and lead to commensurate accuracy between observation locations.
	An example that further highlights this conceptual interpretation is considered next, where the temporal sampling rate is sufficiently low ($U_a \Delta t_m > 2\Lambda_{x,u}$) in order to distinguish the accurately predicted layers associated with different ($x_m,t_m$).

	\begin{figure}
		\centering
		\includegraphics[width=\textwidth]{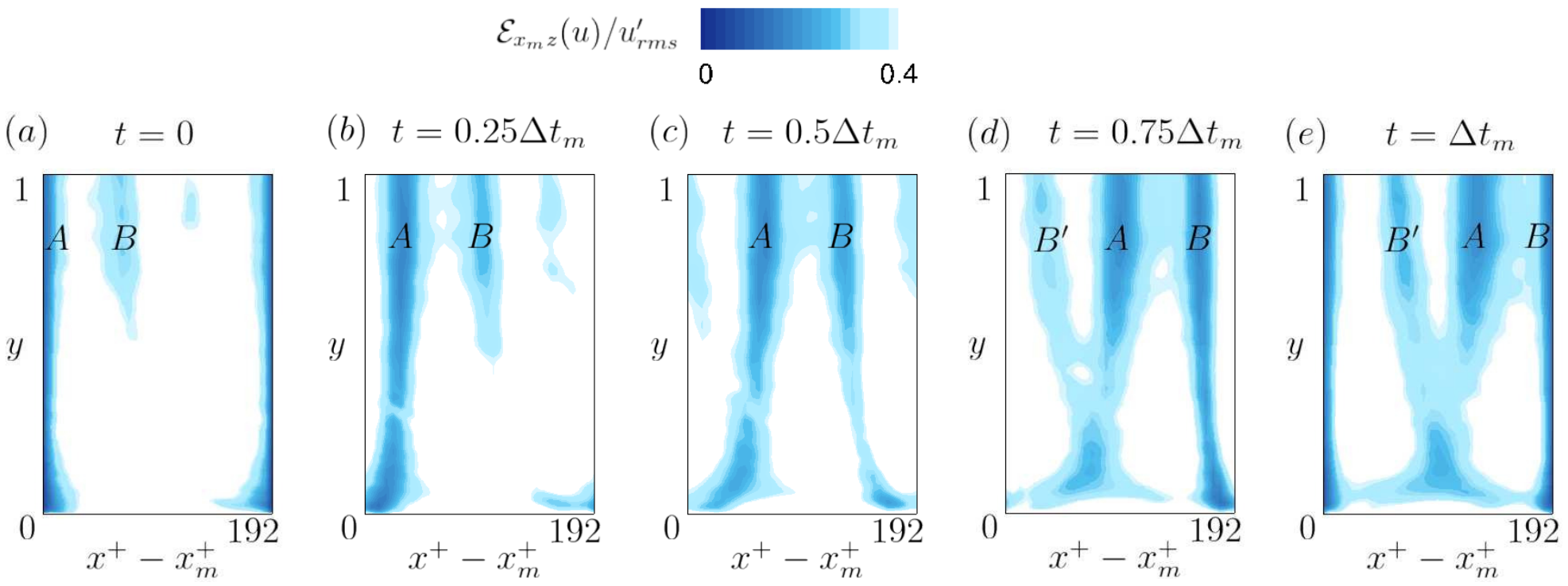}
		\caption{
			Spanwise- and phase-averaged estimation error $\mathcal E_{x_mz}(u) $
			within the first observation time interval ($\Delta t^+_m = 7.4$, $T^+ = 50$, case RT4): ($a$-$e$) $t = \{0.00, 0.25, 0.50, 0.75, 1.00\}\Delta t_m$.
			($A$) regions where error is low at $(x,t) = (x_m,0)$ and advected downstream; ($B$) emergent region of small error that later coincides with observation plane $(x,t)=(x_m,\Delta t_m)$, and similarly ($B^{\prime}$) coincides with $(x,t)=(x_m,2\Delta t_m)$.
		}
		\label{fig:error_nt128_t}
	\end{figure}

	\begin{figure}
		\centering
		\includegraphics[width=0.6\textwidth,trim=0.1in 0.3in 0.1in 1.8in,clip]{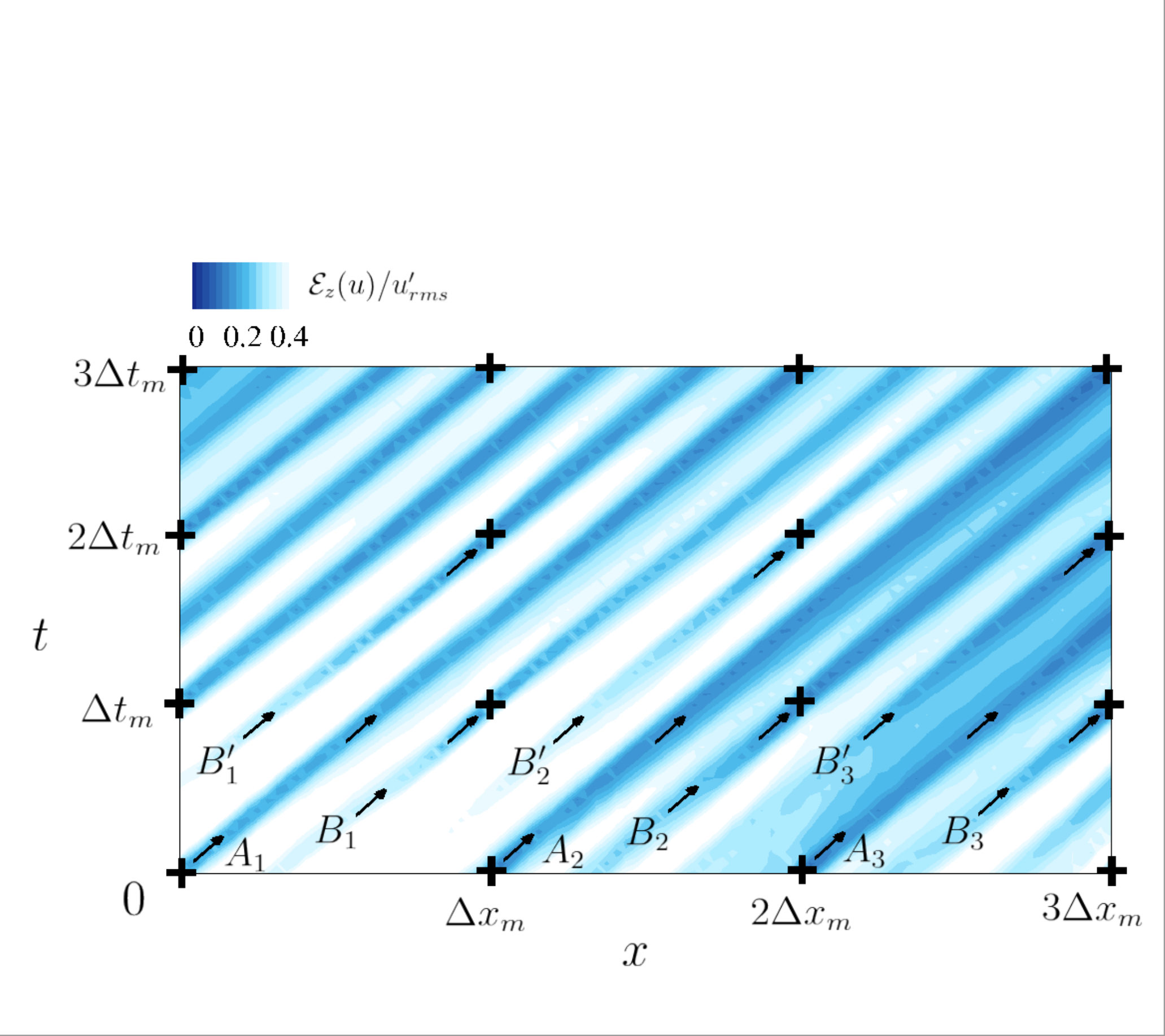}
		\caption{
			    Space-time evolution of the spanwise-averaged estimation error $\mathcal E_z(u)$ at the channel center.
				Black plus: observation location and time; ($A_i, B_i, B_i^{\prime}$) same as ($A, B, B^{\prime}$) in figure \ref{fig:error_nt128_t} and the subscript denotes the $i^{th}$ interval between streamwise observations.
		}
		\label{fig:resolution_nt128_xt}
	\end{figure}
	\begin{figure}
		\centering
		\subfigure{
			\includegraphics[width=0.22\textwidth,trim=0.2in 0.3in 5.2in 3in,clip]{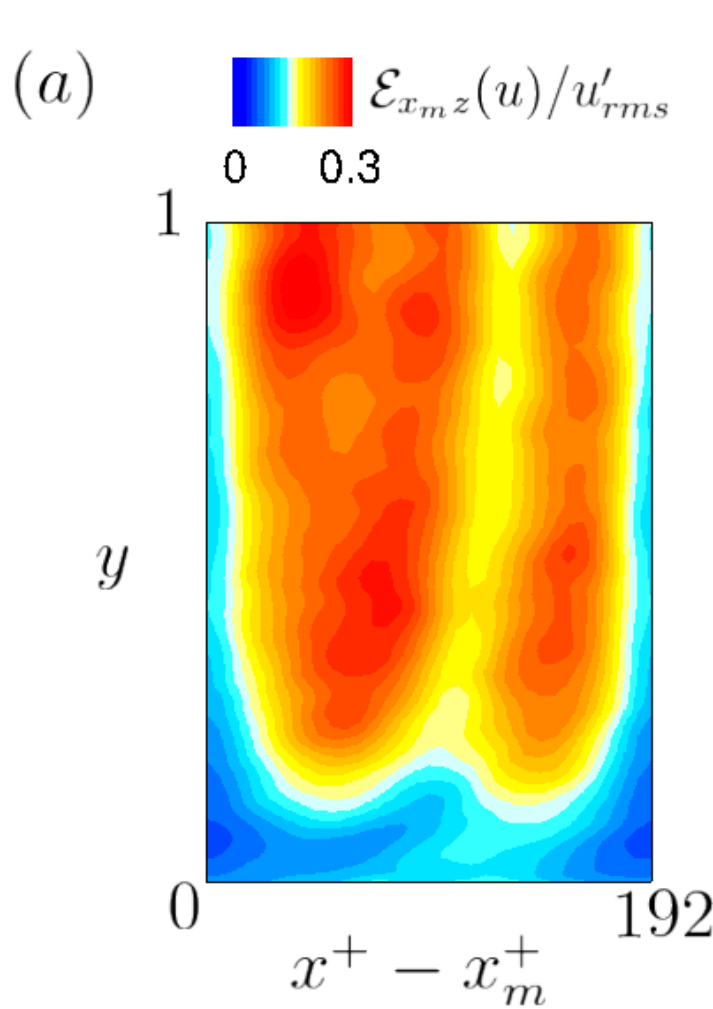}
		}
		\subfigure{
			\includegraphics[width=0.22\textwidth,trim=0.2in 0.3in 5.2in 3in,clip]{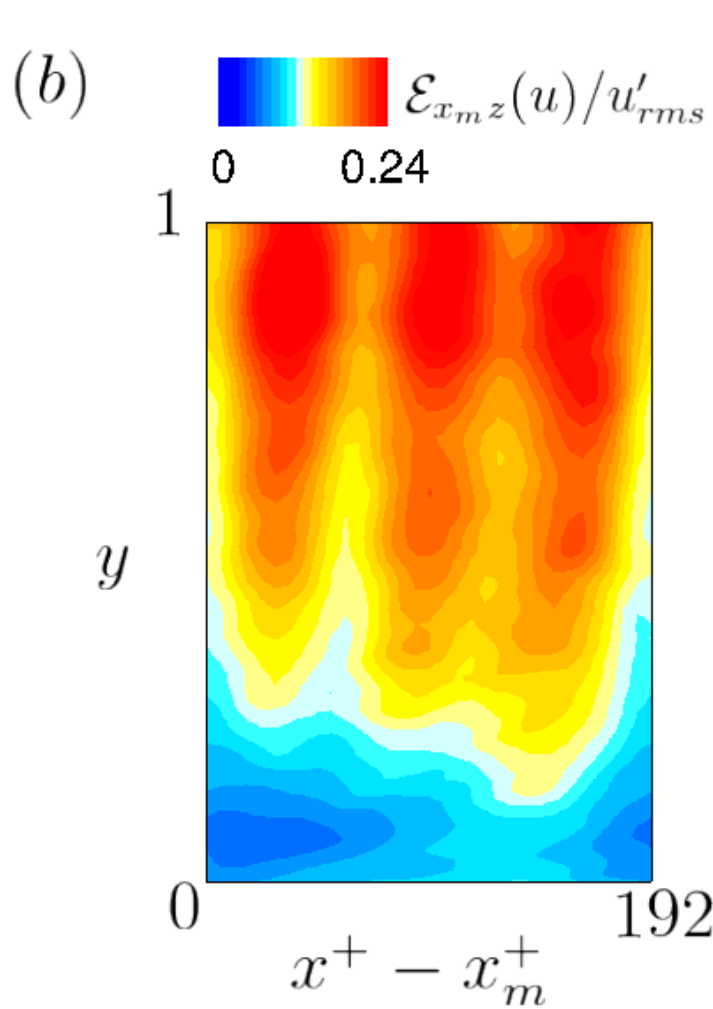}
		}
		\subfigure{
			\includegraphics[width=0.22\textwidth,trim=0.2in 0.3in 5.2in 3in,clip]{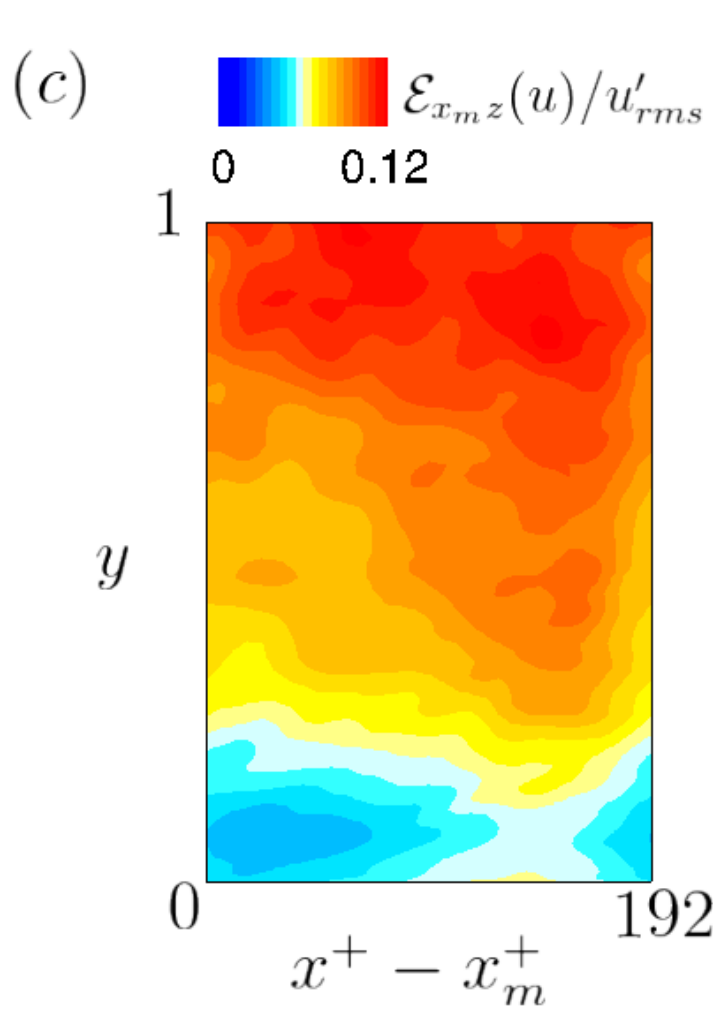}
		}
		\subfigure{
			\includegraphics[width=0.22\textwidth,trim=0.2in 0.3in 5.2in 3in,clip]{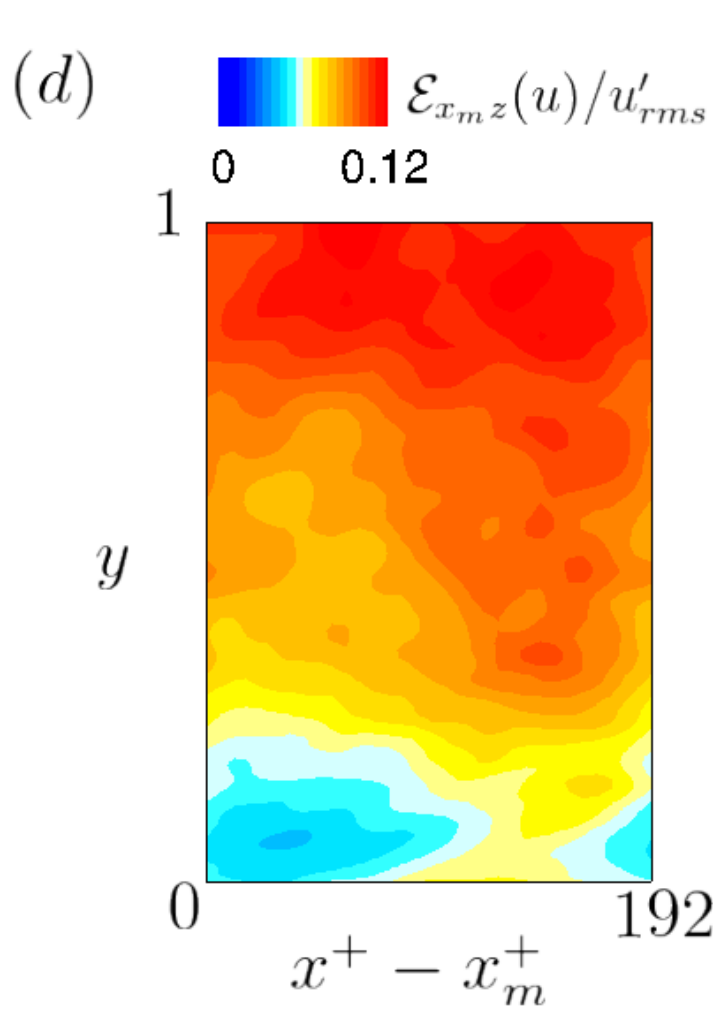}
		}
		\caption{
			Spanwise- and phase-averaged estimation error $\mathcal E_{x_mz}(u)$ at $t=T$ ($T^+ = 50$) with different observation time interval: 
			($a$-$d$):  $\Delta t^+_m = \{7.4, 3.7, 1.8, 0.9\}$.
		}
		\label{fig:error_dtm}
	\end{figure}

	Consider the same coarse spatial resolution from case RX4 ($\Delta x^+_m = 192$), and we now adopt a long time between observations, $\Delta t^+_m = 7.4$.
	This case is denoted RT4, and a temporal evolution of the estimation error from $t = 0$ to $t = \Delta t_m$ is shown in figure \ref{fig:error_nt128_t}.
	Three layers of small errors are observed: 
	($A$) originates at the observation station at $t=0$ and advects downstream;
	($B$) emerges between observation locations and times, such that the field accurately reproduces measurements at a downstream observation point at a subsequent measurement time $t = \Delta t_m$;
	($B^{\prime}$) is similar to ($B$) but reaches the observation station at $t = 2\Delta t_m$.
	Once ($A$) leaves the observation plane, the error increases, at least initially, due to the chaotic nature of turbulence and the absence of a nearby observation station to correct the field\textemdash behaviour later in time is discussed below. 
	By comparison, the error in ($B$) decays with time as it approaches the observation location at time $t=\Delta t_m$.

	A detailed space-time representation of the estimation error at the channel center is shown in figure \ref{fig:resolution_nt128_xt}.
	All the regions of small errors (blue) belong to one of the three classes described above, and can be associated with at least one observation station (black pluses).
	The pattern of the errors is repeated, anchored at $(\Delta x_m, \Delta t_m)$ and inclined according to the advection speed.
	Since $3U_a \Delta t_m \approx 2 \Delta x_m$, regions ($A_i$) that originate from $(x,t) = (x_m, 0)$ will reach another observation station at $(x,t) = (x_m + 2 \Delta x_m, 3 \Delta t_m)$;  the errors in that region thus initially increase moving away from $(x_m,0)$ and reduce as that later observation point is approached. 
	For a general spatiotemporal resolution of observations, regions ($A_i$) may not coincide with another observation position and time, and hence the errors would not undergo the later decay.  
	From figure \ref{fig:resolution_nt128_xt}, it is evident that an accurate estimation of the entire flow state can be achieved by refining either the spatial or temporal data resolution such that the low-error regions overlap.
	This view is demonstrated in figure \ref{fig:error_dtm}, where panels ($a$-$d$) correspond to increasing temporal resolution (cases RT4$-$1 in table \ref{table:obs}).
	When the true state is observed sufficiently frequently, estimation error in the bulk region approaches homogeneity in the streamwise direction (panels $c$, $d$).
	Specifically, accurate reconstruction is achieved when the distance between the thin layers that sample observation points is within the domain of dependence of observations, 
	\begin{equation}
    	\label{eq:crit_t}
    	U_a \Delta t_m \lesssim 2\Lambda_{x,u}.
	\end{equation}
	Recalling that $\Lambda_{x,v} \approx \Lambda_{x,w} < \Lambda_{x,u}$ (c.f. figure \ref{fig:correlation_dz}$c$), the equivalent criterion for accurate estimation of $v$ and $w$ is more restrictive. When the observations only satisfy the bound for $u$, the reconstruction is less accurate for the other two velocity components (e.g. $\mathcal{C}_{x_mz}(u',v,w) = \{0.95, 0.86, 0.79\}$ when $\Delta x^+_m = 192$ and $\Delta t^+_m = 7.4$).
	

	\begin{figure}
		\centering
		\includegraphics[width=0.4\textwidth]{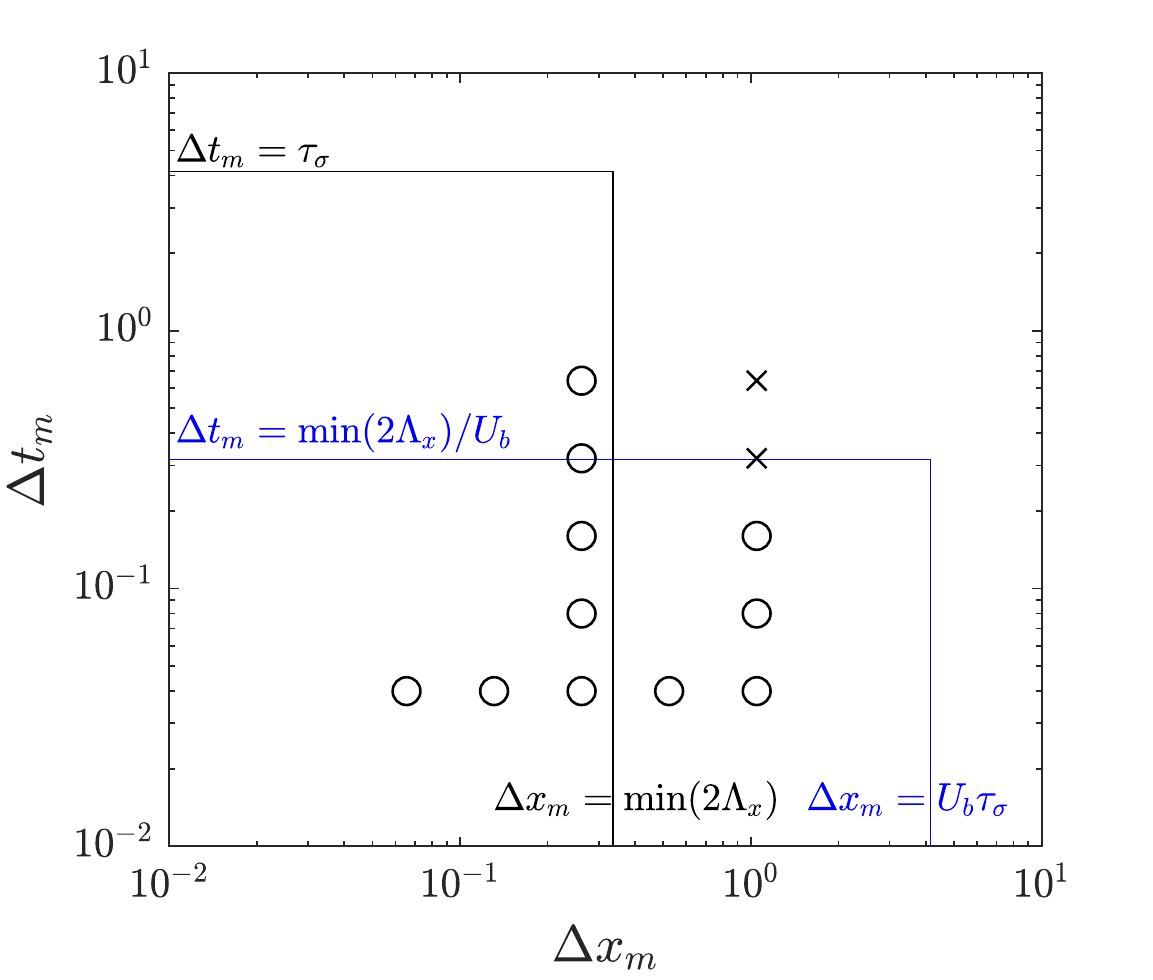}
		\caption{
			Criteria for streamwise and temporal data resolutions.
			(\lline) $\Delta t_m = \tau_{\sigma}$ and $\Delta x_m = \min_{y,i} (2\Lambda_{x,i})$;
			(\textcolor{Blue}{\lline}) $\Delta t_m = \min_{y,i} (2\Lambda_{x,i}) / U_b$ and $\Delta x_m = U_b \tau_{\sigma}$;
			empty circles: successful estimation with correlation coefficient higher than 0.9; 
			crosses: unsuccessful estimation.}	
		\label{fig:bound_xt_v1}
	\end{figure}
	
	In summary, to guarantee an accurate estimation of local velocity component $i$, the streamwise and temporal data resolution must satisfy either one of two conditions:
	$$  \textrm{(i)~} \Delta x_m \lesssim 2\Lambda_{x,i} \textrm{~and~} \Delta t_m \lesssim \tau_{\sigma} 
	\quad \textrm{or} \quad 
	\textrm{(ii)~} \Delta t_m  \lesssim 2\Lambda_{x,i}/U_a  \textrm{~and~}  \Delta x_m  \lesssim U_b\tau_{\sigma} .$$
	The first condition is similar to the one for the spanwise data resolution, namely that the Taylor microscale must be resolved by observations.  Note that it is supplemented by a statement regarding the observation time, which must be smaller than the Lyapunov timescale $\tau_\sigma$ because even a near perfect estimate of the flow state will diverge from the true solution over that period unless additional data are available for assimilation. 
	The second condition is a reinterpretation that accounts for mean advection:
	Should the temporal resolution resolve the advected Taylor scale, the only requirement for spatial resolution becomes a condition based on the travel distance within the Lyapunov time.   
	These two conditions are plotted in figure \ref{fig:bound_xt_v1}.
	We performed additional tests with $\Delta x_m^+ = 48$ and $\Delta t^+_m = \{0.45, 0.90, 1.80, 3.70, 7.40\}$. 
	When the correlation coefficient between the true and estimated state was higher than 90\%, the outcome was deemed successful; these cases are marked by circles. Inaccurate, or unsuccessful, reconstructions are marked by crosses.
	The outcomes of all the cases agree with the above two conditions.

	\subsection{Estimation without wall-layer data}
	\label{sec:wall_layer}
	In all the above cases, observations were distributed throughout the wall-normal extent of the channel, spanning both the inner and outer layers.
	In reality, however, experimental observations become progressively more difficult to obtain near the wall \citep{Hutchins2005,Smits2011}.
	And since most of the turbulence kinetic energy is produced near the wall (peak production at $y^+ \approx 12$) and transported into the bulk region, lack of observation in this region may severely compromise the estimation of the full state. 
	We therefore explore the accuracy of reconstructing the flow within the inner region when observations are only available in the outer layer and directly on the wall.
	Previous efforts for this configuration are limited and have generally focused on use of linear models \citep{Baars2016,Illingworth2018}.
	The present test case is the first attempt to perform the reconstruction based on the full nonlinear Navier-Stokes equations.
	
	Specifically, we consider two types of observations: 
	(i) sub-sampled velocity data available in the outer layer, $y^+>30$, with the same spatial and temporal resolution as the benchmark case; 
	(ii) shear stresses $\boldsymbol{\tau} = (\tau_{xy}, \tau_{zy})$ on both channel walls with the same resolution as velocity data.
	The observation time horizon is the same as the benchmark case, $T^+ = 50$.
	These observations are weighted in the cost function,
	\begin{equation}
	\label{eq:cost_tau}
	J(\boldsymbol u^0) = \frac 12 \sum_{n=0}^N  \| \boldsymbol u^n_m - \mathcal M_u(\boldsymbol u^n) \|^2 + \left(\beta Re / Re_\tau \right)^2 \| \boldsymbol {\tau}^n_m -  \mathcal M_{\tau}(\boldsymbol u^n) \|^2_w,
	\end{equation}
	where $\| \cdot \|_w$ represent integration over top and bottom walls and the weight $\beta = 1/5$.
	The choice of weighting parameter $\beta$ may be motivated by different objectives, such as balancing the two contributions in the cost function or their gradients or minimizing the condition number of the Hessian of $\mathcal{J}$.
	Here we adopt $\beta=1/5$ such that the stress term is commensurate with the velocity at the first fluid grid point that was observed in our benchmark case.
	The estimated state is compared with the benchmark case to highlight the impact of missing observations in the wall layer.

	\begin{figure}
		\centering
		\subfigure{
			\includegraphics[width=0.35\textwidth]{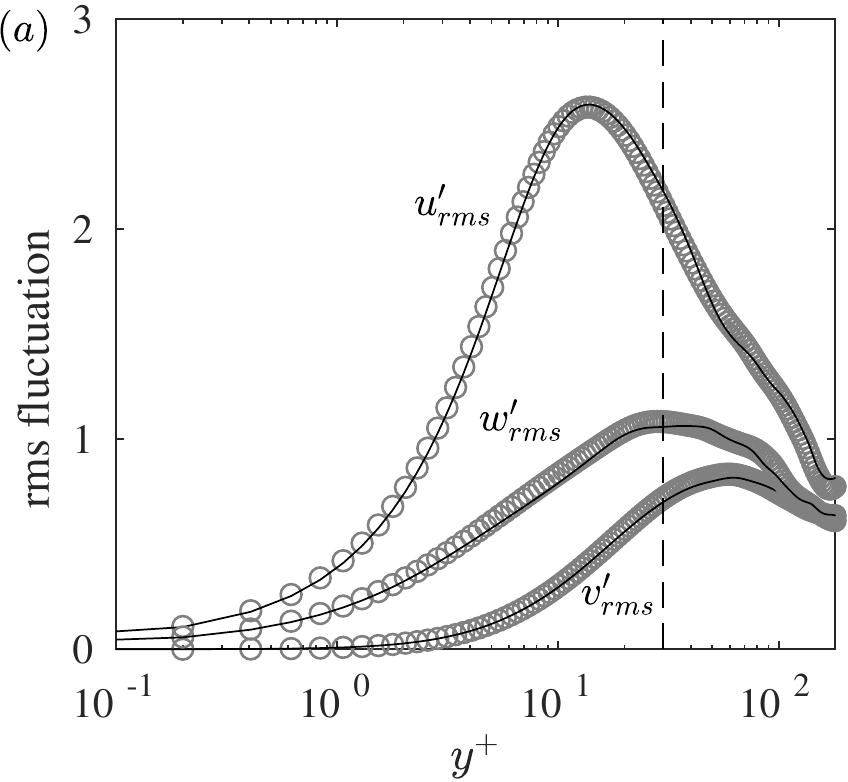}
		}
		\subfigure{
			\hbox{\hspace{0.2in}} 
			\includegraphics[width=0.35\textwidth]{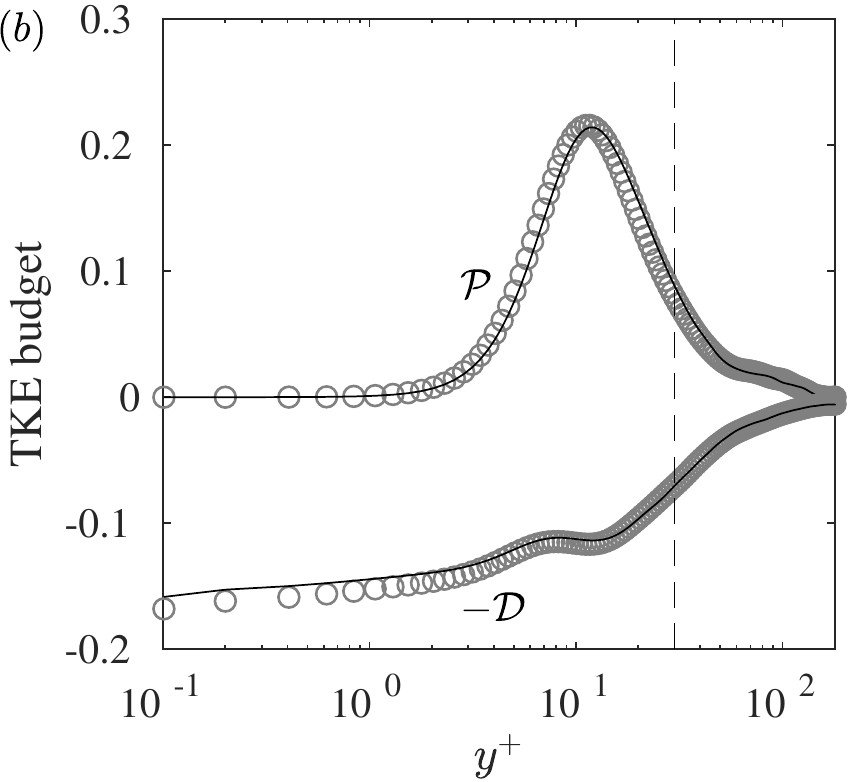}
		}
		\caption{
			Comparison of the instantaneous ($t=T$, $T^+ = 50$), horizontally averaged flow statistics for the (gray circles) true and (black lines) estimated fields, without wall-layer observation data: 
			($a$) root-mean-square fluctuations;
			($b$) turbulent kinetic energy production and dissipation.
			Dashed lines: $y^+ = 30$.
		}
		\label{fig:wall_mean}
	\end{figure}
	
	\begin{figure}
		\centering
		\subfigure{
			\includegraphics[width=0.32\textwidth]{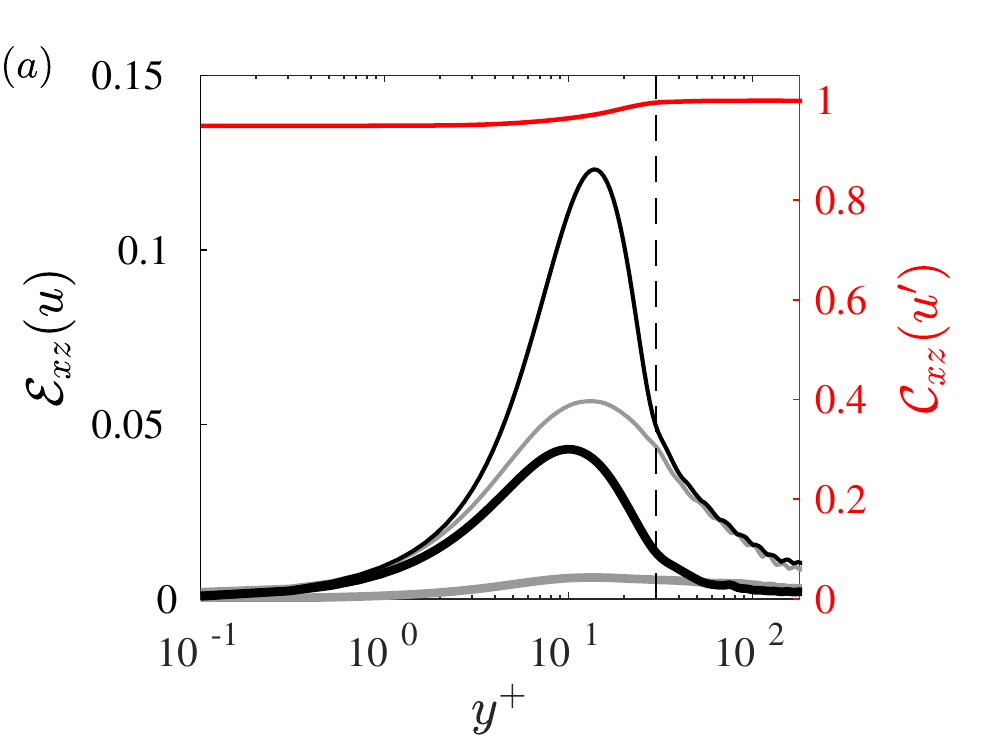}
		}
		\subfigure{
			\hbox{\hspace{-0.25in}} 
			\includegraphics[width=0.32\textwidth]{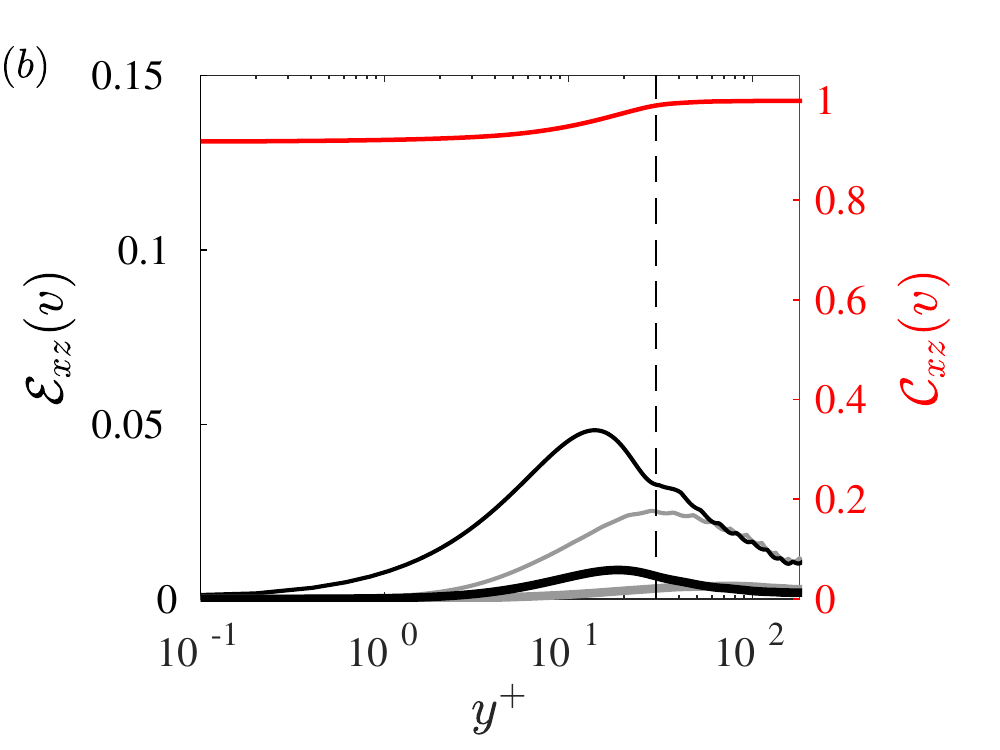}
		}
		\subfigure{
			\hbox{\hspace{-0.25in}} 
			\includegraphics[width=0.32\textwidth]{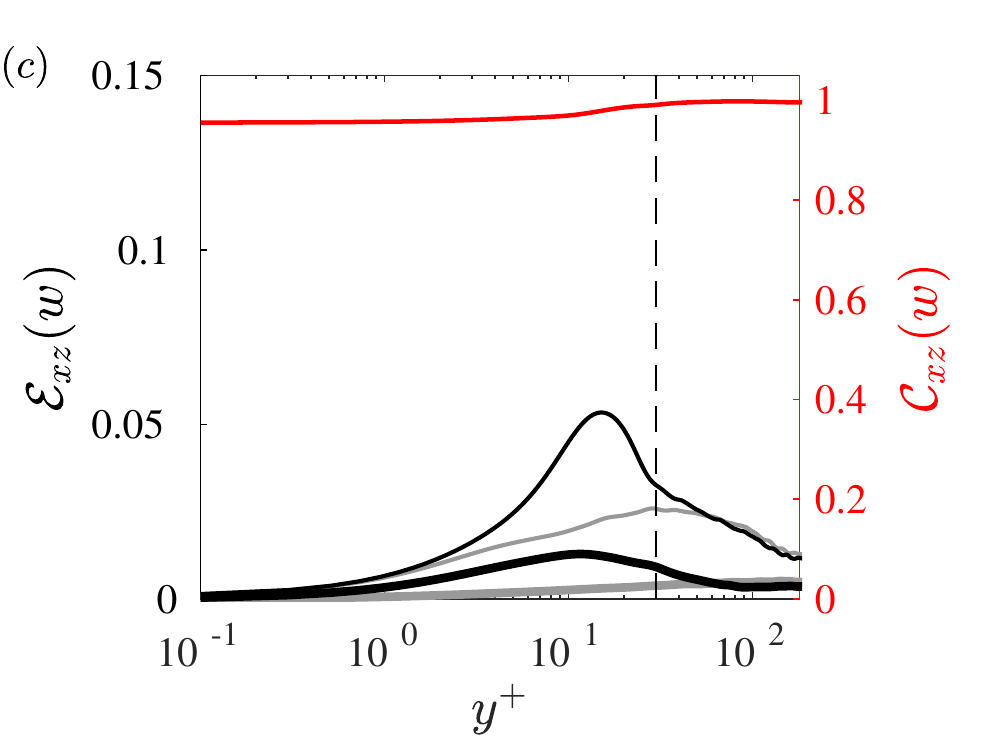}
		}
		\caption{
			Comparison of the estimation errors $\mathcal E_{xz}(q)$ (gray) with and (black) without wall-layer observation data:
			($a$) Streamwise, ($b$) wall-normal and ($c$) spanwise components at (thin) $t=0$ and (thick) $t=T$ ($T^+ = 50$).
			Red: correlation coefficient $\mathcal C_{xz}(q)$ between the true and estimated state without wall-layer observation data at $t=T$.
			Dashed line: $y^+ = 30$.
		}
		\label{fig:wall_error}
	\end{figure}

	\begin{figure}
		\centering
		\subfigure{
			\includegraphics[width=0.35\textwidth]{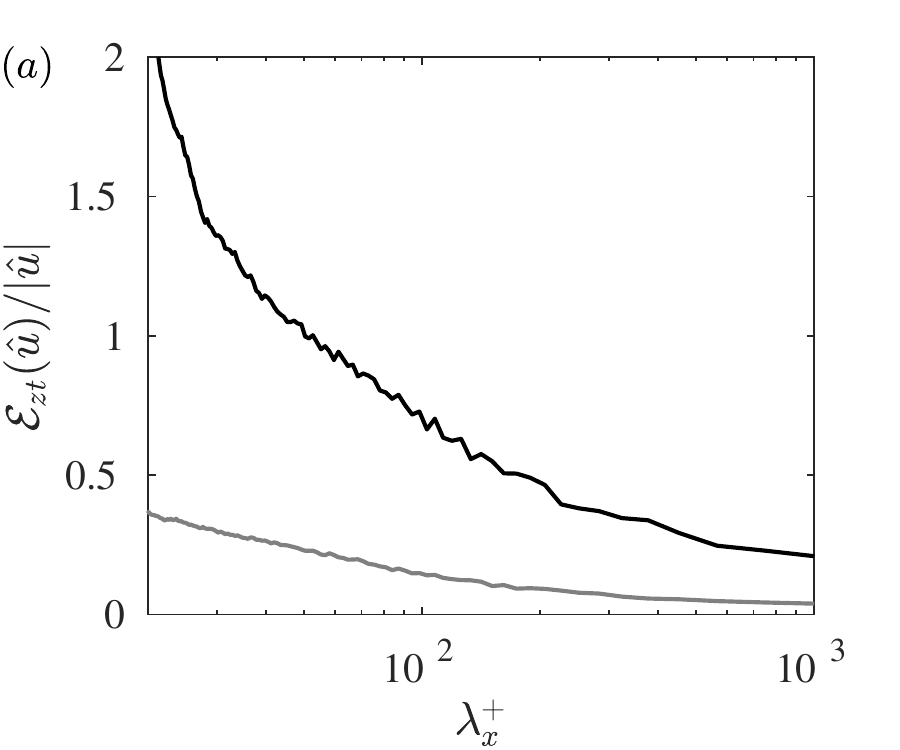}
		}
		\subfigure{
			\hbox{\hspace{0.2in}} 
			\includegraphics[width=0.35\textwidth]{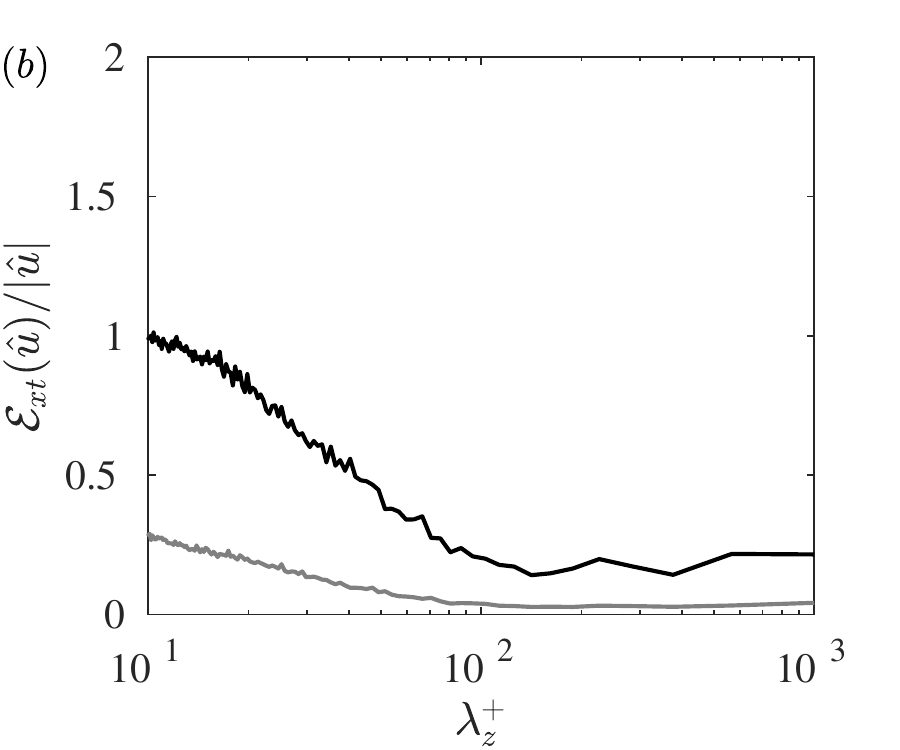}
		}
		\caption{
			One-dimensional Fourier spectra of estimation error at $y^+ = 15$, averaged within $t\in [0.5, 1.5]T$ and normalized by the amplitude of the corresponding Fourier modes: ($a$) streamwise spectrum averaged in the span; ($b$) spanwise spectrum averaged in the streamwise direction. Estimation (\grayline) with and (\lline) without wall-layer observation data.
		}
		\label{fig:wall_spectra}
	\end{figure}

	The predicted statistics without wall-layer observations, evaluated at $t=T$, are plotted in figure \ref{fig:wall_mean} and are compared with the true statistics.
	The shown quantities are horizontally averaged; the corresponding turbulence kinetic energy equation is,
	\begin{equation}
	\label{eq:budget}
		\frac{\partial E}{\partial t} = \underbrace{-\langle u'_i u'_j \rangle \frac{\partial \langle u_i\rangle }{\partial x_j}}_{\mathcal P}  -\frac{\partial }{\partial x_j} \left(\frac{1}{2}\langle u'_i u'_i u_j\rangle + \langle u'_j p\rangle  - \frac{2}{Re}  \langle u'_i s'_{ij} \rangle \right) - \underbrace{\frac{2}{Re} \langle s'_{ij} s'_{ij}\rangle }_{\mathcal D}.
	\end{equation}
	The estimated profiles are accurate, even within the viscous sublayer and buffer layer which are void of observations and where the production and dissipation predominate.
    Although the interpolated initial condition also converges to the true statistics after a long time, our estimated state reaches the statistically stationary state faster, especially for dissipation.
	The errors in the estimated instantaneous velocity fields are shown in figure \ref{fig:wall_error}.
	Compared with the benchmark case (gray lines), the estimation quality in the outer layer is almost unaffected when the wall-layer data are removed. 
	In the near-wall region, however, the estimation error reaches a maximum and exceeds twice the error in the benchmark case.
	The streamwise component is the most poorly reconstructed.
	Without data in the wall layer, the data assimilation algorithm starts with an interpolated velocity field as a first guess, which underestimates the mean flow near the wall; the final prediction of the algorithm retains a similar deficit in the mean, and hence the estimation error is largest in the streamwise component (black lines in figure \ref{fig:wall_error}).
	At $t=T$, the highest estimation error is approximately $4\%$ of the bulk velocity, which is smaller than the $16\%$ local root-mean square streamwise turbulence fluctuations.
	In addition, the correlation coefficient between the true and estimated state at $t=T$ is above 0.8 for all three components of the velocity field.

	The wall layer is host to coherent structures, such as streaks and streamwise vorticies, that play an important role in the dynamics of the near-wall turbulence cycle \citep{Kim_Waleffe_1995} and in flow control. 
	For example, the streak spacing must be resolved by a controller in order to effectively relaminarize turbulent channel flow \citep{Sharma2011}.
	Accurate prediction of these structures is therefore important.  
	The error of our estimation at different lengthscales is reported in figure \ref{fig:wall_spectra}.
	The most accurately reconstructed structures are long in the streamwise direction ($\lambda_x \approx O(L_x)$) and large waves in the span ($\lambda^+_z \in (10^2,10^3)$), which are typical features of the near-wall streaks and streamwise vorticies \citep{Jimenez2018coherent}.
	The estimation error of these structures are mildly affected by removing the wall-layer data, due to their coherence across wall-normal locations, which facilitates the estimation even without near-wall observations. 
	In contrast, reconstruction of small-scale near-wall structures is more sensitive to the lack of observations in that region, which is symptomatic of the weaker sensitivity of the outer flow to these structures.
	
\begin{figure}
		\centering
		\subfigure{
			\includegraphics[width=0.8\textwidth]{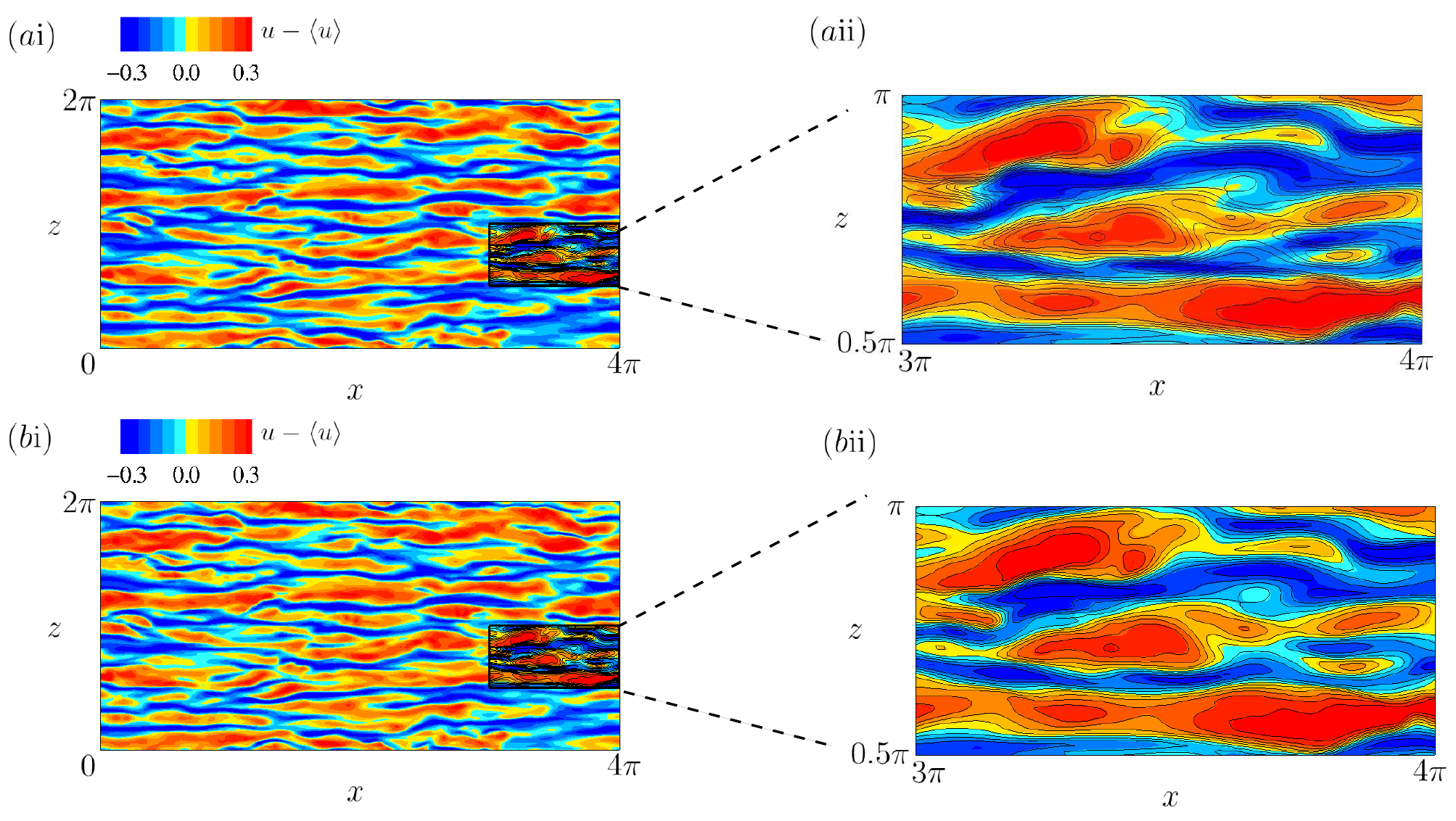}
		}
		\caption{
			Streamwise fluctuations at $y^+ = 15$ and $t=T$ ($T^+ = 50$), calculated by subtracting the true mean velocity. 
			Estimations ($a$) without and ($b$) with wall-layer data;
			Lines: true fluctuation fields. 
		}
		\label{fig:wall_fluct}
	\end{figure}

	Despite the overall increase of estimation errors when wall-layer data are not available, the instantaneous visualization of the coherent structures can still be compelling.
	Figure \ref{fig:wall_fluct} shows a visualization of the predicted instantaneous streamwise fluctuation velocities for two predictions:  without and with observations in the wall layer. 
	The estimated streaky structures without wall-layer observations (panel $a$i) are irregularly spaced in the spanwise direction and meander downstream, as expected from a realistic channel flow. 
	The zoomed in view (panel $a$ii) also includes the true state (line contours), which confirms that the state estimation is predictive:  the true streaks are reproduced.  
	The estimation accuracy is in fact comparable to the case when observations are available in the wall layer (panel $b$).  
	The successful estimation of the missing wall layer is directly tied to the sensitivity of the observations in the outer flow and at the walls to the state in this region at the initial state.
	In other words, accurate reconstruction of the inner layer at the initial time is indispensable to match future observations far from the wall.

	\subsection{Estimation from wall observations}
	\label{sec:Re}
	
	In practice, control of wall-bounded turbulence may rely on sensing and actuation at the wall, and effective strategies are often predicated on decoding the wall signature of outer turbulence structures.  
	It has been demonstrated both mathematically \citep{Lighthill,Constantin2011} and numerically \citep{Eyink2020_theory,Eyink2020_channel} that all the interior vorticity is generated at the wall.
	The converse problem, specifically to what extent the initial flow state can be predicted from wall observations, has not been addressed as comprehensively.
	Previous efforts at $Re_{\tau}=100$, and using a variety of approaches, yielded an estimated state that is nearly uncorrelated with the true flow beyond the buffer layer \citep{Bewley2004,Suzuki2017,Hasegawa2020}.
	The modest Reynolds number in those studies does not support the presence of large-scale outer structures that appreciably influence the wall stress fluctuations \citep{Choi2004}. 
	Linear stochastic estimation (LSE) has been applied at higher $Re_{\tau}$ to reconstruct large-scale structures in the log layer from wall observations \citep{Jimenez2019LSE}. However, the predicted state from LSE does not satisfy the Navier Stokes equations and is not suitable for forecasting \textemdash a further discussion of LSE is provided below.
	Here we focus on adjoint-variational state estimation, specifically on the accuracy of the predicted state and the domain of dependence of wall observations.  The dependence on Reynolds number will be examined for $Re_{\tau} = \{100, 180, 392, 590\}$.
	The computational domain and grid resolution are summarized in table \ref{table:Re}.
	Note that a smaller domain size is adopted for the $Re_{\tau} = \{392, 590\}$ cases due to the limited computational resources.

	The observations are fully-resolved shear stresses $\mathbf{\tau} = (\tau_{xy}, \tau_{yz})$ and pressure $p$ at both walls, similar to those adopted by \citet{Bewley2004} at the lower Reynolds number. 
	The corresponding cost function is defined as,
	\begin{equation}
	\label{eq:cost_fp}
	J(\boldsymbol u^0) = \frac 12 \sum_{n=0}^N \left( \| \boldsymbol {\tau}^n_m - \mathcal  M_{\tau}(\boldsymbol u^n) \|^2_w + 
	\| p^n_m - p^n \|^2_w.\right)
	\end{equation}
	For all the different $Re$ cases, the estimation window is the same in viscous units, $T^+ = 50$,
	which is long enough for perturbation in the bulk region to affect wall signals.
	The first guess of the initial condition is a linear stochastic estimation of the flow using observations at $t=0$.
	The stochastic estimator was constructed with a completely independent time series.
	All the estimated flows are obtained after one hundred L-BFGS iterations of the state-estimation 4DVar algorithm.

	\begin{table}
		\centering
		\begin{tabular}{c c c c c c c c c c c}
			$Re_{\tau}$ & $Re$ & $L_x/h$ & $L_z/h$ & $N_x$ & $N_y$ & $N_z$ & $\Delta x^+$  & $\Delta y^+_{min}$ & $\Delta y^+_{max}$ & $\Delta z^+$ \\
			\hline
			100 & 1429 & $4\pi$ & $2\pi$ & 128 & 128  & 128 & 9.8 & 0.74 & 2.2 & 4.9 \\
			\rowcolor{blue!10}  180 &  2800 & $4\pi$ & $2\pi$ &  384 &  256 &  320 &  5.9 &  0.20 &  3.0 & 3.5\\
			392 & 6875 & $2\pi$ & $\pi$ & 256 & 320 & 192 & 9.6 & 0.34 & 5.1 & 6.4\\
			590 & 10935 & $2\pi$ & $\pi$ & 384 & 384 & 384 & 9.6 & 0.44 & 6.5 & 4.8\\
		\end{tabular}
		\caption{Computational setup of cases with wall observations (\S\ref{sec:Re}).}
		\label{table:Re}
	\end{table}

\begin{figure}
		\centering
		\subfigure{
			\includegraphics[width=0.31\textwidth]{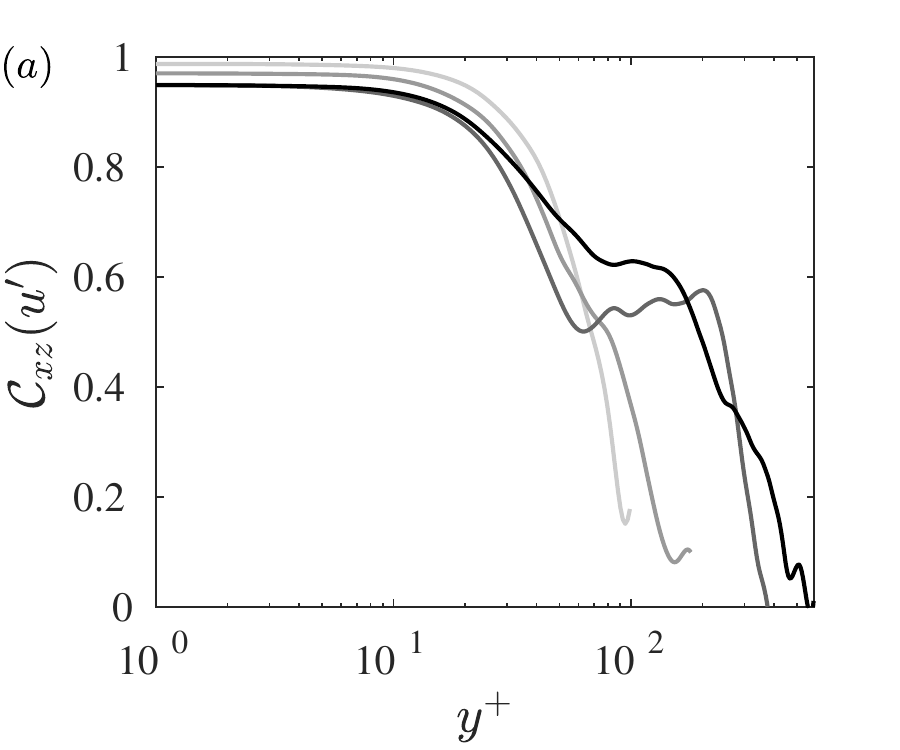}
		}
		\subfigure{
			\includegraphics[width=0.31\textwidth]{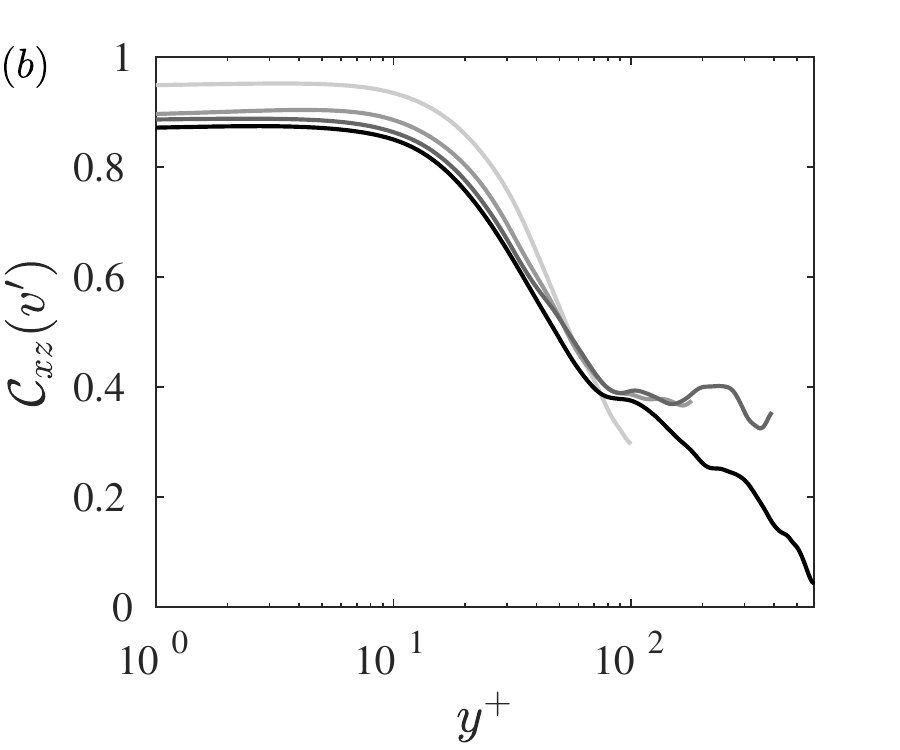}
		}
		\subfigure{
			\includegraphics[width=0.31\textwidth]{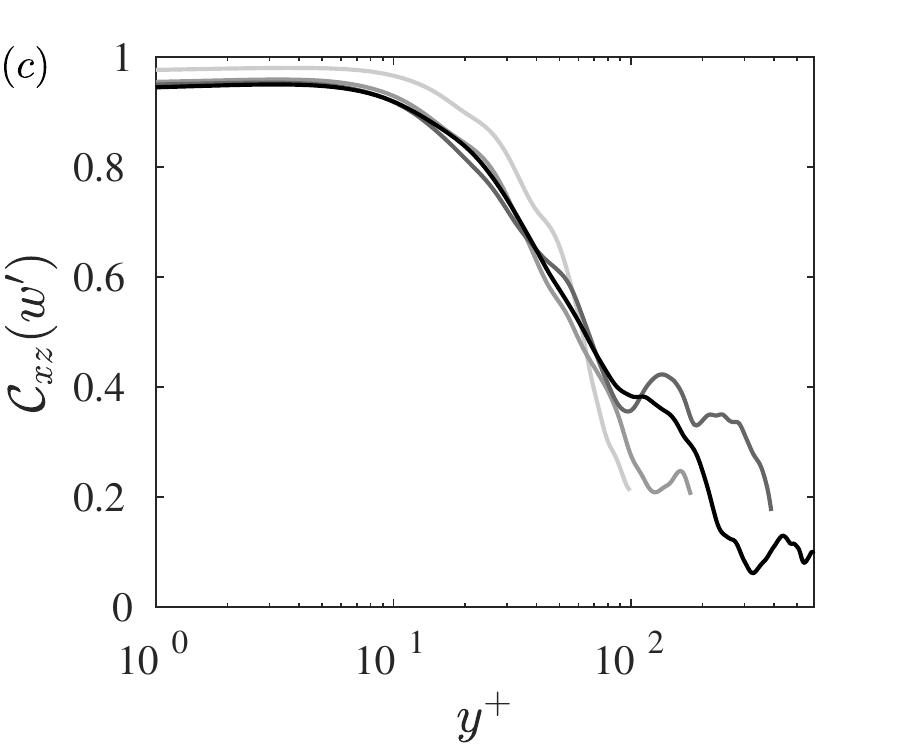}
		}
		\caption{
			Correlation coefficient $\mathcal C_{xz}(q)$ (\ref{eq:cc_y}) between the true and estimated fluctuation fields at $t=T$ ($T^+=50$), evaluated relative to the true mean flow.  
			($a$) Streamwise, ($b$) wall-normal and ($c$) spanwise components. 
			Gray to black: $Re_{\tau}=\{100,180,392,590$\}.
		}
		\label{fig:LSE_Re}
	\end{figure}

	The correlation coefficients of the estimated and true fluctuating velocities at $t=T$ are shown in figure \ref{fig:LSE_Re}.
	The $Re_{\tau} = 100$ results are comparable to those in figure 6$a$ from \citet{Bewley2004} at the same Reynolds number.
	As $Re_{\tau}$ is increased, the correlation near the wall slightly deteriorates but remains sufficiently high to provide confidence in predictions. 
	Precisely, for all Reynolds numbers, $C_{xz} > 0.8$ when $y^+ < 15$.
	The correlations start to decay beyond the buffer layer, and those for $v$ and $w$  (panels $b$\&$c$) do so nearly monotonically and with a similar slope for all $Re_\tau$.  
	This trend highlight the challenge of interpreting turbulent flows from wall observations: 
	the accurately predicted near-wall layer is a diminishingly smaller physical region as the Reynolds number is increased.
	An noteworthy exception is recorded in the correlation coefficients of the streamwise velocity (panel $a$) at $Re_{\tau}=\{392,590\}$: 
	The initial decay outside the buffer layer is followed by a plateau within ($30 < y^+, y < 0.3$) where the reconstruction remains marginally accurate because the large-scale structures in that region superimpose a footprint on wall signals \citep{Choi2004,Marusic2009LSM}.

\begin{figure}
	\centering
	\includegraphics[width=\textwidth]{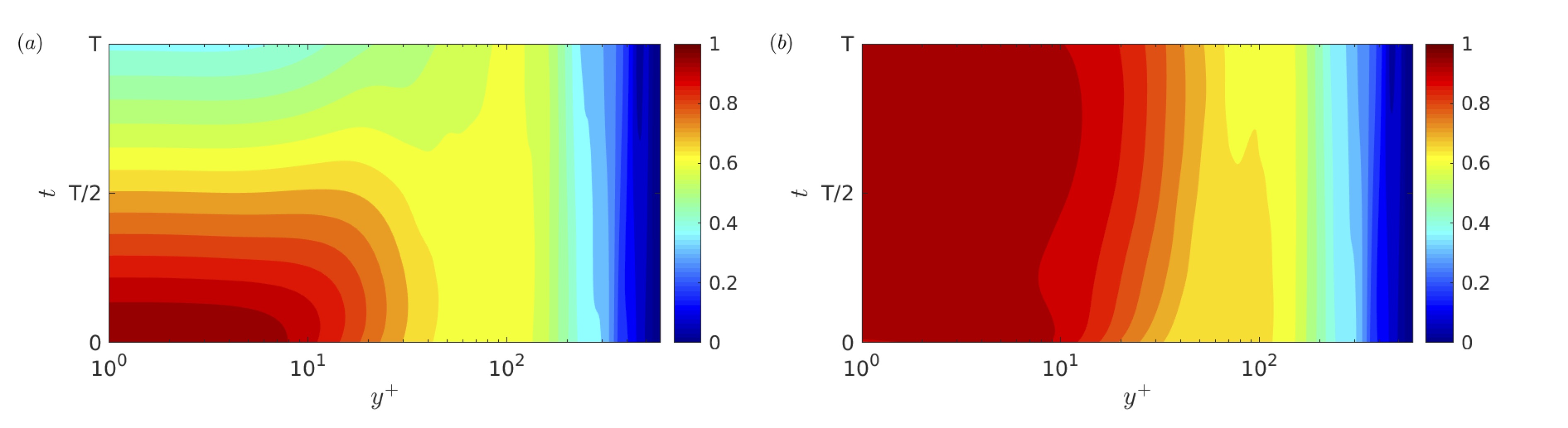}
	\caption{
		Time dependence of the horizontally averaged correlation coefficient $\mathcal C_{xz}(u^{\prime})$ between true state and estimation at $Re_{\tau}=590$ using different strategies:
		($a$) Apply LSE at $t=0$ and advance using Navier-Stokes equations;
		($b$) Adjoint-variational estimation.
	}
	\label{fig:LSE_cc_yt}
    \end{figure}

    A comparison of the adjoint-variational and linear-stochastic estimated initial conditions is provided in figure \ref{fig:LSE_cc_yt}. 
    Both initial states are advanced using the Navier-Stokes equations, and their respective correlation coefficients with the true flow are reported as a function of time. 
    At early times, both accurately reproduce the near-wall layer and outer large-scale $u'$-structures, although the adjoint approach yields slightly better accuracy.
    At later times, the evolution of the LSE state diverges from the true flow, especially in the near-wall region.
    By contrast, the adjoint-variational estimation shadows the true trajectory in the near-wall layer and the large-scale motions in the outer layer.

\begin{figure}
		\centering
		\includegraphics[width=\textwidth]{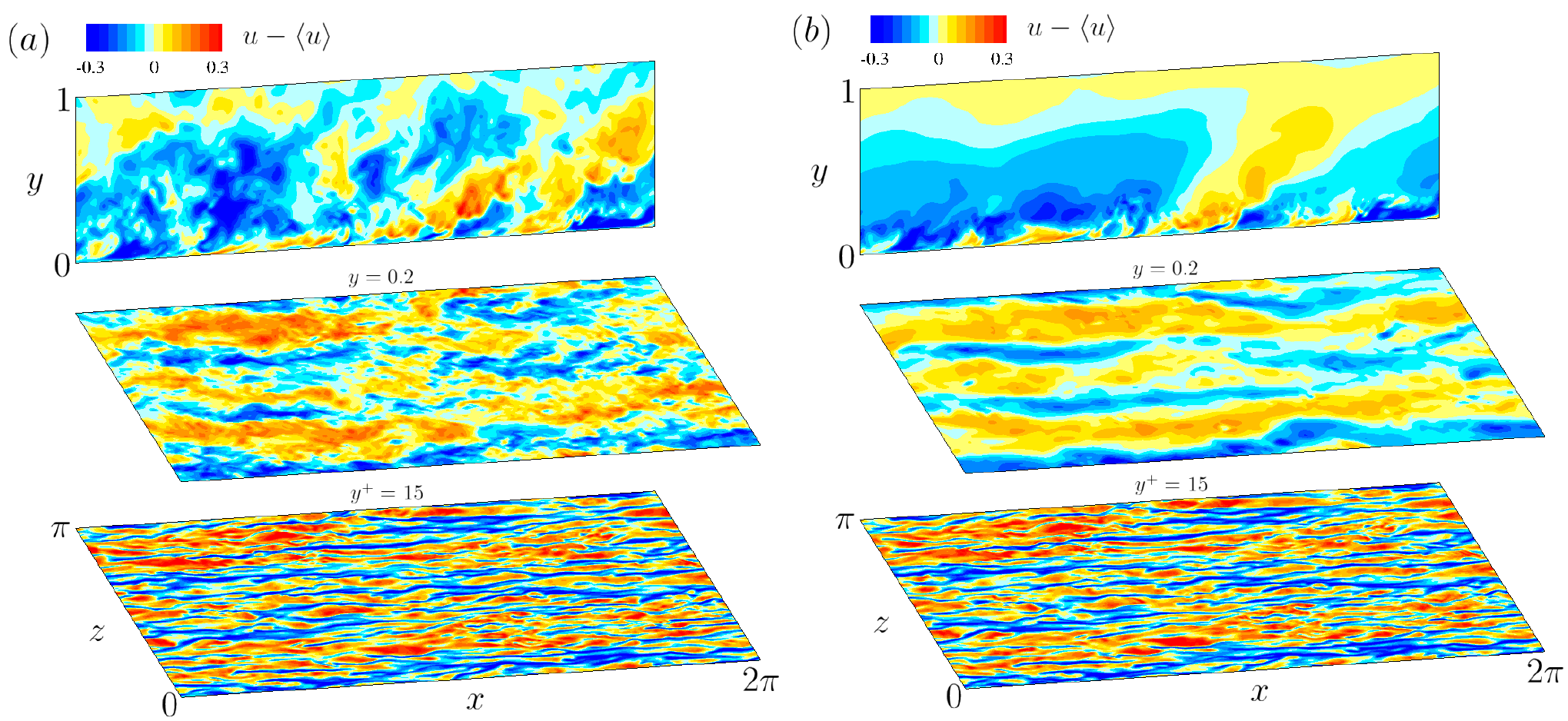}
		\caption{
			Comparison of ($a$) true streamwise fluctuation and ($b$) estimation using wall observations at $t=T$ ($T^+ = 50$), $Re_{\tau}=590$.
		}
		\label{fig:LSEda_u_3D}
	\end{figure}

    Instantaneous views of the perturbation field for $Re_{\tau} = 590$ are provided in figure \ref{fig:LSEda_u_3D}, at $t=T$.  
    In the outer layer, the estimated field matches the mean flow and large-scale motions, while the small-scale fluctuations in the bulk and vorticies that are detached from the wall are not captured.
	These results indicate that the wall signature at these Reynolds numbers is not sensitive to the wall-detached motions in the initial condition\textemdash a demonstration of the inherent difficulty of turbulence reconstruction from wall observations.

\begin{figure}
		\centering
		\includegraphics[width=\textwidth]{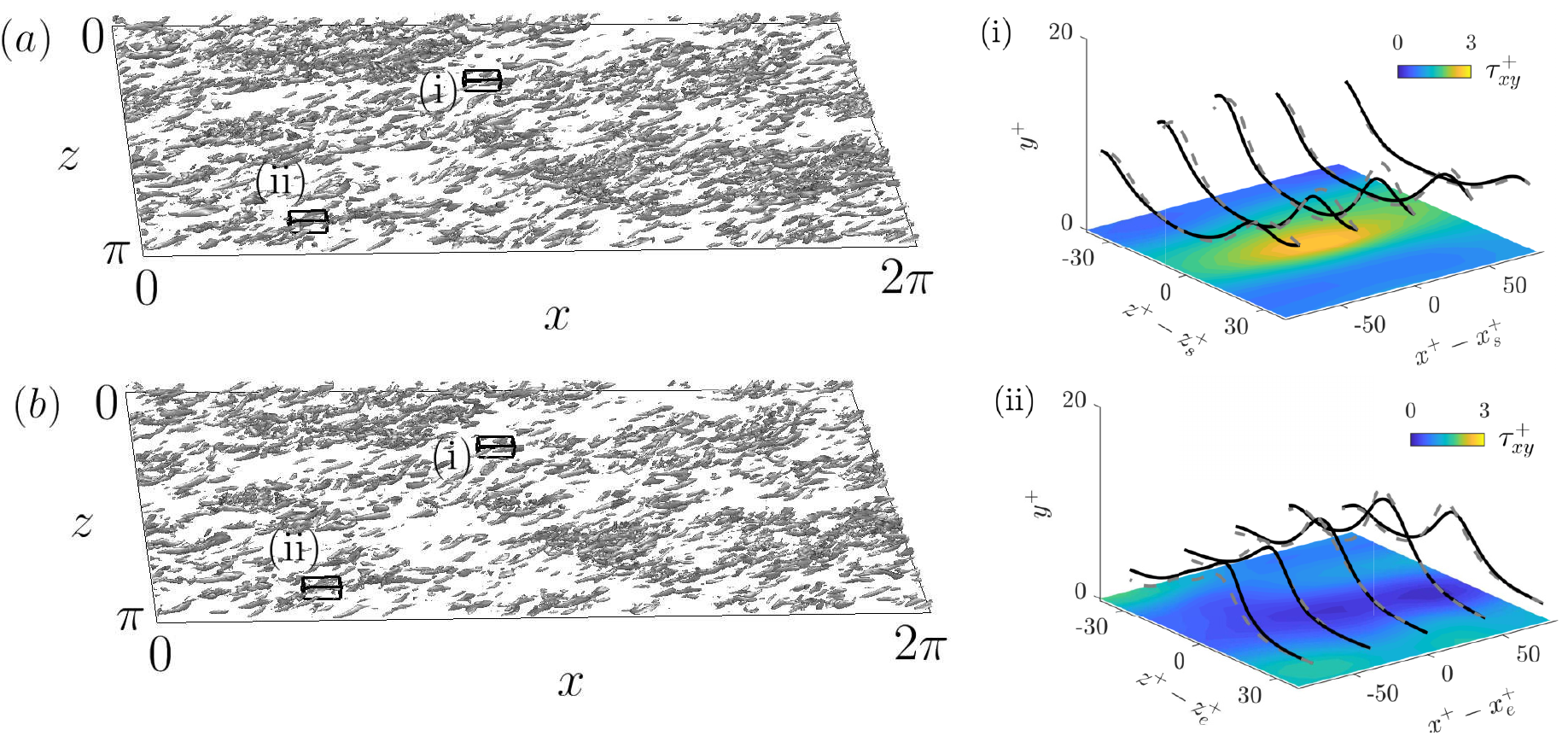}
		\caption{
			($a$) True and ($b$) estimated vortical structures at $t=T$ ($T^+ = 50$) within $y^+ \in [0, 30]$ ($Re_{\tau}=590$), visualized using the $\lambda_2$ vortex-identification criterion with threshold $\lambda_2 = -15$. 
			(i,ii)  (\dashedgray) True and (\lline) estimated vortex lines associated with extreme events: (i) sweep, $(x_s, z_s) = (3.0,0.9)$, vortex lines initiated at $y^+ = 7.6$; (ii) ejection, $(x_e, z_e) = (1.3, 2.8)$, vortex lines initiated at $y^+ = 2.3$. 
		}
		\label{fig:fric_vort_events}
	\end{figure}

	In light of the importance of the near-wall turbulence regeneration cycle, we turn to the detail of the true and estimated states in that region. 
	The near-wall streaks (bottom $x-z$ planes in figure \ref{fig:LSEda_u_3D}) are accurately reconstructed, and many of the vortical structures (gray isosurfaces in figure \ref{fig:fric_vort_events}) are reproduced in the estimated field.
	\citet{Sheng2009} reported that violent near-wall ejections and sweeps contribute appreciably to the local turbulence energy.  
    Examples of both events are shown in panels (i) and (ii), in figure \ref{fig:fric_vort_events}: 
	the true vortex lines above the stress local extrema are closely followed by the estimation, which demonstrates that the extreme events in the buffer layer are encoded in the wall signatures.
	

	The accuracy of the estimated state can be viewed against the backdrop of the near-wall ``autonomous cycle'' \citep{Jimenez1999,Jimenez2007}.
	By artificially removing the outer flow in direct numerical simulations, \citet{Jimenez1999} demonstrated that the near-wall dynamics, especially the regeneration cycle of streaks and vorticies, is self-sustained.  
	And as observed in experiments, ``ejection'' and ``sweep'' events in that cycle have a wall signature in the form of minima and maxima of the wall shear stress \citep{Sheng2009}.
	Here we demonstrate that the wall stress is in fact encoded with the entire dynamics of the regeneration cycle, and our state estimation algorithm decodes these observations to discover the entire flow within the wall layer. The algorithm can also reconstruct the outer large-scale motions when their footprint is encoded in the wall measurements.

	\section{Conclusions}
	\label{sec:conclusion}
	Starting from sparse observations, we attempted to reconstruct the initial state of turbulence in channel flow.  
	The problem was formulated as an adjoint-variational minimization of a cost function that is defined in terms of the difference between the available observations and their estimates from fully-resolved Navier-Stokes simulations. 
	The gradient of the cost function was computed by solving the forward equations and their discrete adjoints, and an L-BFGS algorithm was adopted to update the estimate of the initial state during successive iteration;  this step was supplemented by a symmetric projector that constrains both the search direction and estimated initial state to be solenoidal.

	The performance of the algorithm was evaluated in a benchmark case, where the observations were low-resolution data, at 1/4096 of the required sampling to spatially and temporally resolve the flow in direct numerical simulations (DNS).  The variational state estimation algorithm achieved more than $80\%$ error reduction compared to interpolating the coarse-resolution velocity data, and ensured that the predicted flow not only satisfies the Navier-Stokes equations but also tracks the evolution of the true field in state space over the observation time horizon.
	The estimation errors are initially high wavenumber, and decay within the assimilation time window which was designed to be on the order of the Lyapunov timescale.  The error characteristics were explained in terms of the sensitivity of observations to the initial state.  
	Specifically, observations are insensitive to high-wavenumber content of the initial condition.  In addition, the estimation error decays because the optimization problem is dominated by late observations since any mismatch with available data amplifies exponentially in the adjoint reverse time.  We cautioned, however, that longer time horizons than the Lyapunov timescale would lead to diverging trajectories because small unstable errors in the initial condition would amplify sufficiently and compromise accuracy.

	Using the same benchmark configuration, the observations were contaminated with Gaussian noise and yet the variational state-estimation algorithm successfully reconstructed a noise-free Navier-Stokes solution.
	The correlation between the estimated and true states exceeds 95\%, even when observation noise reaches 10\% of local velocity.
	The vortical structures, which are generally difficult to reproduce from noisy data, were also accurately reconstructed.
	Quantitatively, the error of the estimated vorticity field is within $4\%$ of the wall vorticity.  
	That the noise in the observations was not amplified in the estimated state demonstrates the robustness of the method to reconstruct velocity gradients.

	Criteria for the density of observations in the horizontal plane and in time were identified, and are related to the Taylor microscale of the turbulence.  
	Physically, in order to ensure accurate reconstruction, the separation of observation stations cannot exceed their domain of dependence.
	The criteria are consistent with the one obtained in homogeneous isotropic turbulence \citep{Yoshida2005,Eyink2013,Li2020}, and can accommodate anisotropy of wall turbulence.   
	The presence of mean advection can be exploited to relax the critical streamwise data resolution, when the frequency of temporal sampling can resolve the advected Taylor scale.  
	These criteria were also supplemented with a condition that accounts for divergence of trajectories during the Lyapunov timescale due to the stochasticity of the flow.  
	
	Another important configuration was considered where no observations were available in the wall layer. Instead, observations were only comprised of sub-sampled velocities in the outer flow and wall shear stresses.  
	This test case is to date the first attempt to reconstruct the instantaneous flow field in the wall layer $y^+ < 30$ from such observations and using the full nonlinear Navier-Stokes equations.  
	In spite of the lack of observations in the region of peak turbulence kinetic-energy production, the estimated profiles of flow statistics and the streaky structures in the wall layer were almost indistinguishable from the true state. 
    These results demonstrate the sensitivity of the outer flow and wall shear stress to the turbulence field in the wall layer at the initial state\textemdash a result consistent with our notion that the turbulence produced in that region imprints onto the wall stresses \citep{Sheng2009} and extends and is transported into the outer flow \citep{Jimenez2007}.

	One final configuration that is of both theoretical and practical interest is state estimation from wall observations. 
	From a theoretical perspective, it is known that all the vorticity in the field can be traced back in time to its origin at the wall \citep{Lighthill,Constantin2011,Eyink2020_theory,Eyink2020_channel}; here the converse problem is examined where the wall vorticity is traced back to the initial state of the flow.  
	From a practical perspective, the capacity to control wall turbulence is often reliant on our ability to predict its state from wall measurements.  
	State estimation was performed using fully resolved wall observations of shear stresses and pressure, at four Reynolds numbers.
	The first guess of the initial condition was constructed from a linear stochastic estimation (LSE), and was updated using the iterative adjoint-variational approach.  
    Both the LSE-based initial condition and the outcome of the adjoint algoirthm were compared.
	When the former is advanced using the Navier-Stokes equations, it diverges from the true state especially in the near-wall region.
	By contrast, the initial condition from the adjoint-variational approach reproduces, or shadows, the true trajectories in the near-wall region and also captures the outer large-scale motions when their footprint reaches the wall.
	Despite the accurate estimation of $u'$ large-scale structures in the log layer, overall the estimation quality deteriorates appreciably beyond the buffer layer.
	This deterioration is due to a lack of sensitivity of wall observations to the wall-detached outer structures, which is consistent with the autonomy of the near-wall regeneration cycle that was demonstrated by \citet{Jimenez1999}.
	Finally, it is important to recall that the accurately predicted region below the buffer layer corresponds to a progressively smaller physical height at higher Reynolds numbers which highlights the challenge of estimating wall-bounded turbulence in that regime.


	\bibliographystyle{jfm}
	\bibliography{reference}

\begin{thebibliography}{59}
\expandafter\ifx\csname natexlab\endcsname\relax\def\natexlab#1{#1}\fi
\def\au#1{#1} \def\ed#1{#1} \def\yr#1{#1}\def\at#1{#1}\def\jt#1{\textit{#1}}
  \def\bt#1{#1}\def\bvol#1{\textbf{#1}} \def\vol#1{#1} \def\pg#1{#1}
  \def\publ#1{#1}\def\arxiv#1{#1}\def\org#1{#1}\def\st#1{\textit{#1}}

\bibitem[Abe {\em et~al.\/}(2004)Abe, Kawamura \& Choi]{Choi2004}
{\sc \au{Abe, Hiroyuki}, \au{Kawamura, Hiroshi} \& \au{Choi, Haecheon}}
  \yr{2004}  \at{Very large-scale structures and their effects on the wall
  shear-stress fluctuations in a turbulent channel flow up to
  $\mathit{Re}_{{\it\tau}}$= 640}.  \jt{J. Fluids Eng.}  \bvol{126}~(5),
  \pg{835--843}.

\bibitem[Abrahamson \& Lonnes(1995)]{Abrahamson1995}
{\sc \au{Abrahamson, S} \& \au{Lonnes, S}} \yr{1995}  \at{Uncertainty in
  calculating vorticity from {2D} velocity fields using circulation and
  least-squares approaches}.  \jt{Exp. Fluids}  \bvol{20}~(1),  \pg{10--20}.

\bibitem[Adrian \& Moin(1988)]{Adrian1988LSE}
{\sc \au{Adrian, Ronald~J} \& \au{Moin, Parviz}} \yr{1988}  \at{Stochastic
  estimation of organized turbulent structure: homogeneous shear flow}.
  \jt{J.~Fluid Mech.}  \bvol{190},  \pg{531--559}.

\bibitem[Baars {\em et~al.\/}(2016)Baars, Hutchins \& Marusic]{Baars2016}
{\sc \au{Baars, Woutijn~J.}, \au{Hutchins, Nicholas} \& \au{Marusic, Ivan}}
  \yr{2016}  \at{Spectral stochastic estimation of high-reynolds-number
  wall-bounded turbulence for a refined inner-outer interaction model}.
  \jt{Phys. Rev. Fluids}  \bvol{1},  \pg{054406}.

\bibitem[Bardet {\em et~al.\/}(2010)Bardet, Peterson \&
  Sava{\c{s}}]{Bardet2010PIV}
{\sc \au{Bardet, Philippe~M}, \au{Peterson, Per~F} \& \au{Sava{\c{s}},
  {\"O}mer}} \yr{2010}  \at{Split-screen single-camera stereoscopic piv
  application to a turbulent confined swirling layer with free surface}.
  \jt{Experiments in fluids}  \bvol{49}~(2),  \pg{513--524}.

\bibitem[Bewley \& Protas(2004)]{Bewley2004}
{\sc \au{Bewley, T.~R.} \& \au{Protas, B.}} \yr{2004}  \at{Skin friction and
  pressure: the ``footprints'' of turbulence}.  \jt{Physica D}  \bvol{196},
  \pg{28--44}.

\bibitem[Cerizza {\em et~al.\/}(2016)Cerizza, Sekiguchi, Tsukahara, Zaki \&
  Hasegawa]{Cerizza_Zaki_2016}
{\sc \au{Cerizza, D}, \au{Sekiguchi, W}, \au{Tsukahara, Takahiro}, \au{Zaki,
  TA} \& \au{Hasegawa, Y}} \yr{2016}  \at{Reconstruction of scalar source
  intensity based on sensor signal in turbulent channel flow}.  \jt{Flow,
  Turbulence and Combustion}  \bvol{97}~(4),  \pg{1211--1233}.

\bibitem[Chandramouli {\em et~al.\/}(2020)Chandramouli, M{\'e}min \&
  Heitz]{Chandramouli2020}
{\sc \au{Chandramouli, Pranav}, \au{M{\'e}min, Etienne} \& \au{Heitz,
  Dominique}} \yr{2020}  \at{4{D} large scale variational data assimilation of
  a turbulent flow with a dynamics error model}.  \jt{J.~Comput. Phys.}  \pg{p.
  109446}.

\bibitem[Chevalier {\em et~al.\/}(2006)Chevalier, Hœpffner, Bewley \&
  Henningson]{Bewley_part2}
{\sc \au{Chevalier, Mattias}, \au{Hœpffner, Jérôme}, \au{Bewley, Thomas~R.}
  \& \au{Henningson, Dan~S.}} \yr{2006}  \at{State estimation in wall-bounded
  flow systems. part 2. turbulent flows}.  \jt{J.~Fluid Mech.}  \bvol{552},
  \pg{167–187}.

\bibitem[Colburn {\em et~al.\/}(2011)Colburn, Cessna \& Bewley]{Bewley_part3}
{\sc \au{Colburn, C.H.}, \au{Cessna, J.~B.} \& \au{Bewley, T.~R.}} \yr{2011}
  \at{State estimation in wall-bounded flow systems. part 3. the ensemble
  kalman filter}.  \jt{J.~Fluid Mech.}  \bvol{682},  \pg{289–303}.

\bibitem[Constantin {\em et~al.\/}(2011)Constantin, Iyer {\em
  et~al.\/}]{Constantin2011}
{\sc \au{Constantin, Peter}, \au{Iyer, Gautam} \& \au{others}} \yr{2011}  \at{A
  stochastic-lagrangian approach to the navier--stokes equations in domains
  with boundary}.  \jt{Ann. Probab.}  \bvol{21}~(4),  \pg{1466--1492}.

\bibitem[Deissler(1986)]{Deissler1986chaotic}
{\sc \au{Deissler, RG}} \yr{1986}  \at{Is navier--stokes turbulence chaotic?}
  \jt{Phys. Fluids}  \bvol{29}~(5),  \pg{1453--1457}.

\bibitem[Di~Leoni {\em et~al.\/}(2019)Di~Leoni, Mazzino \& Biferale]{PCDL2019}
{\sc \au{Di~Leoni, P~Clark}, \au{Mazzino, Andrea} \& \au{Biferale, Luca}}
  \yr{2019}  \at{Synchronization to big-data: Nudging the navier-stokes
  equations for data assimilation of turbulent flows}.  \jt{arXiv preprint
  arXiv:1905.05860} .

\bibitem[Dimet \& Talagrand(1986)]{Dimet1986_4dvar}
{\sc \au{Dimet, FranÇois-Xavier~Le} \& \au{Talagrand, Olivier}} \yr{1986}
  \at{Variational algorithms for analysis and assimilation of meteorological
  observations: theoretical aspects}.  \jt{Tellus A: Dyn. Meteorol. Oceanogr.}
  \bvol{38}~(2),  \pg{97--110}.

\bibitem[Encinar \& Jim{\'e}nez(2019)]{Jimenez2019LSE}
{\sc \au{Encinar, Miguel~P} \& \au{Jim{\'e}nez, Javier}} \yr{2019}
  \at{Logarithmic-layer turbulence: {A} view from the wall}.  \jt{Phys. Rev.
  Fluids}  \bvol{4}~(11),  \pg{114603}.

\bibitem[Evensen(1994)]{Evensen1994EnKF}
{\sc \au{Evensen, Geir}} \yr{1994}  \at{Sequential data assimilation with a
  nonlinear quasi-geostrophic model using monte carlo methods to forecast error
  statistics}.  \jt{J. Geophys. Res. Oceans}  \bvol{99}~(C5),
  \pg{10143--10162}.

\bibitem[Eyink {\em et~al.\/}(2020{\natexlab{{\em a\/}}})Eyink, Gupta \&
  Zaki]{Eyink2020_theory}
{\sc \au{Eyink, Gregory~L.}, \au{Gupta, Akshat} \& \au{Zaki, Tamer~A.}}
  \yr{2020{\natexlab{{\em a\/}}}}  \at{Stochastic lagrangian dynamics of
  vorticity. part 1. general theory for viscous, incompressible fluids}.
  \jt{Journal of Fluid Mechanics}  \bvol{901},  \pg{A2}.

\bibitem[Eyink {\em et~al.\/}(2020{\natexlab{{\em b\/}}})Eyink, Gupta \&
  Zaki]{Eyink2020_channel}
{\sc \au{Eyink, Gregory~L.}, \au{Gupta, Akshat} \& \au{Zaki, Tamer~A.}}
  \yr{2020{\natexlab{{\em b\/}}}}  \at{Stochastic lagrangian dynamics of
  vorticity. part 2. application to near-wall channel-flow turbulence}.
  \jt{Journal of Fluid Mechanics}  \bvol{901},  \pg{A3}.

\bibitem[Hamilton {\em et~al.\/}(1995)Hamilton, Kim \&
  Waleffe]{Kim_Waleffe_1995}
{\sc \au{Hamilton, James~M}, \au{Kim, John} \& \au{Waleffe, Fabian}} \yr{1995}
  \at{Regeneration mechanisms of near-wall turbulence structures}.  \jt{J.
  ~Fluid Mech.}  \bvol{287},  \pg{317--348}.

\bibitem[Hutchins {\em et~al.\/}(2005)Hutchins, Hambleton \&
  Marusic]{Hutchins2005}
{\sc \au{Hutchins, N.}, \au{Hambleton, W.~T.} \& \au{Marusic, Ivan}} \yr{2005}
  \at{Inclined cross-stream stereo particle image velocimetry measurements in
  turbulent boundary layers}.  \jt{J. ~Fluid Mech.}  \bvol{541},  \pg{21–54}.

\bibitem[Hœpffner {\em et~al.\/}(2005)Hœpffner, Chevalier, Bewley \&
  Henningson]{Bewley_part1}
{\sc \au{Hœpffner, Jérôme}, \au{Chevalier, Mattias}, \au{Bewley, Thomas~R.}
  \& \au{Henningson, Dan~S.}} \yr{2005}  \at{State estimation in wall-bounded
  flow systems. part 1. perturbed laminar flows}.  \jt{J.~Fluid Mech.}
  \bvol{534},  \pg{263–294}.

\bibitem[Illingworth {\em et~al.\/}(2018)Illingworth, Monty \&
  Marusic]{Illingworth2018}
{\sc \au{Illingworth, Simon~J.}, \au{Monty, Jason~P.} \& \au{Marusic, Ivan}}
  \yr{2018}  \at{Estimating large-scale structures in wall turbulence using
  linear models}.  \jt{J. ~Fluid Mech.}  \bvol{842},  \pg{146–162}.

\bibitem[Jelly {\em et~al.\/}(2014)Jelly, Jung \& Zaki]{Jelly2014}
{\sc \au{Jelly, TO}, \au{Jung, SY} \& \au{Zaki, TA}} \yr{2014}  \at{Turbulence
  and skin friction modification in channel flow with streamwise-aligned
  superhydrophobic surface texture}.  \jt{Phys. Fluids}  \bvol{26}~(9),
  \pg{095102}.

\bibitem[Jeong \& Hussain(1995)]{lambda2}
{\sc \au{Jeong, Jinhee} \& \au{Hussain, Fazle}} \yr{1995}  \at{On the
  identification of a vortex}.  \jt{J.~Fluid Mech.}  \bvol{285},  \pg{69–94}.

\bibitem[Jim{\'e}nez(2018)]{Jimenez2018coherent}
{\sc \au{Jim{\'e}nez, Javier}} \yr{2018}  \at{Coherent structures in
  wall-bounded turbulence}.  \jt{J. ~Fluid Mech.}  \bvol{842}.

\bibitem[Jim{\'e}nez \& Moser(2007)]{Jimenez2007}
{\sc \au{Jim{\'e}nez, Javier} \& \au{Moser, Robert~D}} \yr{2007}  \at{What are
  we learning from simulating wall turbulence?}  \jt{Philosophical Transactions
  of the Royal Society A: Mathematical, Physical and Engineering Sciences}
  \bvol{365}~(1852),  \pg{715--732}.

\bibitem[Jim{\'e}nez \& Pinelli(1999)]{Jimenez1999}
{\sc \au{Jim{\'e}nez, Javier} \& \au{Pinelli, Alfredo}} \yr{1999}  \at{The
  autonomous cycle of near-wall turbulence}.  \jt{J.~Fluid Mech.}  \bvol{389},
  \pg{335--359}.

\bibitem[Jones {\em et~al.\/}(2015)Jones, Heins, Kerrigan, Morrison \&
  Sharma]{Jones2015}
{\sc \au{Jones, Bryn~Ll.}, \au{Heins, P.~H.}, \au{Kerrigan, E.~C.},
  \au{Morrison, J.~F.} \& \au{Sharma, A.~S.}} \yr{2015}  \at{Modelling for
  robust feedback control of fluid flows}.  \jt{J.~Fluid Mech.}  \bvol{769},
  \pg{687--722}.

\bibitem[Jones {\em et~al.\/}(2011)Jones, Kerrigan, Morrison \&
  Zaki]{Jones2011}
{\sc \au{Jones, Bryn~L}, \au{Kerrigan, Eric~C}, \au{Morrison, Jonathan~F} \&
  \au{Zaki, Tamer~A}} \yr{2011}  \at{Flow estimation of boundary layers using
  {DNS}-based wall shear information}.  \jt{Int. J. Control}  \bvol{84}~(8),
  \pg{1310--1325}.

\bibitem[Kim \& Bewley(2007)]{Bewley_review}
{\sc \au{Kim, John} \& \au{Bewley, Thomas~R.}} \yr{2007}  \at{A linear systems
  approach to flow control}.  \jt{Annu. Rev. Fluid Mech.}  \bvol{39}~(1),
  \pg{383--417}.

\bibitem[Kim {\em et~al.\/}(1987)Kim, Moin \& Moser]{Moin_1987}
{\sc \au{Kim, John}, \au{Moin, Parviz} \& \au{Moser, Robert}} \yr{1987}
  \at{Turbulence statistics in fully developed channel flow at low {R}eynolds
  number}.  \jt{J.~Fluid Mech.}  \bvol{177},  \pg{133–166}.

\bibitem[Lalescu {\em et~al.\/}(2013)Lalescu, Meneveau \& Eyink]{Eyink2013}
{\sc \au{Lalescu, Cristian~C}, \au{Meneveau, Charles} \& \au{Eyink, Gregory~L}}
  \yr{2013}  \at{Synchronization of chaos in fully developed turbulence}.
  \jt{Phys. Rev. Lett.}  \bvol{110}~(8),  \pg{084102}.

\bibitem[Lee \& Zaki(2017)]{Sangjin}
{\sc \au{Lee, S.~J.} \& \au{Zaki, T.~A.}} \yr{2017}  \at{Simulations of natural
  transition in viscoelastic channel flow}.  \jt{J.~Fluid Mech.}  \bvol{820},
  \pg{232--262}.

\bibitem[Li {\em et~al.\/}(2020)Li, Zhang, Dong \& Abdullah]{Li2020}
{\sc \au{Li, Yi}, \au{Zhang, Jianlei}, \au{Dong, Gang} \& \au{Abdullah,
  Naseer~S}} \yr{2020}  \at{Small-scale reconstruction in three-dimensional
  kolmogorov flows using four-dimensional variational data assimilation}.
  \jt{J.~Fluid Mech.}  \bvol{885}.

\bibitem[Lighthill(1963)]{Lighthill}
{\sc \au{Lighthill, M.J.}} \yr{1963}  \at{Boundary layer theory}.  \bt{In {\em
  Laminar boundary layers\/} (ed. \ed{L.~Rosenhead})},  \pg{pp. 46--113}.
  \publ{Oxford University Press, Oxford}.

\bibitem[Liu \& Hasegawa(2020)]{Hasegawa2020}
{\sc \au{Liu, Zhuchen} \& \au{Hasegawa, Yosuke}} \yr{2020}  \at{Estimation of
  turbulent channel flow based on time-series wall measurements}.  \jt{SEISAN
  KENKYU}  \bvol{72}~(1),  \pg{5--8}.

\bibitem[Mao {\em et~al.\/}(2013)Mao, Blackburn \& Sherwin]{Mao2013}
{\sc \au{Mao, Xuerui}, \au{Blackburn, Hugh~M} \& \au{Sherwin, Spencer~J}}
  \yr{2013}  \at{Calculation of global optimal initial and boundary
  perturbations for the linearised incompressible navier--stokes equations}.
  \jt{J. Comput. Phys.}  \bvol{235},  \pg{258--273}.

\bibitem[Mao {\em et~al.\/}(2017)Mao, Zaki, Sherwin \& Blackburn]{Mao2017}
{\sc \au{Mao, X.}, \au{Zaki, T.~A.}, \au{Sherwin, S.~J.} \& \au{Blackburn,
  H.~M.}} \yr{2017}  \at{Transition induced by linear and nonlinear
  perturbation growth in flow past a compressor blade}.  \jt{J.~Fluid Mech.}
  \bvol{820},  \pg{604–632}.

\bibitem[Mathis {\em et~al.\/}(2009)Mathis, Hutchins \&
  Marusic]{Marusic2009LSM}
{\sc \au{Mathis, Romain}, \au{Hutchins, Nicholas} \& \au{Marusic, Ivan}}
  \yr{2009}  \at{Large-scale amplitude modulation of the small-scale structures
  in turbulent boundary layers}.  \jt{J.~Fluid Mech.}  \bvol{628},
  \pg{311--337}.

\bibitem[Mons {\em et~al.\/}(2016)Mons, Chassaing, Gomez \& Sagaut]{Mons2016}
{\sc \au{Mons, V.}, \au{Chassaing, J.~C.}, \au{Gomez, T.} \& \au{Sagaut, P.}}
  \yr{2016}  \at{Reconstruction of unsteady viscous flows using data
  assimilation schemes}.  \jt{J. Comput. Phys.}  \bvol{316},  \pg{255--280}.

\bibitem[Mons {\em et~al.\/}(2017)Mons, Chassaing \& Sagaut]{Mons2017sensor}
{\sc \au{Mons, Vincent}, \au{Chassaing, Jean-Camille} \& \au{Sagaut, Pierre}}
  \yr{2017}  \at{Optimal sensor placement for variational data assimilation of
  unsteady flows past a rotationally oscillating cylinder}.  \jt{Journal of
  Fluid Mechanics}  \bvol{823},  \pg{230–277}.

\bibitem[Mor{\'e} \& Thuente(1994)]{linesearch}
{\sc \au{Mor{\'e}, J.~J.} \& \au{Thuente, D.~J.}} \yr{1994}  \at{Line search
  algorithms with guaranteed sufficient decrease}.  \jt{ACM Trans. Math.
  Softw.}  \bvol{20}~(3),  \pg{286--307}.

\bibitem[Naguib {\em et~al.\/}(2010)Naguib, Morrison \&
  Zaki]{Naguib_Morrison_Zaki_2010}
{\sc \au{Naguib, A.M.}, \au{Morrison, J.F.} \& \au{Zaki, T.A.}} \yr{2010}
  \at{On the relationship between the wall-shear-stress and transient-growth
  disturbances in a laminar boundary layer}.  \jt{Phys. Fluids}  \bvol{22}~(5),
   \pg{054103}.

\bibitem[Nikitin(2018)]{Nikitin2018}
{\sc \au{Nikitin, N.}} \yr{2018}  \at{Characteristics of the leading {Lyapunov}
  vector in a turbulent channel flow}.  \jt{J.~Fluid Mech.}  \bvol{849},
  \pg{942--967}.

\bibitem[Nocedal(1980)]{LBFGS}
{\sc \au{Nocedal, J.}} \yr{1980}  \at{Updating {quasi-Newton} matrices with
  limited storage}.  \jt{Math. Comput.}  \bvol{35}~(151),  \pg{773--782}.

\bibitem[Rosenfeld {\em et~al.\/}(1991)Rosenfeld, Kwak \&
  Vinokur]{Rosenfeld1991}
{\sc \au{Rosenfeld, M.}, \au{Kwak, D.} \& \au{Vinokur, M.}} \yr{1991}  \at{A
  fractional step solution method for the unsteady incompressible
  {Navier-Stokes} equations in generalized coordinate systems}.  \jt{J.~Comput.
  Phys.}  \bvol{94},  \pg{102--137}.

\bibitem[Sharma {\em et~al.\/}(2011)Sharma, Morrison, McKeon, Limebeer, Koberg
  \& Sherwin]{Sharma2011}
{\sc \au{Sharma, AS}, \au{Morrison, JF}, \au{McKeon, BJ}, \au{Limebeer, DJN},
  \au{Koberg, WH} \& \au{Sherwin, SJ}} \yr{2011}  \at{Relaminarisation of
  {$Re_{\tau} = 100$} channel flow with globally stabilising linear feedback
  control}.  \jt{Phys. Fluids}  \bvol{23}~(12),  \pg{125105}.

\bibitem[Sheng {\em et~al.\/}(2009)Sheng, Malkiel \& Katz]{Sheng2009}
{\sc \au{Sheng, J}, \au{Malkiel, E} \& \au{Katz, J}} \yr{2009}  \at{Buffer
  layer structures associated with extreme wall stress events in a smooth wall
  turbulent boundary layer}.  \jt{J. ~Fluid Mech.}  \bvol{633},  \pg{17--60}.

\bibitem[Smits {\em et~al.\/}(2011)Smits, McKeon \& Marusic]{Smits2011}
{\sc \au{Smits, Alexander~J.}, \au{McKeon, Beverley~J.} \& \au{Marusic, Ivan}}
  \yr{2011}  \at{High reynolds number wall turbulence}.  \jt{Annu. Rev. Fluid
  Mech.}  \bvol{43}~(1),  \pg{353--375}.

\bibitem[Stuart \& Zygalakis(2015)]{Stuart2015}
{\sc \au{Stuart, Andrew} \& \au{Zygalakis, Kostas}} \yr{2015}  \bt{Data
  assimilation: A mathematical introduction}. {\em Tech. Rep.\/}.  \org{Oak
  Ridge National Lab.(ORNL), Oak Ridge, TN (United States)}.

\bibitem[Suzuki(2012)]{Suzuki2012}
{\sc \au{Suzuki, Takao}} \yr{2012}  \at{Reduced-order kalman-filtered hybrid
  simulation combining particle tracking velocimetry and direct numerical
  simulation}.  \jt{J.~Fluid Mech.}  \bvol{709},  \pg{249–288}.

\bibitem[Suzuki \& Hasegawa(2017)]{Suzuki2017}
{\sc \au{Suzuki, Takao} \& \au{Hasegawa, Yosuke}} \yr{2017}  \at{Estimation of
  turbulent channel flow at {$Re_{\tau}=100$} based on the wall measurement
  using a simple sequential approach}.  \jt{J.~Fluid Mech.}  \bvol{830},
  \pg{760--796}.

\bibitem[Vishnampet {\em et~al.\/}(2015)Vishnampet, Bodony \&
  Freund]{Vishnampet2015}
{\sc \au{Vishnampet, Ramanathan}, \au{Bodony, Daniel~J} \& \au{Freund,
  Jonathan~B}} \yr{2015}  \at{A practical discrete-adjoint method for
  high-fidelity compressible turbulence simulations}.  \jt{J.~Comput. Phys.}
  \bvol{285},  \pg{173--192}.

\bibitem[Wang {\em et~al.\/}(2019{\natexlab{{\em a\/}}})Wang, Wang \&
  Zaki]{Wang2019}
{\sc \au{Wang, Mengze}, \au{Wang, Qi} \& \au{Zaki, Tamer~A}}
  \yr{2019{\natexlab{{\em a\/}}}}  \at{Discrete adjoint of fractional-step
  incompressible navier-stokes solver in curvilinear coordinates and
  application to data assimilation}.  \jt{J.~Comput. Phys.}  \bvol{396},
  \pg{427--450}.

\bibitem[Wang {\em et~al.\/}(2019{\natexlab{{\em b\/}}})Wang, Hasegawa \&
  Zaki]{Wang_hasegawa_zaki_2019}
{\sc \au{Wang, Qi}, \au{Hasegawa, Yosuke} \& \au{Zaki, Tamer~A.}}
  \yr{2019{\natexlab{{\em b\/}}}}  \at{Spatial reconstruction of steady scalar
  sources from remote measurements in turbulent flow}.  \jt{J.~Fluid Mech.}
  \bvol{870},  \pg{316–352}.

\bibitem[Yoshida {\em et~al.\/}(2005)Yoshida, Yamaguchi \& Kaneda]{Yoshida2005}
{\sc \au{Yoshida, Kyo}, \au{Yamaguchi, Junzo} \& \au{Kaneda, Yukio}} \yr{2005}
  \at{Regeneration of small eddies by data assimilation in turbulence}.
  \jt{Phys. Rev. Lett.}  \bvol{94}~(1),  \pg{014501}.

\bibitem[Zaki(2013)]{Zaki2013}
{\sc \au{Zaki, Tamer~A.}} \yr{2013}  \at{From streaks to spots and on to
  turbulence: Exploring the dynamics of boundary layer transition}.  \jt{Flow,
  Turb. \& Comb.}  \bvol{91}~(3),  \pg{451--473}.

\bibitem[Zaki \& Durbin(2005)]{zaki_durbin_2005}
{\sc \au{Zaki, Tamer~A.} \& \au{Durbin, Paul~A.}} \yr{2005}  \at{Mode
  interaction and the bypass route to transition}.  \jt{J.~Fluid Mech.}
  \bvol{531},  \pg{85--111}.

\bibitem[Zaki {\em et~al.\/}(2010)Zaki, Wissink, Rodi \& Durbin]{Zaki2010}
{\sc \au{Zaki, Tamer~A.}, \au{Wissink, Jan~G.}, \au{Rodi, Wolfgang} \&
  \au{Durbin, Paul~A.}} \yr{2010}  \at{Direct numerical simulations of
  transition in a compressor cascade: the influence of free-stream turbulence}.
   \jt{J.~Fluid Mech.}  \bvol{665},  \pg{57--98}.

\end{thebibliography}
	
\end{document}